\documentclass[prd,aps,a4paper,preprintnumbers,superscriptaddress,nofootinbib]{revtex4}
\usepackage{graphicx}
\usepackage{color}
\usepackage{dcolumn}
\usepackage{bm}
\usepackage{slashed}
\usepackage{amsmath}
\usepackage{latexsym}
\usepackage{amssymb}
\usepackage{mathrsfs}
\usepackage{mathtools}
\usepackage{amsfonts}
\usepackage{array,booktabs,longtable}
\usepackage{enumitem}   
\usepackage{physics}
\usepackage{url}
\usepackage{stackengine}
\usepackage{hyperref}

\allowdisplaybreaks
\begin{document}

\preprint{IFT-UAM/CSIC-26-54}

\def\SH{\scriptscriptstyle{\mathrm{SH}}}
\def\MH{\scriptscriptstyle{\mathrm{MH}}}
\def\ADM{\scriptscriptstyle{\mathrm{ADM}}}
\def\EOB{\scriptscriptstyle{\mathrm{EOB}}}
\def\tt{\scriptscriptstyle{\mathrm{TT}}}

\def\Ai{\mathrm{Ai}}
\def\Gi{\mathrm{Gi}}
\def\DD{\mathrm{D}}
\def\calJ{\mathcal{J}}

\def\Cilog{\mathrm{Ci}^{\mathrm{log}}}
\def\Silog{\mathrm{Si}^{\mathrm{log}}}

\def\inner{\mathrm{inner}}
\def\main{\mathrm{main}}
\newcommand{\fPN}[1]{\tiny{{#1}\mathrm{PN}}}
\newcommand{\ivc}[1]{\frac{1}{c^{#1}}}

\newcommand{\idx}[1]{(#1)}

\newcommand{\poch}[1]{(#1)}

\newcommand{\feJ}[1]{\mathrm{J}_{#1}}
\newcommand{\feK}[1]{\mathrm{K}_{#1}}

\newcommand{\fej}[1]{\hat{j}_{#1}}
\newcommand{\fek}[1]{\hat{k}_{#1}}

\def\intV{\mathcal{V}}
\def\tintV{\tilde{\mathcal{V}}}
\def\intU{\mathcal{U}}
\def\tintU{\tilde{\mathcal{U}}}
\def\intW{\mathcal{W}}
\def\tintW{\tilde{\mathcal{W}}}

\def\lgxOxp{\ln\left(\frac{v^2}{x_0'}\right)}
\def\ivdelta{\check\delta}
\def\pr{p_r}
\def\prt{p_{r_*}}
\def\rdot{\dot{r}}
\def\pnvec{\hat{\pmb{n}}}
\def\pvvec{\pmb{v}}
\def\ppvec{\pmb{p}}

\def\vavg{\bar{v}}
\def\eavg{\bar{e}}
\def\lavg{\bar{l}}
\def\lbdavg{\bar{\lambda}}

\def\vpa{\tilde{v}}
\def\epa{\tilde{e}}
\def\lpa{\tilde{l}}
\def\lbdpa{\tilde{\lambda}}

\def\vphi{v_{\phi}}
\def\xom{x_{\omega}}
\def\vom{v_{\omega}}
\def\et{e_{t}}
\def\betae{\beta_e}

\def\SplusPM{\mathrm{S}^{(+)}}
\def\SminusPM{\mathrm{S}^{(-)}}

\def\ii{\mathrm{i}}
\def\ee{\mathrm{e}}
\def\FinitePart{\mathrm{FP}}

\def\Jint{\mathrm{J}^{(0)}}
\def\Jintd{\mathrm{J}^{(\delta)}}
\def\Jintdd{\mathrm{J}^{(\delta^2)}}
\def\Jintddd{\mathrm{J}^{(\delta^3)}}
\def\Jintln{\mathrm{J}^{(\ln)}}
\def\Jintdln{\mathrm{J}^{(\delta\ln)}}
\newcommand{\fJint}[2]{\mathrm{J}^{(#1)}_{#2}}
\newcommand{\fJintln}[2]{\mathrm{K}^{(#1)}_{#2}}
\newcommand{\fJintlnln}[2]{\mathrm{L}^{(#1)}_{#2}}

\newcommand{\fJints}[3]{{}^{(-)}\mathrm{J}^{(#2,#1)}_{#3}}
\newcommand{\fJintc}[3]{{}^{(+)}\mathrm{J}^{(#2,#1)}_{#3}}

\newcommand{\fcjJint}[2]{\bar{\mathrm{J}}^{(#1)}_{#2}}
\newcommand{\fcjJintln}[2]{\bar{\mathrm{K}}^{(#1)}_{#2}}
\newcommand{\fcjJintlnln}[2]{\bar{\mathrm{L}}^{(#1)}_{#2}}

\def\cK{\mathcal{K}}
\def\cJ{\mathcal{J}}
\def\cL{\mathcal{L}}

\def\hypf{[{}_{2}\mathrm{F}_{1}]}
\def\hypff{[{}_{3}\mathrm{F}_{2}]}

\newcommand{\hypfunc}[4]{[{}_{2}\mathrm{F}_{1}]\bigg(\begin{array}{c}#1,\ #2\\#3\end{array};#4\bigg)}

\newcommand{\DHarmN}[2]{\Delta{\mathrm{H}}_{#2}^{#1}}

\newcommand{\fRLIntegral}[3]{{}_{#2}\pmb{\mathcal{I}}^{#1}_{#3} }
\newcommand{\fRLDeriv}[3]{{}_{#2}\pmb{\mathcal{D}}^{#1}_{#3}}
\def\sgn{\mathrm{sgn}}
\newcommand{\eexp}[1]{\mathrm{e}^{#1}}

\newcommand{\fmodexIp}[2]{\mathop{\mathbb{I}}\limits_{\mathclap{(#1)}}\!_{#2}}
\newcommand{\fmodexJp}[2]{\mathop{\mathbb{J}}\limits_{\mathclap{(#1)}}\!_{#2}}

\newcommand{\fmodeyIp}[2]{\mathop{\tilde{\mathbb{I}}}\limits_{\mathclap{(#1)}}\!_{#2}}
\newcommand{\fmodeyJp}[2]{\mathop{\tilde{\mathbb{J}}}\limits_{\mathclap{(#1)}}\!_{#2}}

\newcommand{\fmodexaIp}[3]{\big[{}_{(#3)}\mathop{\mathbb{I}}\limits_{\mathclap{(#1)}}\!_{#2}\big]}
\newcommand{\fmodexaJp}[3]{\big[{}_{(#3)}\mathop{\mathbb{J}}\limits_{\mathclap{(#1)}}\!_{#2}\big]}

\newcommand{\fmodeyaIp}[3]{\big[{}_{(#3)}\mathop{\tilde{\mathbb{I}}}\limits_{\mathclap{(#1)}}\!_{#2}\big]}
\newcommand{\fmodeyaJp}[3]{\big[{}_{(#3)}\mathop{\tilde{\mathbb{J}}}\limits_{\mathclap{(#1)}}\!_{#2}\big]}

\newcommand{\fmodeI}[3]{\underset{(#1,#2)}{\mathrm{I}_{#3}}}
\newcommand{\fmodeIpn}[4]{\underset{(#1,#2)}{\mathrm{I}_{#3}^{#4}}}

\newcommand{\fmodeIp}[2]{\mathop{\mathrm{I}}\limits_{\mathclap{(#1)}}\!_{#2}}

\newcommand{\fmodeIppn}[3]{\mathop{\mathrm{I}}\limits_{\mathclap{(#1)}}\!_{#2}^{#3}}

\newcommand{\fmodeconjIp}[2]{\mathop{\mathrm{I}}\limits_{\mathclap{(#1)}}\!_{#2}^*}

\newcommand{\fmodeconjIppn}[3]{\mathop{\mathrm{I}}\limits_{\mathclap{(#1)}}\!_{#2}^{*,#3}}

\newcommand{\fmodeJ}[3]{\underset{(#1,#2)}{\mathrm{J}_{#3}}}
\newcommand{\fmodeJpn}[4]{\underset{(#1,#2)}{\mathrm{J}_{#3}^{#4}}}

\newcommand{\fmodeJp}[2]{\mathop{\mathrm{J}}\limits_{\mathclap{(#1)}}\!_{#2}}

\newcommand{\fmodeJppn}[3]{\mathop{\mathrm{J}}\limits_{\mathclap{(#1)}}\!_{#2}^{#3}}

\newcommand{\fmodeconjJp}[2]{\mathop{\mathrm{J}}\limits_{\mathclap{(#1)}}\!_{#2}^*}

\newcommand{\fmodeconjJppn}[3]{\mathop{\mathrm{J}}\limits_{\mathclap{(#1)}}\!_{#2}^{*,#3}}

\newcommand{\fmodemIp}[2]{\big[{}_{(1)}\mathop{\mathrm{I}}\limits_{\mathclap{(#1)}}\!_{#2}\big]}
\newcommand{\fmodemnIp}[3]{\big[{}_{(#3)}\mathop{\mathrm{I}}\limits_{\mathclap{(#1)}}\!_{#2}\big]}
\newcommand{\fmodemIppn}[3]{\big[{}_{(1)}\mathop{\mathrm{I}}\limits_{\mathclap{(#1)}}\!_{#2}^{#3}\big]}
\newcommand{\fmodemnIppn}[4]{\big[{}_{(#4)}\mathop{\mathrm{I}}\limits_{\mathclap{(#1)}}\!_{#2}^{#3}\big]}

\newcommand{\fmodeconjmIp}[2]{\big[{}_{(1)}\mathop{\mathrm{I}}\limits_{\mathclap{(#1)}}\!_{#2}^*\big]}
\newcommand{\fmodeconjmnIp}[3]{\big[{}_{(#3)}\mathop{\mathrm{I}}\limits_{\mathclap{(#1)}}\!_{#2}^*\big]}
\newcommand{\fmodeconjmIppn}[3]{\big[{}_{(1)}\mathop{\mathrm{I}}\limits_{\mathclap{(#1)}}\!_{#2}^{*,#3}\big]}
\newcommand{\fmodeconjmnIppn}[4]{\big[{}_{(#4)}\mathop{\mathrm{I}}\limits_{\mathclap{(#1)}}\!_{#2}^{*,#3}\big]}

\newcommand{\fmodemJp}[2]{\big[{}_{(1)}\mathop{\mathrm{J}}\limits_{\mathclap{(#1)}}\!_{#2}\big]}
\newcommand{\fmodemnJp}[3]{\big[{}_{(#3)}\mathop{\mathrm{J}}\limits_{\mathclap{(#1)}}\!_{#2}\big]}
\newcommand{\fmodemJppn}[3]{\big[{}_{(1)}\mathop{\mathrm{J}}\limits_{\mathclap{(#1)}}\!_{#2}^{#3}\big]}
\newcommand{\fmodemnJppn}[4]{\big[{}_{(#4)}\mathop{\mathrm{J}}\limits_{\mathclap{(#1)}}\!_{#2}^{#3}\big]}

\newcommand{\fmodeconjmJp}[2]{\big[{}_{(1)}\mathop{\mathrm{J}}\limits_{\mathclap{(#1)}}\!_{#2}^*\big]}
\newcommand{\fmodeconjmnJp}[3]{\big[{}_{(#3)}\mathop{\mathrm{J}}\limits_{\mathclap{(#1)}}\!_{#2}^*\big]}
\newcommand{\fmodeconjmJppn}[3]{\big[{}_{(1)}\mathop{\mathrm{J}}\limits_{\mathclap{(#1)}}\!_{#2}^{*,#3}\big]}
\newcommand{\fmodeconjmnJppn}[4]{\big[{}_{(#4)}\mathop{\mathrm{J}}\limits_{\mathclap{(#1)}}\!_{#2}^{*,#3}\big]}

\newcommand{\fmodeM}[3]{\underset{(#1,#2)}{\mathrm{M}_{#3}}}
\newcommand{\fmodeMpn}[4]{\underset{(#1,#2)}{\mathrm{M}_{#3}^{#4}}}
\newcommand{\fmodeMp}[2]{\mathop{\mathrm{M}}\limits_{\mathclap{(#1)}}\!_{#2}}
\newcommand{\fmodemnMp}[3]{\big[{}_{(#3)}\mathop{\mathrm{M}}\limits_{\mathclap{(#1)}}\!_{#2}\big]}
\newcommand{\fmodeconjMp}[2]{\mathop{\mathrm{M}}\limits_{\mathclap{(#1)}}\!_{#2}^*}
\newcommand{\fmodeconjmnMp}[3]{\big[{}_{(#3)}\mathop{\mathrm{M}}\limits_{\mathclap{(#1)}}\!_{#2}^*\big]}

\newcommand{\fmodexMp}[2]{\mathop{\mathbb{M}}\limits_{\mathclap{(#1)}}\!_{#2}}
\newcommand{\fmodexaMp}[3]{\big[{}_{(#3)}\mathop{\mathbb{M}}\limits_{\mathclap{(#1)}}\!_{#2}\big]}

\newcommand{\fmodeyMp}[2]{\mathop{\tilde{\mathbb{M}}}\limits_{\mathclap{(#1)}}\!_{#2}}
\newcommand{\fmodeyaMp}[3]{\big[{}_{(#3)}\mathop{\tilde{\mathbb{M}}}\limits_{\mathclap{(#1)}}\!_{#2}\big]}

\newcommand{\fmodeS}[3]{\underset{(#1,#2)}{\mathrm{S}_{#3}}}
\newcommand{\fmodeSp}[2]{\mathop{\mathrm{S}}\limits_{\mathclap{(#1)}}\!_{#2}}
\newcommand{\fmodeSpn}[4]{\underset{(#1,#2)}{\mathrm{S}_{#3}^{#4}}}

\newcommand{\fmodeconjSp}[2]{\mathop{\mathrm{S}}\limits_{\mathclap{(#1)}}\!_{#2}^*}

\newcommand{\avg}[1]{\left\langle{#1}\right\rangle}

\def\Eflux{\mathcal{F}}
\def\Jflux{\mathcal{G}}

\def\head{\mathrm{head}}

\def\MADM{M_{\ADM}}
\def\ege{\gamma_E}
\def\tail{\mathrm{tail}}
\def\ttail{\tail(\tail)}
\def\tttail{\tail(\tail(\tail))}
\def\tailsq{(\tail)^2}
\def\tailcb{(\tail)^3}

\def\Etflux{\mathcal{F}_{\tail}}
\def\Ettflux{\mathcal{F}_{\ttail}}
\def\Etttflux{\mathcal{F}_{\tttail}}
\def\Etsqflux{\mathcal{F}_{\tailsq}}
\def\Etcbflux{\mathcal{F}_{\tailcb}}

\newcommand{\fEtfluxn}[1]{\mathcal{F}_{\tail(#1) }}
\newcommand{\fEtnflux}[1]{\mathcal{F}_{(\tail)^{#1}}}

\def\Jtflux{\mathcal{G}_{\tail}}
\def\Jttflux{\mathcal{G}_{\ttail}}
\def\Jtttflux{\mathcal{G}_{\tttail}}
\def\Jtsqflux{\mathcal{G}_{\tailsq}}
\def\Jtcbflux{\mathcal{G}_{\tailcb}}

\newcommand{\fJtfluxn}[1]{\mathcal{G}_{\tail(#1) }}
\newcommand{\fJtnflux}[1]{\mathcal{G}_{(\tail)^{#1}}}

\def\Bell{\mathrm{B}}
\def\sigPoly{\mathcal{P}}

\newcommand\fehphi[2]{\varphi_{#1}^{(#2)}}
\newcommand\ftehphi[2]{\tilde{\varphi}_{#1}^{(#2)}}

\newcommand\fehbeta[2]{\beta_{#1}^{(#2)}}
\newcommand\ftehbeta[2]{\tilde{\beta}_{#1}^{(#2)}}

\newcommand\fehgamma[2]{\gamma_{#1}^{(#2)}}
\newcommand\ftehgamma[2]{\tilde{\gamma}_{#1}^{(#2)}}

\newcommand\fehF[2]{F_{#1}^{(#2)}}
\newcommand\ftehF[2]{\tilde{F}_{#1}^{(#2)}}

\newcommand\fehchi[2]{\chi_{#1}^{(#2)}}
\newcommand\ftehchi[2]{\tilde{\chi}_{#1}^{(#2)}}

\newcommand{\feh}[2]{\eta_{#1}^{(#2)}}
\newcommand{\fteh}[2]{\tilde\eta_{#1}^{(#2)}}

\newcommand{\cfinttailU}[1]{\mathfrak{i}^{(\ln)}_{#1}}
\newcommand{\cfinttailV}[1]{\mathfrak{j}^{(\ln)}_{#1}}

\newcommand{\cfgm}[1]{{\bar{\pmb{\gamma}}_{#1}}}

\newcommand{\IContr}[2]{\mathrm{C}^{(#1)}_{#2}}

\newcommand{\cgcoeffs}[2]{\mathrm{CG}^{#1}_{#2}}

\newcommand{\fIehlpn}[2]{\varphi^{(#1)}_{#2}}
\newcommand{\ftIehlpn}[2]{\tilde\varphi^{(#1)}_{#2}}

\newcommand{\fJehlpn}[2]{\gamma^{(#1)}_{#2}}
\newcommand{\ftJehlpn}[2]{\tilde\gamma^{(#1)}_{#2}}

\newcommand{\red}[1]{\textcolor{red}{#1}}

\title{Large-Eccentricity Asymptotics and Fast Analytic Approximation \\for Fourier modes of Post-Newtonian Eccentric Waveforms}

\author{Xiaolin Liu}\email[Xiaolin Liu: ]{shallyn.liu@foxmail.com}
\affiliation{Instituto de Física Téorica UAM-CSIC, Universidad Autónoma de Madrid, Cantoblanco 28049 Madrid, Spain}
\author{Zhoujian Cao\footnote{corresponding author}}\email[Zhoujian Cao: ]{zjcao@bnu.edu.cn}
\affiliation{Institute for Frontiers in Astronomy and Astrophysics, Beijing Normal University, Beijing 102206, China}
\affiliation{School of Physics and Astronomy, Beijing Normal University, Beijing 100875, China}
\affiliation{School of Fundamental Physics and Mathematical Sciences, Hangzhou Institute for Advanced Study, UCAS, Hangzhou 310024, China}

\begin{abstract}
In this work, we develop analytic asymptotic methods for computing the Fourier modes of gravitational waves from post-Newtonian binary systems in the quasi-Keplerian parametrization in the high eccentricity regime. 
We also derive the large-eccentricity asymptotic expansion of the eccentricity enhancement function appearing in the tail contributions to the radiation. 
Furthermore, based on these results, we construct an endpoint-constrained analytic approximation that significantly accelerates the computation of the Fourier modes at large eccentricity.
The overall error of this analytic approximation is controlled within $10^{-3}$, and it remains valid for Fourier modes with $p\le200$ and $e\le 0.9$.
This approach provides analytic building blocks for modeling frequency-domain gravitational waves from highly eccentric binaries.
\end{abstract}

\maketitle

\section{Introduction}
Since the first direct detection of gravitational waves (GWs) from the binary-black-hole merger GW150914 in 2015~\cite{1602.03837}, gravitational-wave astronomy has developed into a precision observational science. Over the past decade, the LIGO-Virgo-KAGRA network has accumulated an increasingly rich catalog of compact-binary signals, including binary black holes, binary neutron stars, and neutron-star--black-hole mergers~\cite{1710.05832,GWTC_2,2111.03606,2508.18082}. In parallel, pulsar-timing-array observations have reported evidence for a nanohertz stochastic gravitational-wave background, opening a complementary low-frequency window onto the gravitational-wave universe~\cite{2306.16213,2306.16214,2306.16215,2306.16216}. Looking ahead, the next generation of ground- and space-based detectors---including the Einstein Telescope~\cite{2503.12263}, Cosmic Explorer~\cite{2109.09882}, LISA~\cite{1702.00786}, TianQin~\cite{Tianqin}, Taiji~\cite{1807.09495}, and decihertz missions such as DECIGO/B-DECIGO~\cite{1802.06977}---will greatly expand the accessible source population and improve the precision of source characterization.

Coalescing compact binaries are among the primary sources of GWs, and theoretical waveform templates play a crucial role in GW data analysis. The construction of accurate waveform models relies on several complementary approaches, including the post-Newtonian (PN) expansion~\cite{blanchet_PNReview}, the post-Minkowskian (PM) framework~\cite{1609.00354,2405.19181}, black-hole perturbation theory and gravitational self-force methods~\cite{1805.10385,2101.04592}, together with extensive numerical-relativity simulations~\cite{1605.03204,1703.03423,SXS_2019}. Based on these ingredients, a number of fast and accurate waveform families have been developed, among which the IMRPhenom family~\cite{0704.3764,1508.07253,2001.11412,2004.06503} is one of the most widely used. 
More recently, \texttt{IMRPhenomXODE}~\cite{2306.08774} improved the treatment of spin precession, while \texttt{IMRPhenomXE}~\cite{IMRPhenomXE} further extended the framework to eccentric aligned-spin binaries, \texttt{IMRPhenomXO4a}~\cite{Thompson:2023ase} and \texttt{IMRPhenomXPNR}~\cite{Hamilton:2025xru} provide significant improvements in the modeling of spin precession. 
Another major class of waveform models is based on the effective-one-body (EOB) formalism~\cite{Buonanno_EOB_1999,Buonanno_2000}, which has been substantially advanced in recent years through the \texttt{SEOBNRv5}~\cite{2303.18039,EOBv5PHM,EOBv5EHM} and \texttt{TEOBResumS}~\cite{1806.01772,2503.14580} families.

These developments make the problem of eccentric frequency-domain modeling particularly relevant for future low-frequency observations. Next-generation detectors will probe lower-frequency bands, where compact binaries remain in band for much longer and may still retain significant orbital eccentricity. However, the current status of frequency-domain eccentric modeling remains limited. Among recent frequency-domain eccentric models, \texttt{pyEFPE}~\cite{Morras_2025,Morras:2026fho} improves the treatment of moderately large eccentricities by exploiting a Fourier--Bessel representation of the inspiral waveform, but it remains restricted to an inspiral-only PN framework. By contrast, \texttt{IMRPhenomXE}~\cite{IMRPhenomXE} incorporates eccentric corrections based on waveform information expanded in small eccentricity, and therefore loses accuracy in the large-eccentricity regime.

In our previous work~\cite{work1}, we introduced a set of integral representations for the Fourier modes of post-Newtonian eccentric waveforms within the quasi-Keplerian framework. These integrals generalize the usual Bessel-function structure and provide an effective way to overcome the loss of accuracy suffered by small-eccentricity expansions in the high-eccentricity regime. However, their direct numerical evaluation remains computationally expensive, especially when the eccentricity becomes large. 
The problem is not merely the absence of closed forms, but the lack of controlled asymptotics in the joint high-eccentricity and high-harmonic regime.
In this work, we derive their analytic expansions in the large-eccentricity, low-frequency regime, and convert them into a practically usable approximation for frequency-domain waveforms generation that is reliable across a broad eccentricity range.

The remainder of this paper is organized as follows. 
In Sec.~\ref{sec_Review}, we review the construction of post-Newtonian frequency-domain gravitational-wave templates. 
In Sec.~\ref{sec_LEA}, we develop two methods for deriving the large-eccentricity expansions of the integrals introduced in Ref.~\cite{work1}. 
In Secs.~\ref{sec_LEA_gw} and \ref{sec_LEA_eef}, we illustrate the applications of these expansions to the gravitational-wave waveform and to the tail contribution to the gravitational-wave flux, respectively.

\section{Review of the theory of post-Newtonian frequency-domain gravitational waveform}\label{sec_Review}
The construction of post-Newtonian frequency-domain gravitational-wave templates mainly relies on two techniques: the quasi-Keplerian parametrization~\cite{Damour_1985_1PN,Damour_1988_RPA,Schafer_1993_2PN,Memmesheimer_2004_QK} and the stationary phase approximation (SPA)~\cite{Yunes_2009,Klein_2013}.

The post-Newtonian (PN) approximation~\cite{RevModPhys.52.299,blanchet1986radiative,blanchet_PNReview} provides a systematic method for solving the Einstein field equations in the weak-field, slow-motion regime. 
In this framework, the equations of motion are expanded as a series in powers of $c^{-1}$, and the acceleration can be decomposed into a conservative part $\pmb{a}_{\mathrm{cons}}$ and a $\pmb{a}_{\mathrm{diss}}$ part,
\begin{align}
\pmb{a} = \pmb{a}_{\mathrm{cons}} + \pmb{a}_{\mathrm{diss}}.
\end{align}
Within the multipolar post-Minkowskian (MPM) framework, the gravitational waveform in the radiation zone can also be expressed in terms of the same set of orbital variables $(r,v)$. 
The quasi-Keplerian parametrization instead describes the orbit in terms of quantities such as the eccentricity $e$ and the semi-major axis $a$, which provide a more transparent physical description of eccentric motion, especially from the viewpoint of the adiabatic approximation.

We write the orbital separation in the center-of-mass frame as
\begin{align}
& r = a_r(1-e_r\cos{\chi}),
\end{align}
where $a_r$ and $e_r$ denote the semi-major axis and the radial eccentricity, respectively. 
They are obtained by solving $\dot{r}:=\mathrm{d}r/\mathrm{d}t=0$. 
For conservative motion, both quantities can be expressed in terms of the orbital energy $E$ and angular momentum $J$. 
The angle $\chi$ is the eccentric anomaly.

For a conservative eccentric orbit, the variable $\chi$ does not increase uniformly with time. 
A more convenient angular variable is the mean anomaly $l:=2\pi(t-t_0)/P$, where $P$ is the orbital period. 
The variables $l$ and $\chi$ are related through the Kepler equation,
\begin{align}
& l = \chi - e_t\sin{\chi} + \order{c^{-2}},
\end{align}
where the new parameter $e_t$ is referred to as the time eccentricity. 
The orbital phase $\phi$ can be expressed as
\begin{align}
& \phi = \lambda + W,
\end{align}
where $\lambda:=(1+k)l$ and $k$ is the periastron advance, defined as the excess of the orbital phase over $2\pi$ during one radial period:
\begin{align}
& 1+k = \frac{1}{2\pi}\int_0^{2\pi} \dv{\phi}{l}\dd{l}.
\end{align}
It is also convenient to introduce the PN expansion parameter
\begin{align}
& v := \qty(\frac{2\pi(1+k)}{Pc^3})^{1/3},
\end{align}
which we will use in the following to label PN orders instead of $c^{-1}$.
The function 
$W$ can be written as
\begin{align}
& W = \delta_\chi + e_t\sin{\chi} + \order{v^{2}},
\end{align}
with
\begin{align}
& \delta_\chi = 2\arctan{\bigg( \frac{\betae\sin{\chi} }{1-\betae\cos{\chi} } \bigg)},
\end{align}
where $\betae:=(1-\sqrt{1-e^2})/e$.

Using these parameter transformations, the gravitational waveform $h(r,\dot{r},\phi)$ obtained from PN-MPM theory can be rewritten as $h(v,e,l)$, where the eccentricity $e$ may be chosen as either $e_t$ or $e_r$, depending on convenience. 
Different choices lead to different explicit forms, and in the following we will omit the eccentricity subscript whenever no confusion arises. 
Formally, the waveform can then be expanded as a Fourier series,
\begin{align}
& h(v,e,l) = \sum_{m,p} h_{mp}(v,e) \ee^{\ii (m\lambda + pl)},
\end{align}
where the Fourier coefficients $h_{mp}$ are given by
\begin{align}
& h_{mp}(v,e) = \frac{1}{(2\pi)^2}\iint_{-\pi}^\pi h(v,e,l)\ee^{-\ii(m\lambda + pl)} \dd{l}\dd{\lambda}. \label{eq_FourierExpansionOfGW}
\end{align}

The discussion above applies only to conservative dynamics. Once radiation reaction is included, one must take into account the post-adiabatic corrections to the evolution of these orbital parameters~\cite{Pound_2008,Miller_2021}. 
The dissipative force is sourced by the gravitational-wave energy flux $\Eflux$ and angular-momentum flux $\Jflux$, which, upon orbital averaging, are equal to the losses $\dot{E}$ and $\dot{J}$ of the orbital energy and angular momentum. 
As a result, the parameters $(v,e)$ are no longer conserved, but instead evolve secularly. 
Within the adiabatic approximation, this evolution is non-oscillatory at leading order, and one usually decomposes the orbital elements as
\begin{align}
    v = \bar{v} + \tilde{v},\ e=\bar{e} + \tilde{e},
\end{align}
while the phases are similarly written as
\begin{align}
    l = \bar{l} + \tilde{l},\ \lambda=\bar{\lambda} + \tilde{\lambda}.
\end{align}
The oscillatory pieces $(\tilde{v},\tilde{e},\tilde{l},\tilde{v})$ depend on the specific form of the radiation-reaction force, and there is some freedom in how this force is chosen~\cite{Bini_2012,Fumagalli_2025}. 
Once a particular prescription is fixed, these oscillatory contributions can be expressed in terms of the secular variables~\cite{K_nigsd_rffer_2006,Boetzel_2019}.

Because the secular phases grow approximately linearly in time, the SPA can be applied to obtain analytic Fourier transforms of the individual modes $h_{mp}\ee^{\ii (m\lambda+pl)}$,
\begin{align}
& \tilde{h}_{mp}(v,e,f) = \int h_{mp}(v,e)\ee^{\ii (m\lambda + pl - 2\pi ft)}\dd{t}.
\end{align}
Expanding the phase
\begin{align}
& \phi_{mp}:=m\lambda + pl - 2\pi ft
\end{align}
around the stationary point $t_{mp}$, one has
\begin{align}
& \phi_{mp}(t)\approx\phi_{mp}(t_{mp})
+ \frac{1}{2}(t-t_{mp})^2\ddot\phi_{mp},
\end{align}
where the stationary point is determined by the condition $\dot{\phi}_{mp}(t_{mp})=0$. This yields the relation between the secular PN parameter $\bar{v}$ and the Fourier frequency $f$. 
Combined with the adiabatic evolution law relating $v$ and $e$~\cite{PhysRev.136.B1224}, one may then obtain an approximate analytic frequency-domain waveform.

The main difficulty, however, is that the Fourier coefficients $h_{mp}(v,e)$ have a highly nontrivial dependence on the eccentricity. 
In~\cite{work1}, we derived the Fourier coefficients of the leading-order spherical-harmonic modes $h_{\ell m}$, where the polarization waveform is decomposed in spin-weighted $-2$ spherical harmonics~\cite{Blanchet_2008},
\begin{align}
h_+-\ii h_\times = \sum_{l=2}^\infty \sum_{m=-\ell}^\ell h_{\ell m} {}_{-2}Y_{\ell m}.
\end{align}
The Fourier coefficients obtained by expanding $h_{\ell m}(r,\dot{r},\phi)$ according to Eq.~(\ref{eq_FourierExpansionOfGW}) depend on several integrals that cannot be evaluated analytically in closed form.
For example, for the leading $(\ell,m)=(2,2)$ mode, the waveform takes the form up to 1PN order
\begin{align}
    & h_{22} = \sqrt{\frac{\pi}{5}}\frac{4\nu v^2}{c^4R}\eexp{\ii 2(l-\lambda)}\sum_{p=-\infty}^\infty \Big( \tilde{H}^{N}_{2(-2)p} + v^2\tilde{H}_{2(-2)p}^{\fPN{1}} + \order{v^3} \Big)\eexp{\ii p l},\label{eq_MHh22} \\
    & \tilde{H}^{N}_{2(-2)p} = -\frac{2 \left(1-e^2\right)^{3/2} p-e^2+2}{e^2}J_p(ep) + \frac{2 \sqrt{1-e^2}}{e}J'_p(ep) + \bigg( 2-\frac{2}{e^2} \bigg)\feJ{(p,1,0)} + \frac{2 \left(e^2-1\right)^2}{e^2} \feJ{(p,2,0)},\\
    & \tilde{H}_{2(-2)p}^{\fPN{1}} = \frac{1}{42pe^2(1-e^2)}\Big[ -504 \left(e^2-1\right)^2 p^2+252 \left(e^2-2\right)-3 \left(19e^4+195 e^2-242\right) p \nonumber\\
    &\qquad +2\sqrt{1-e^2} \left(22 e^4-387 e^2+365\right) p^2 \Big]J_p(ep) + \frac{1}{7pe(1-e^2)}\Big[ \sqrt{1-e^2} \left(\left(37 e^2-121\right) p+84\right) \nonumber\\
    &\qquad -42 \left(e^2-2\right) p \Big]J'_{p}(ep) -\frac{12 p}{e^2 \sqrt{1-e^2}}\feJ{(p,-2,1)} + \frac{6(2-e^2)}{e^2(1-e^2)}\feJ{(p,-1,1)} -\frac{6 \left(2 \left(1-e^2\right)^{3/2} p-e^2+2\right)}{e^2\left(1-e^2\right)}\feJ{(p,0,1)} \nonumber\\
    &\qquad + \frac{1}{21pe^2\sqrt{1-e^2}} \Big[ -124 \left(e^2-1\right)^2 p^2+252 \left(e^2-1\right)p+\sqrt{1-e^2} \left(\left(113-23 e^2\right) p+252\right)-126\left(e^2-2\right) \Big]\feJ{(p,1,0)} \nonumber\\
    &\qquad + \frac{1}{21pe^2\sqrt{1-e^2}}\Big[ 2 \left(e^2-1\right)^3 p^2+252 \left(e^2-1\right)^2 p+126\left(e^2-2\right)+\sqrt{1-e^2} \left(\left(-19 e^4+884e^2-865\right) p-252\right) \Big]\feJ{(p,2,0)} \nonumber\\
    &\qquad + 12 \left(\frac{1}{e^2}-1\right)\feJ{(p,2,1)} + \frac{242 \left(e^2-1\right)^2}{21 e^2}\feJ{(p,3,0)} -\frac{2 \left(e^2-1\right)^3}{7 e^2} \feJ{(p,4,0)} + \frac{6 \left(\sqrt{1-e^2}+1\right)^2}{e \sqrt{1-e^2} p}\partial_e\feJ{(p,1,0)} \nonumber\\
    &\qquad + \nu \bigg[ -\frac{e^2 \left(22 \sqrt{1-e^2} p+17\right)+2 \left(73\sqrt{1-e^2} p+67\right)}{42 e^2}J_p(ep) + \frac{67-25 e^2}{21 e \sqrt{1-e^2}}J'_p(ep) \nonumber\\
    &\qquad -\frac{6 \left(1-e^2\right)^{3/2} p+8 e^2+73}{21 e^2} \feJ{(p,1,0)} + \frac{\left(e^2-1\right) \left(e^2 \left(6 \sqrt{1-e^2}p-20\right)-6 \sqrt{1-e^2} p-61\right)}{21 e^2} \feJ{(p,2,0)} \nonumber\\
    &\qquad + \frac{10 \left(e^2-1\right)^2}{7 e^2}\feJ{(p,3,0)} + \frac{6 \left(e^2-1\right)^3}{7 e^2}\feJ{(p,4,0)} \bigg],
\end{align}
where $\nu:=m_1m_2/(m_1+m_2)^2$ is the mass ratio, and one encounters the quantities $\feJ{(p,a,b)}$, defined by
\begin{align}
& \feJ{(p,a,b)}(e) := \frac{1}{2\pi}\int_{-\pi}^\pi \frac{\big(\ii\delta_{\chi}(x,e)\big)^b}{(1-e\cos{x})^a}\ee^{\ii p(x-e\sin{x})} \dd{x}. \label{eq_PNEllipticJ}
\end{align}
At 3PN order, another class of logarithmic integrals also appears,
\begin{align}
& \feK{(p,a,b)}(e) := \frac{1}{2\pi}\int_{-\pi}^\pi \frac{\big(\ii\delta_{\chi}(x,e)\big)^b}{(1-e\cos{x})^a}\ee^{\ii p(x-e\sin{x})}\ln{(1-e\cos{x})} \dd{x}. \label{eq_PNEllipticK}
\end{align}

We summarize all elliptic integrals appearing within 3PN order in TABLE~\ref{tab_Integrals}.
\renewcommand{\arraystretch}{1.2}
\setlist[itemize]{nosep,leftmargin=*} 
\begin{longtable}{|c|c|}
\caption{The list of elliptic integrals appearing in each PN order.}\label{tab_Integrals} \\

\toprule
\hline
Order & Integrals \\
\hline
\midrule
\endfirsthead



\bottomrule
\hline
\endlastfoot

$\order{v^0}$ & 
$\begin{array}{ll}
	\feJ{(p,1,0)}(e) & \feJ{(p,2,0)}(e)
\end{array}$ \\
\hline
$\order{v^1}$ & 
$\begin{array}{ll}
	\feJ{(p,3,0)}(e) & \feJ{(p,4,0)}(e)
\end{array}$ \\
\hline
$\order{v^2}$ & 
$\begin{array}{lll}
	\feJ{(p,-1,0)}(e) 	& \feJ{(p,5,0)}(e)    		& \feJ{(p,6,0)}(e) \\
    \feJ{(p,-2,1)}(e)   & \feJ{(p,-1,1)}(e) 	    & \feJ{(p,0,1)}(e) \\ 
    \feJ{(p,1,1)}(e)	& \feJ{(p,2,1)}(e)          & \partial_e\feJ{(p,1,0)}(e)
\end{array}$\\
\hline
$\order{v^3}$ & 
$\begin{array}{llll}
	\feJ{(p,7,0)}(e) 		& \feJ{(p,8,0)}(e)    		& \feJ{(p,3,1)}(e) 		& \feJ{(p,4,1)}(e)
\end{array}$\\
\hline
$\order{v^4}$ & 
$\begin{array}{llll}
	\feJ{(p,-2,0)}(e)		& \feJ{(p,9,0)}(e) 				& \feJ{(p,10,0)}(e)   	& \feJ{(p,-3,1)}(e)\\
	\feJ{(p,5,1)}(e)		& \feJ{(p,6,1)}(e)				& \feJ{(p,-2,2)}(e)		& \feJ{(p,-1,2)}(e)\\
	\feJ{(p,0,2)}(e) 		& \feJ{(p,1,2)}(e)				& \feJ{(p,2,2)}(e)			& \partial_e\feJ{(p,1,1)}(e)
\end{array}$\\
\hline
$\order{v^5}$ & 
$\begin{array}{lll}
	\feJ{(p,11,0)}(e) 	& \feJ{(p,12,0)}(e)    	& \feJ{(p,7,1)}(e) \\
	\feJ{(p,8,1)}(e) 		& \feJ{(p,3,2)}(e)			& \feJ{(p,4,2)}(e)
\end{array}$\\
\hline
$\order{v^6}$ & 
$\begin{array}{llll}
	\feJ{(p,-3,0)}(e)		& \feJ{(p,13,0)}(e) 			& \feJ{(p,14,0)}(e) 	& \feJ{(p,-4,1)}(e) \\
	\feJ{(p,9,1)}(e)		& \feJ{(p,10,1)}(e) 			& \feJ{(p,-3,2)}(e)		& \feJ{(p,5,2)}(e) \\
	\feJ{(p,6,2)}(e) 		& \feJ{(p,-1,3)}(e)				& \feJ{(p,-2,3)}(e)		& \feJ{(p,0,3)}(e) \\
	\feJ{(p,1,3)}(e)		& \feJ{(p,2,3)}(e)				& \partial_e\feJ{(p,1,2)}(e) & \feK{(p,1,0)}(e)	\\	
    \feK{(p,2,0)}(e)        &\feK{(p,3,0)}(e)		         & \feK{(p,4,0)}(e)		& \feK{(p,5,0)}(e)	\\ 
    \feK{(p,6,0)}(e)        & \ & \  & \ 
\end{array}$
\end{longtable}
It is precisely the presence of such integrals that has long forced frequency-domain eccentric waveforms to rely on small-eccentricity expansion.
Such expansion can be easily evaluated. 
We found~\cite{work1},
\begin{align}
	& \feJ{(p,a,b)}(e) = \bigg(\frac{e}{2}\bigg)^p\sum_{n=0}^\infty \fej{(p,a,b,n)}\bigg(\frac{e}{2}\bigg)^{2n}, \label{eq_JTaylor}\\
	& \feK{(p,a,0)}(e) = \bigg(\frac{e}{2}\bigg)^p\sum_{n=0}^\infty \fek{(p,a,n)}\bigg(\frac{e}{2}\bigg)^{2n}, \label{eq_KTaylor}
\end{align}
where
\begin{align}
	& \fej{(p,a,0,n)} = \frac{p^p}{n!(p+n)!}\sum_{k=0}^{n} p^{2k}(-1)^k a_{(2n-2k)}\binom{n}{k}\sum_{m=0}^p p^{-m}\big(a+2(n-k)\big)_{(m)}\binom{p}{m}, \label{eq_feja0n}\\
	& \fej{(p,a,1,n)} = \sum_{m=0}^n\sum_{k=0}^{2m-1+p}\sum_{j=0}^{2m-1+p-k}\frac{p^j(k+2n-2m)!a_{(2m-1-j-k+p)}}{(n-m)!j!(2m-1-j-k+p)!(k+n-m+1)!} \nonumber\\
    &\quad\times \sum_{j'=0}^{\min(j,m)}(-1)^{j'}\binom{j}{j'}\bigg[\binom{2m+p-1-k-j}{m-1-k-j'} - \binom{2m+p-1-k-j}{m-j'} \bigg], \label{eq_feja1n}\\
	& \fej{(p,a,2,n)} = \sum_{m=0}^n\sum_{k=1}^{2m-1+p}\sum_{j=0}^{2m-1+p-k}\sum_{k'=1}^k\frac{(k+1)p^j(k+2n-2m)!a_{(2m-1-j-k+p)}}{k'(k-k'+1)(n-m)!j!(2m-1-j-k+p)!(k+n-m+1)!} \nonumber\\
    & \quad \times \sum_{j'=0}^{\min(j,m)}(-1)^{j'}\binom{j}{j'}\bigg[ \binom{2m-1+p-k-j}{m-1-k-j'} - \binom{2m-1+p-k-j}{m-1-k-j'+k'} + \binom{2m-1+p-k-j}{m-j'} \nonumber\\
    &\quad - \binom{2m-1+p-k-j}{m-j'-k'} \bigg] \label{eq_feja2n}\\
	& \fej{(p,a,3,n)} = \sum_{m=0}^n\sum_{k=1}^{2m-1+p}\sum_{j=0}^{2m-1+p-k}\sum_{k'=1}^k\sum_{k''=1}^{k-k'}\frac{(k+1)p^j(k+2n-2m)!a_{(2m-1-j-k+p)}}{k'k''(k-k'-k''+1)(n-m)!j!(2m-1-j-k+p)!(k+n-m+1)!} \nonumber\\
    & \quad \times \sum_{j'=0}^{\min(j,m)}(-1)^{j'}\binom{j}{j'}\bigg[ \binom{2m-1+p-k-j}{m-1-k-j'} + \binom{2m-1+p-k-j}{m-1-j'+k+k'+k''} - \binom{2m-1+p-k-j}{m-1-j'-k+k''} \nonumber\\
    & \quad - \binom{2m-1+p-k-j}{m-1-j'-k+k'} + \binom{2m-1+p-k-j}{m-j'-k'} + \binom{2m-1+p-k-j}{m-j'-k''} - \binom{2m-1+p-k-j}{m-j'-k'-k''} \nonumber\\
    & \quad - \binom{2m-1+p-k-j}{m-j'} \bigg], \label{eq_feja3n}\\
	& \fek{(p,a,n)} = \sum_{k=0}^{2n+p-1}\sum_{k'=0}^k\sum_{k''=0}^{k-k'}\frac{p^{k-k'}(-1)^{k''+1}a_{(2n+p-1-k)}}{(k-k')!(2n+p-1-k)!}\binom{k-k'}{k''}\binom{2n+p-k+k'}{n-k''}, \label{eq_feka0n}
\end{align}
where $a_{(b)}:=\Gamma(a+b)/\Gamma(a)$ denotes upper factorial or Pochhammer symbol.

However, for large eccentricities the required expansion order becomes extremely high. 
Moreover, to achieve waveform accuracy comparable to that of circular orbit cases, one must include a much larger number of Fourier modes~\cite{Morras_2025_1PN}. 
If one could instead find simple and accurate analytic approximations to these integrals, the speed and accuracy of frequency-domain eccentric waveform calculations could be substantially improved. 
This, however, requires a sufficiently detailed understanding of their mathematical structure when $e\to1$ and $p\to\infty$. 

In the following section, we will present two methods to derive the asymptotic expansions of these integrals in the high-eccentricity regime. 
The first method is more direct, providing a large-eccentricity asymptotic expansion at fixed $p$. 
The second method is the uniform asymptotic expansion that applies in the limit where $p\to\infty$.

These two methods can be cross-checked against each other. We will also demonstrate their distinct roles in gravitational-wave calculations in section~\ref{sec_LEA_gw} and section~\ref{sec_LEA_eef}.

\section{Large eccentricity asymptotic expansion of PN-elliptic integrals}\label{sec_LEA}
In this section, we present a detailed account of the evaluation procedures for the two asymptotic methods. 
In Section~\ref{sec_MAE}, we derive the asymptotic expansion in the limit $e\to1$ at fixed $p$. 
In Section~\ref{sec_UAE}, we introduce the uniform asymptotic expansion valid in the joint limit $p\to\infty$ and $e\to1$. 
Finally, in Section~\ref{sec_Check}, we perform a cross-check between the two methods.

\subsection{Asymptotic approximation} \label{sec_MAE}
Inspecting these two types of integrals (\ref{eq_PNEllipticJ}) and (\ref{eq_PNEllipticK}), it is clear that when $a\le0$ and $b=0$, 
\begin{align}
    & \feJ{(p,-2,0)} = -\frac{e}{p}J'_p(ep), \\
    & \feJ{(p,-3,0)} = -\frac{2}{p^2}J_p(ep).
\end{align}
Due to the recurrence relations given below, it is not necessary to compute all cases with $a<0$,
\begin{align}
    & \feJ{(p,-2,1)} = -\frac{\Delta_e}{p}J_p(ep) - \frac{1}{\Delta_ep^3}\feJ{(p,1,0)} + \frac{1}{\Delta_ep^3}\feJ{(p,2,0)} - \frac{e}{p^2}\partial_{e}\feJ{(p,0,1)} + \frac{e}{\Delta_e}\partial_{e}\feJ{(p,1,0)}, \\
    & \feJ{(p,-3,1)} = \frac{1}{p}\bigg(1+\frac{2}{\Delta_e}\bigg)\feJ{(p,-2,0)} + \frac{2}{p^2}\feJ{(p,-1,1)} - \frac{2}{p^2}\feJ{(p,0,1)} - \frac{2e}{p^2}\partial_e\feJ{(p,-1,1)}, \\
    & \feJ{(p,-4,1)} = \frac{1}{2p}\bigg(2+\frac{3}{\Delta_e}\bigg)\feJ{(p,-3,0)} - \frac{\Delta_e}{p}\feJ{(p,-2,0)} + \frac{6}{p^2}\feJ{(p,-2,1)}-\frac{6}{p^2}\feJ{(p,-1,1)} - \frac{3e}{p^2}\partial_e\feJ{(p,-2,1)}, \\
    & \feJ{(p,-2,2)} = \frac{2}{p}\feJ{(p,-1,1)} - \frac{2}{p^2\Delta_e^2}J_p(ep) - \frac{2\Delta_e}{p}\feJ{(p,0,1)} - \frac{2}{\Delta_ep^3}\feJ{(p,1,1)} + \frac{2}{\Delta_ep^3}\feJ{(p,2,1)} \nonumber\\
    &\qquad - \frac{e}{p^2}\partial_e\feJ{(p,0,2)} - \frac{2e}{\Delta_ep}\partial_e\feJ{(p,1,1)}, \\
    & \feJ{(p,-3,2)} = \frac{1}{p}\bigg(2 + \frac{4}{\Delta_e}\bigg)\feJ{(p,-2,1)} - \frac{4}{\Delta_ep^2}\feJ{(p,-1,0)} - \frac{2\Delta_e}{p}\feJ{(p,-1,1)} + \frac{2}{p^2}\feJ{(p,-1,2)} + \frac{4}{p^2}J_p(ep) \nonumber\\
    &\qquad - \frac{2}{p^2}\feJ{(p,0,2)} - \frac{2e}{p^2}\partial_e\feJ{(p,-1,2)}, \\
    & \feJ{(p,-2,3)} = \frac{3}{p}\feJ{(p,-1,2)} - \frac{6}{\Delta_e^2p^2}\feJ{(p,0,1)} - \frac{3\Delta_e}{p}\feJ{(p,0,2)} + \frac{6}{\Delta_e^2p^3}\feJ{(p,1,0)} - \frac{3}{\Delta_ep^3}\feJ{(p,1,2)} - \frac{6}{\Delta_ep^3}\feJ{(p,2,0)} \nonumber\\
    &\qquad + \frac{3}{\Delta_ep^3}\feJ{(p,2,2)} - \frac{e}{p^2}\partial_e\feJ{(p,0,3)} + \frac{3e}{\Delta_ep^3}\partial_e\feJ{(p,1,2)},
\end{align}
where $\Delta_e:=\sqrt{1-e^2}$.
When $a>0$, however, the integrand develops singular behavior in the limits $z\to0$ and $\Delta_e\to0$, leading to a divergence of the integral. 
Moreover, when $b>0$, even if $a<0$, a direct expansion of the integrand gives rise to higher-order derivatives of $\delta_\chi$, which also introduce singular behavior into the integrand.
The problem must therefore be treated as an asymptotic analysis of a parameter-dependent integral~\cite{Bleistein_SUAE,olver1997asymptotics}.
Noting that the leading divergence of the denominator is governed by $(1-e\cos{z})^{-1}\sim (z^2+\Delta_e^2)^{-1} = \Delta_e^{-2}\big( (z/\Delta_e)^2 + 1 \big)$, it is natural to introduce the new variable $t=z/\Delta_e$. 
In this way, the leading-order divergence
$\int (1-e\cos{z})^{-a}\dd{z} \sim \Delta_e^{1-2a}\int (t^2+1)^{-a} \dd{t}$
can be factored out explicitly, leaving a regular remainder.

In general, consider the integral
\begin{align}
I = \int_0^\pi \frac{f(\Delta_e,z)}{(1-e\cos{z})^a}\dd{z}.
\end{align}
Here the function $f(\Delta_e,z)$ is analytic over the full domain $0\le \Delta_e\le 1$ and $0\le z\le \pi$. When $a>0$, this integral may diverge in the limit $\Delta_e\to0$.
After introducing the change of variables $t=z/\Delta_e$, the divergent part can be extracted explicitly. 
As a result, the $\Delta_e$-expansion of the integrand becomes uniformly convergent, and one obtains
\begin{align}
I = \Delta_e^{1-2a} \left( \int_0^{\pi/\Delta_e} \frac{ 2 g(t) }{ (t^2+1)^a}\dd{t} + \order{\Delta_e} \right),\label{eq_MAE_I}
\end{align}
where
\begin{align}
g(t):=\lim_{\Delta_e\to0} f(\Delta_e, \Delta_e t).
\end{align}
However, it would be incorrect to proceed by directly performing a small-$\Delta_e$ asymptotic expansion of (\ref{eq_MAE_I}). 
The reason is that this expression captures only the local behavior near $z\to0$, while the contribution from the region near $z\to\pi$ is lost. 
The problem is therefore, in essence, a multiscale matching problem; see, for example, the textbook discussions in \cite{holmes2012introduction,bender1999advanced}.

The correct procedure is to split the integration domain into two regions,
$\int_0^\pi\Rightarrow\int_0^\epsilon + \int_\epsilon^\pi$,
where $\epsilon\ll1$ is a small constant. 
For the interval $(0,\epsilon)$, one may follow the previous steps, except that the upper limit in (\ref{eq_MAE_I}) is replaced by $\epsilon/\Delta_e$, and the resulting integrals are then expanded in the limit $\Delta_e\to0$. 
For the interval $(\epsilon,\pi)$, the integrand is uniformly convergent, so it can be expanded directly into a series. 
Since the sum of these two parts must be independent of the arbitrary constant $\epsilon$, expanding each part in powers of $\epsilon$ must lead to a cancellation of all $\epsilon$-dependent terms at any given order in $\Delta_e$.

Meanwhile, there is another approach. 
Since we only need the expansion up to finite order—in fact, in this section we only retain terms up to $\Delta_e^4$, more precisely, it is $\Delta_e^{\alpha+4}$, where $\alpha$ is the leading asymptotic power. 
It is sufficient to construct a suitable matching function $g_{\mathrm{match}}(e,z)$ whose divergent asymptotic behavior in the joint limit $z\to0$, $\Delta_e\to0$ agrees, through order $\order{\Delta_e^4}$, with that of the integrands in (\ref{eq_PNEllipticJ}) and (\ref{eq_PNEllipticK}). 
One may then decompose the integral as
\begin{align}
I = \int_0^\pi \bigg( \frac{f(e,z)}{(1-e\cos{z})^a} - g_{\mathrm{match}}(e,z) \bigg)\dd{z} + \int_0^\pi g_{\mathrm{match}}(e,z) \dd{z}.
\end{align}
The first term may then be treated as approximately uniformly convergent up to order $\Delta_e^4$, and can therefore be expanded straightforwardly into a series.

In the following subsections, we will apply these two methods respectively to $b=0$ and $b\ne0$ cases respectively.

\subsubsection{$b=0$ case}
Let's start from the simplest case $\feJ{(p,a,0)}$.
We found it is sufficient to choose
\begin{align}
g_{\mathrm{match}}(e,z) = \frac{1}{(1-e\cos{z})^a},
\end{align}
which already yields the expansion up to order $\Delta_e^4$.
This integral can be evaluated by
\begin{align}
    \frac{1}{\pi}\int_0^\pi \frac{1}{(1-e\cos{z})^n} \dd{z} = \Delta_e^{-n}P_{n-1}(\Delta_e^{-1}),
\end{align}
where $P_n(x)$ is the $n$th order Legendre polynomial.
We obtain the following expansion
\begin{align}
    & \Delta_e \feJ{(p,1,0)} = 1 + \frac{\Delta_e}{\pi}\int_0^\pi \frac{\cos{p(z-\sin{z}) - 1}}{1-\cos{z}}\dd{z} + \frac{\Delta^3_e}{\pi}\int_0^\pi \frac{1}{2(1-\cos{z})}\bigg[ \frac{\cos{z}\big(1 - \cos{p(z-\sin{z})}\big) }{1-\cos{z}} \nonumber\\
    &\qquad - p\sin{z}\sin{p(z-\sin{z})} \bigg]\dd{z} + \order{\Delta_e^5}, \\
    & \Delta^{3}_e \feJ{(p,2,0)} = 1 + \frac{\Delta_e^3}{\pi}\int_0^\pi \frac{\cos{p(z-\sin{z}) - 1}}{(1-\cos{z})^2}\dd{z} + \order{\Delta_e^5}.
\end{align}
One can verify, using trigonometric identities, that these two integrals are actually equal to each other,
\begin{align}
    &\int_0^\pi \frac{\cos{p(z-\sin{z}) - 1}}{(1-\cos{z})^2}\dd{z} \equiv \int_0^\pi \frac{1}{2(1-\cos{z})}\bigg[ \frac{\cos{z}\big(1 - \cos{p(z-\sin{z})}\big) }{1-\cos{z}} - p\sin{z}\sin{p(z-\sin{z})} \bigg]\dd{z},
\end{align}
Nevertheless, we retain their original forms for clarity, and also to facilitate comparison with the subsequent discussion of the expansion of $\feK{(p,a,0)}$.

When $a>2$, one has $\feJ{(p,a,0)}=\Delta_e^{-a}P_{a-1}(1/\Delta_e) + \order{\Delta_e^{6-2a}}$. 
This implies that the complicated coefficients—those that can only be expressed in terms of integrals—do not appear at lower orders, and the overall structure is therefore significantly simplified.

The advantage of choosing this form is that, using the relation
\begin{align}
    & \feK{(p,a,0)} = -\pdv{a}\feJ{(p,a,0)} = \frac{1}{\pi}\int_0^\pi \frac{\cos{p(z-e\sin{z})}-1}{(1-e\cos{z})^a}\ln{(1-e\cos{z})}\dd{z} - \pdv{a}\frac{1}{\Delta_e^{a}}P_{a-1}\big(\Delta_e^{-1}\big),
\end{align}
one can straightforwardly obtain the expansion of $\feK{(p,a,0)}$.
The second term is given by
\begin{align}
    &\pdv{a}\frac{1}{\Delta_e^{a}}P_{a-1}\big(\Delta_e^{-1}\big) = -\frac{\ln\Delta_e}{\Delta_e^a}P_{a-1}\big(\Delta_e^{-1}\big) + \frac{1}{\Delta_e^{a}} \pdv{a}P_{a-1}\big(\Delta_e^{-1}\big).
\end{align}
The derivative with respective to the order of Legendre polynomial reads
\begin{align}
    \pdv{n}P_{n}(x) = \ln{\bigg(\frac{x+1}{2}\bigg)}P_n(x) + 2\Delta{H}_n^{2n}P_n(x) + 2\sum_{k=0}^{n-1} \frac{ (-1)^{n+k}(2k+1) }{ (n-k)(n+k+1) }P_k(x),
\end{align}
where $\Delta{H}^a_b:=H_a - H_b$ denotes the difference between the $a$th harmonic number $H_a$ and the $b$th harmonic number $H_b$.
One would obtain the expansions
\begin{align}
    & \Delta_e\feK{(p,1,0)} = \ln{\big( 2\Delta_e^2 \big)} + \Delta_e\bigg( -1 + \frac{1}{\pi}\int_0^\pi \frac{\cos{p(z-\sin{z})}-1}{(1-\cos{z})}\ln{(1-\cos{z})}\dd{z} \bigg) + \nonumber\\
    &\qquad + \frac{\Delta_e^2}{2} + \Delta_e^3\Bigg[ -\frac{1}{3} + \frac{1}{\pi}\int_0^\pi \frac{1}{2(1-\cos{z})} \bigg[ \ln{(1-\cos{z})}\bigg( \frac{\cos{z}\big( 1-\cos{p(z-\sin{z})} \big)}{1-\cos{z}}-p\sin{z}\sin{p(z-\sin{z})} \bigg) \nonumber\\
    &\qquad - \frac{\cos{z}\big( 1 - \cos{p(z-\sin{z})} \big)}{1-\cos{z}} \bigg]\dd{z} \Bigg] + \frac{\Delta_e^4}{4} + \order{\Delta_e^5}, \\
    & \Delta_e^3\feK{(p,2,0)} = -1+\ln{\big( 2\Delta_e^2 \big)} + \frac{\Delta_e^2}{2} + \Delta_e^3\bigg( -\frac{1}{3} + \frac{1}{\pi}\int_0^\pi \frac{\cos{p(z-\sin{z})}-1}{(1-\cos{z})^2}\ln{(1-\cos{z})}\dd{z} \bigg) + \frac{\Delta_e^4}{4} + \order{\Delta_e^5}.
\end{align}
Similarly, when $a>2$, $\feK{(p,a,0)} = \partial_a \Delta_e^{-1}P_{a-1}(1/\Delta_e)$. Now we've finished the expansion of all $b=0$ cases. 
Although some of these coefficients can only be expressed in terms of integrals, we will show in the following sections how to evaluate them.

\subsubsection{$b\ne0$ case} \label{sec_Jpab}
When $b\ne0$, the integrals involve structures such as $\delta_\chi$ that are not amenable to analytic integration. 
We therefore directly employ the matched asymptotic approximation method to compute the expansion. 
Due to the length of the resulting expressions, we present them in Appendix ~\ref{app_UAERes_Deltae}.

At this point, we have completed the analysis of the asymptotic behavior in the limit $e\to1$ at fixed $p$. 
However, from a global perspective this description is incomplete. 
We therefore proceed to analyze the asymptotic behavior of these integrals in the limit $p\to\infty$.

\subsection{Uniform asymptotic expansion}\label{sec_UAE}
We now introduce the integral approximation method of uniform asymptotic expansion (UAE).
More details can be found in~\cite{Bleistein_SUAE,UAETextbook,temme2013uniformasymptoticmethodsintegrals}.

Consider the integral
\begin{align}
I_p(e;F) := \frac{1}{2\pi}\int_{-\pi}^{\pi}F(e,z)\ee^{\ii p \phi(e,z)}, \label{eq_InitIpeF}
\end{align}
where $\phi(e,z)=z-e\sin{z}$, $F(e,z)$ is an analytical function when $|z|<\pi$ and $0\le e<1$. 
We decompose the integrand function $F$ into the even part and odd part,
\begin{align}
F(e,z) = \frac{1}{2}\big(F(e,z) + F(e,-z)\big) + \frac{1}{2}\big(F(e,z) - F(e,-z)\big) := F_E(e,z) + F_O(e,z).
\end{align}
The integral $I_p(e;F)$ is decomposed as
\begin{align}
I_p(e;F) = I^{(E)}_p(e;F) + I^{(O)}_p(e;F),
\end{align}
where
\begin{align}
& I^{(E)}_p(e;F) = \frac{1}{\pi}\int_0^\pi F_E(e,z)\cos{\big( p\phi(e,z) \big)}\dd{z} ,\nonumber\\
& I^{(O)}_p(e;F) = \frac{1}{\pi}\int_0^\pi \ii F_O(e,z)\sin{\big( p\phi(e,z) \big)}\dd{z}.
\end{align}
The phase function $\phi(e,z)$ gives two saddle points or stationary phase points, which are the two roots of $\partial_z\phi(e,z)=0$, denotes by $z_\pm=\pm\arccos{1/e}$.
Since $0\le e<1$, $z_\pm$ are complex numbers, therefore we introduce parameter transformation $t=t(z)$ such that
\begin{align}
\phi(e,z(t)) = \frac{t^3}{3} + \zeta t,\label{eq_UAE_param_transform}
\end{align}
the solution in real axis reads
\begin{align}
t(z) = \frac{2^{1/3}\zeta}{\big(-3\phi(e,z)+\sqrt{4\zeta^3+9\phi(e,z)^2}\big)^{1/3}} - \frac{1}{2^{1/3}}\big(-3\phi(e,z)+\sqrt{4\zeta^3+9\phi(e,z)^2}\big)^{1/3}.
\end{align}
Therefore $\zeta$ is given by
\begin{align}
\zeta^{3/2} = \frac{3\ii}{4}\big( \phi(e,z_-) - \phi(e,z_+) \big) = \frac{3}{2}\big( \ln{\betae^{-1}} - \sqrt{1-e^2} \big).
\end{align}
One can check that $\zeta=\Delta_e^2/2^{2/3} + \order{\Delta_e^4}$.
The saddle points $t_\pm=\pm\ii\zeta^{1/2}$. In~\cite{UAETextbook}, the whole process are performed on the complex plane. 
Here, we simply transfer the variables to the real axis, but in fact, the two are equivalent.
It shows that when $e\to1$, the two saddle points $t_\pm\to0$ along the imaginary axis. 
The oscillating integral (\ref{eq_InitIpeF}) is mainly contributed by the portion close to the saddle points, so when $e\to1$, the contribution around $t\to0$ will become increasingly significant. Suppose a scale transformation $s=p^{1/3}t$, the integrals in terms of $s$ read
\begin{align}
& I^{(E)}_p(e;F) = \frac{1}{p^{1/3}\pi}\int_0^{t_{\pi}p^{1/3}} G_E(e,s)\cos{(s^3/3+ys)}\dd{s} ,\label{eq_intEven}\\
& I^{(O)}_p(e;F) = \frac{1}{p^{1/3}\pi}\int_0^{t_{\pi}p^{1/3}} G_O(e,s)\sin{(s^3/3+ys)}\dd{s}, \label{eq_intOdd}
\end{align}
where $y:=p^{2/3}\zeta$, and $t_\pi:=t(z=\pi)$ is the real root of the equation $t_\pi^3/3 + \zeta t_\pi=\pi$. We define
\begin{align}
& G_E(e,s) := F_E(e, z(s/p^{1/3}))z'(s/p^{1/3}), \\
& G_O(e,s) := \ii F_O(e, z(s/p^{1/3}))z'(s/p^{1/3}).
\end{align}
Since $z'(t)=(t^2+\zeta)/(1-e\cos{z})$ is even, $G_E,G_O$ are still even and odd respectively.

The above process is nothing but the UAE for the case of two merging saddle points, as mentioned in~\cite{UAETextbook}.
In this case the phase of the integral is of the type of Airy function $\Ai(y)$ and Scorer function $\Gi(y)$, which take the real axis integral forms
\begin{align}
& \Ai(y) := \frac{1}{\pi}\int_{0}^\infty \cos{(s^3/3+ys)} \dd{s}, \\
& \Gi(y) := \frac{1}{\pi}\int_{0}^\infty \sin{(s^3/3+ys)} \dd{s}.
\end{align}
In the limit of $p\to\infty$, the integrals (\ref{eq_intEven})-(\ref{eq_intOdd}) can be expressed asymptotically in terms of $\Ai(y)$ and $\Gi(y)$. The resulting error arises from the contribution of the tail integral over $s\in(p^{1/3}t_\pi,\infty)$.
The error of the tail term can be estimated by thinking about a repeated integration by parts: since the oscillatory kernel effectively vanishes as $s\to\infty$, only the boundary term $s=p^{1/3}t_\pi$ remains. 
Moreover, each integration by parts introduces an additional factor of order $s^{-2}\sim p^{-2/3}$. 
Because of the structure of the trigonometric function itself, the first boundary term does not appear until after twice integration by parts. 
Combined with the scaling of the integrand, this implies that the resulting error is of order $\order{p^{-5/3}}$.
In the subsequent calculations, we will keep terms only up to the order preceding the tail contribution, meaning the tail itself will be omitted.
We have
\begin{align}
& I^{(E)}_p(e;F) = \frac{1}{p^{1/3}\pi}\int_0^{\infty} G_E(e,s)\cos{(s^3/3+ys)}\dd{s} + \order{p^{-5/3}} ,\label{eq_intEven_extended}\\
& I^{(O)}_p(e;F) = \frac{1}{p^{1/3}\pi}\int_0^{\infty} G_O(e,s)\sin{(s^3/3+ys)}\dd{s} + \order{p^{-5/3}} \label{eq_intOdd_extended}
\end{align}
The fundamental framework of UAE is established so far.
Follow this procedure, one can easily reproduce the UAE of Bessel function $J_p(ep)$ by substituting $F\equiv 1$~\cite{NIST:DLMF}
\begin{align}
J_p(ep) = \left(\frac{4\zeta}{1-e^2}\right)^{1/4} p^{-1/3}\Ai\left( p^{2/3}\zeta \right) + \order{p^{-5/3}},
\end{align}

With these mathematical preparations, we can now begin to discuss how to use UAE to construct a proper approximation for PN-elliptic (\ref{eq_PNEllipticJ}) and (\ref{eq_PNEllipticK}).
In the following subsections, we will describe how to compute the UAE for $\feJ{(p,a,0)}$, $\feK{(p,a,0)}$, and $\feJ{(p,a,b>0)}$, respectively.

\subsubsection{$\feJ{(p,a,0)}$}
The simplest case is $\feJ{(p,a,0)}$ and we only consider $a>0$.
In this case, the integrand $F(e,z)=F_E(e,z)=1/(1-e\cos{z})^{a}$ is even.
The UAE procedure is expressing $F_E(e,t)$ in terms of the following Laurent series~\cite{temme2013uniformasymptoticmethodsintegrals,lopez2002twopointtaylorexpansionsanalytic}, 
\begin{align}
F_E(e,t) = \sum_{j=-a}^\infty a_{j}(t^2+\zeta)^j. \label{eq_EvenLaurrentExpansion}
\end{align}
Substituting it it into the integral (\ref{eq_intOdd_extended}), one would obtain the integrals that take the form of
\begin{align}
V_j(y):=\frac{1}{\pi}\int_0^{\infty} (s^2+y)^j \cos{(s^3/3+ys)} \dd{s}.
\end{align}
With the help of the expansion,
\begin{align}
& \frac{1}{1-e\cos{z(t)}} = \frac{h_0}{t^2+\zeta} + \frac{h_0(h_0^3-2)}{12\zeta} - \frac{h_0(20-5h_0^6+12h_0^2\zeta)}{384\zeta^2}(t^2+\zeta) + \order{(t^2+\zeta)^2},
\end{align}
where $h_0:=(4\zeta/\Delta_e^2)^{1/4}$. 
This yields
\begin{align}
& \feJ{(p,a,0)}(e) = h_0^{a+1}p^{2a/3} \bigg[ \frac{1}{p^{1/3}} V_{-a}(p^{2/3}\zeta) - \frac{(a+1)(h_0^3-2)}{12p\zeta}V_{1-a}(p^{2/3}\zeta) + \order{p^{-5/3}}  \bigg], \label{eq_Jpa0_VUAE}
\end{align}
We will encounter many integrals of the form $V_j(y)$, but these integrals can be expressed in terms of a small set of elementary types. 
For the case where $j>0$, one can expand $(s^2+y)^j$ and obtain
\begin{align}
V_j(y) = \sum_{j'=0}^j\binom{j}{j'} (-1)^{j'} y^{j-j'}\dv[2j']{y}\Ai(y).
\end{align}
For the case $j$ is negative, using integration by parts and the property that the oscillation phase vanishes at infinity, we obtain the recurrences,
\begin{align}
& V_{-(j+1)} = \frac{1}{4yj(j-2)}\big( 2(j-2)(2j-1)V_{-j} + V_{-(j-3)} \big), \\
& \dv{y}V_{-j} = (j+1)\big( V_{-(j+1)} - 2yV_{-(j+2)} \big).
\end{align}
Obviously, $V_{-1}(y)$ diverges when $y\to0$. 
In the end, we obtain all the required UAE of $\feJ{(p,a,0)}$ in terms of $V_0,V_{-1}$. 
Because their explicit forms are rather cumbersome, we present them in the Appendix~\ref{app_UAE_Integrals_Jpa0}.

\subsubsection{$\feK{(p,a,0)}$}
There is a logarithmic term $\ln(1-e\cos{z})$ in the integral $\feK{(p,a,0)}$, which is even but also diverges when $e\to1$ and $z\to0$.
We expand the integrand as the following form
\begin{align}
\frac{\ln(1-e\cos{z})}{(1-e\cos{z})^a} = \sum_{j=-a}^\infty\sum_{k=0}^1 c_{j,k}(t^2+\zeta)^j\ln{(t^2+\zeta)}^k.
\end{align}
There will be a new kind of integral
\begin{align}
U_j(y) := \frac{1}{\pi}\int_0^\infty \ln{(s^2+y)} (s^2+y)^j \cos{(s^3/3+sy)} \dd{s} = \pdv{j} V_j.
\end{align}
Similarly, for $j>0$ the recurrences
\begin{align}
& U_j(y) = 4(j-1)(j-3)yU_{j-4}(y) - 2(j-1)(2j-5)U_{j-3}(y) + 8(j-2)yV_{j-4}(y) - 2(4j-7)V_{j-3}(y), \\
& U_{-(j+1)}(y) = \frac{1}{4j(j-2)y}\Big( 2(j-2)(2j-1)U_{-j}(y) + U_{-(j-3)}(y) + 8(j-1)yV_{-(j+1)}(y) + 2(5-4j)V_{-j}(y) \Big),
\end{align}
and the algebraic relations
\begin{align}
& U_{-2}(y) = -\frac{1}{2y}\Big( U'_0(y) - U_{-1}(y) + V_{-1}(y) - 2y V_{-1}(y) \Big), \\
& U_{-3}(y) = -\frac{1}{4y^2}\Big( U'_0(y) + U'_{-1}(y) - U_{-1}(y) + V_{-1}(y) - yV_{-2}(y) - 2y^2V_{-3}(y) \Big)
\end{align}
can replace $U_j(y)$ with $U_0(y), U_{-1}(y)$ and their derivatives. 

Similarly, one can obtain the UAE of $\feK{(p,a,0)}$. We show the results in appendix~\ref{app_UAE_Integrals_Kpa0}.

\subsubsection{$\feJ{(p,a,b)}$}
When $b>0$, it becomes more complicated. 
The reason is that $\delta_\chi(e=1,z)$ is discontinuous at $z=2k\pi$. 
When expanding it into the series of $t^2+\zeta$ according to the steps of UAE described above, one would find the leading order is $\ln(\zeta+t^2)$, and the higher-order coefficients of the expansion diverge at $e\to1$. 
This divergence is precisely because of the discontinuity of $\delta_\chi(1,z)$, or rather, because $\partial_z \delta_\chi(e,z)$ diverges at $e\to1$. 
Therefore, the correct approach is to first subtract this part that leads to singularities. We denote the singular part as
\begin{align}
\delta^{\mathrm{sing}}_\chi(e, z) = 2\arctan\big(z/\tilde\zeta\big). \label{eq_deltachiSingular}
\end{align}
where $\tilde\zeta:=\ln1/\betae=(2/3)\zeta^{3/2}+\sqrt{1-e^2}$.
The rest part $\delta_\chi(e,z)-\delta^{\mathrm{sing}}_\chi(e,z)$ is regular and can be processed normally by the UAE method.
However, $\delta^{\mathrm{sing}}_\chi(e,z)$ does not have similar expansion structure, so we need to handle it separately. 

We expand $z(t)$ around $z_\pm$
\begin{align}
& z(t) = \frac{\tilde\zeta}{\zeta^{1/2}}t + \frac{(\tilde\zeta - h_0\zeta^{1/2})}{2\zeta^{3/2}}t(t^2+\zeta) + \order{(t^2+\zeta)^2},
\end{align}
therefore $\arctan{(z/\tilde\zeta)}\sim\arctan{\big(t/\zeta^{1/2}\big)}$, one can expand $\delta_\chi(e,z)$ as
\begin{align}
&\delta_\chi(e,z) = 2\arctan{\big(t/\zeta^{1/2}\big)} + t\sum_{j=0}^\infty (t^2+\zeta)^j\Delta_{\chi,j}(e) \nonumber\\
& \quad = 2\arctan{\big(t/\zeta^{1/2}\big)} - \frac{t}{\zeta^{1/2}}\ln{\bigg(\frac{h_0(h_0^2+2\zeta^{1/2})}{2}\bigg)} - t(t^2+\zeta)\bigg[ \frac{1}{2\zeta^{3/2}}\ln{\bigg(\frac{h_0(h_0^2+2\zeta^{1/2})}{2}\bigg)} \nonumber\\
& \quad + \frac{2h_0\zeta^{1/2}(3+\betae^2)(1+\betae^2)^2 - h_0^3(5-\betae^2)(1-\betae^2)^2 - 8(1-\betae^4)}{12h_0^2(1-\betae^2)^2\zeta} \bigg] + \order{(t^2+\zeta)^2} \label{eq_deltachi_wExpansion}
\end{align}
where $\Delta_{\chi,j}(e)$ are analytic functions among $0\le e\le 1$ up to $\order{(t^2+\zeta^2)^2}$.
The leading term $\arctan{(t/\zeta^{1/2})}$ cannot be expanded anymore, so we define 4 new integrals here,
\begin{align}
& \intV_{j,k}(y) := \frac{1}{\pi}\int_0^{\infty} (s^2+y)^j\Big(\mathrm{arctan}\big(y^{1/2}/s\big)\Big)^k \cos{(s^3/3+ys)} \dd{s}, \\
& \tintV_{j,k}(y) := \frac{1}{\pi}\int_0^{\infty} s(s^2+y)^j\Big(\mathrm{arctan}\big(y^{1/2}/s\big)\Big)^k \cos{(s^3/3+ys)} \dd{s}, \\
& \intW_{j,k}(y) := \frac{1}{\pi}\int_0^{\infty} (s^2+y)^j\Big(\mathrm{arctan}\big(y^{1/2}/s\big)\Big)^k \sin{(s^3/3+ys)} \dd{s}, \\
& \tintW_{j,k}(y) := \frac{1}{\pi}\int_0^{\infty} s(s^2+y)^j\Big(\mathrm{arctan}\big(y^{1/2}/s\big)\Big)^k \sin{(s^3/3+ys)} \dd{s},
\end{align}
Obviously, $\intV_{j,0}(y)=V_j(y),\tintV_{0,0}(y)=\Ai'(y),\intW_{0,0}(y)=\Gi(y),\tintW_{0,0}(y)=-\Gi'(y)$. The tilde integrals $(\tintV,\tintW)$ are equivalent to the derivative of $(\intV,\intW)$ via,
\begin{align}
    & \intV'_{j,k}(y) = - y^{1/2}\tintW_{j,k}(y) + jy^{1/2}\intV_{j-1,k}(y) + \frac{k}{2y^{1/2}}\tintV_{j-1,k-1}(y) ,\label{eq_intVDeriv} \\
    & \intW'_{j,k}(y) = y^{1/2}\tintV_{j,k}(y) + jy^{1/2}\intW_{j-1,k}(y) + \frac{k}{2y^{1/2}} \tintW_{j-1,k-1}(y). \label{eq_intWDeriv}
\end{align}
For the sake of simplicity, we keep the tilde integrals.
These integrals also satisfy some recurrences relations,
\begin{align}
& \intV_{j+1,k}(y) - ky^{1/2}\intW_{j-1,k-1}(y) + 2j\tintW_{j-1,k}(y) = 0, \\
& \tintV_{j+1,k}(y) - 2j y \intW_{j-1,k}(y) + (2j+1)\intW_{j,k}(y) - ky^{1/2}\tintW_{j-1,k-1}(y) = 0, \\
& \intW_{j+1,k}(y) - 2j \tintV_{j-1,k}(y) + ky^{1/2}\intV_{j-1,k-1} = y^j\bigg(\frac{\pi}{2}\bigg)^k, \\
& \tintW_{j+1,k}(y) + 2jy\intV_{j-1,k}(y) - (2j+1)\intV_{j,k}(y) + ky^{1/2}\tintV_{j-1,k-1}(y) = 0.
\end{align}
At last only $\big(\intV_{0,k}(y), \intV_{-1,k}(y), \intW_{0,k}(y), \intW_{-1,k}(y)\big)$ and $\big(\tintV_{0,k}(y), \tintV_{-1,k}(y), \tintW_{0,k}(y), \tintW_{-1,k}(y)\big)$ would remain.
We show the results in Appendix~\ref{app_UAE_Integrals_Jpab}.

\subsection{Cross-check of the two methods}\label{sec_Check}
So far, we have, in a rather straightforward manner, obtained formal asymptotic expansions of these integrals in the two limits $e\to1$ and $p\to\infty$. 
However, several issues remain to be addressed:
\begin{enumerate}
    \item Although we have derived explicit mathematical expressions, they are composed of many non-trivial integrals, which still limits their direct applicability in practice.
    \item What is the precise relationship between these two kinds of asymptotic expansions?
\end{enumerate}
This can be interpreted as an asymptotic expansion in two parameters, $\Delta_e$ and $1/p^{1/3}$. The two approaches yield
\begin{align}
J &= \Delta_e^{\alpha} \Big( c_0(p) + c_1(p)\Delta_e + c_2(p)\Delta_e^2 + \order{\Delta_e^3} \Big) \nonumber\\
&= p^{\beta}\Big( d_0(e,y) + p^{-1/3}d_1(e,y) + p^{-2/3}d_2(e,y) + \order{p^{-1}} \Big).
\end{align}
Therefore, by performing a secondary expansion of the coefficients $c_k(p)$ and $d_k(e,y\equiv p^{2/3}\zeta)$ in the limits $p\to\infty$ and $\Delta_e\to0$, respectively, and reorganizing the result into a double series in $\Delta_e^a/p^{b/3}$, the coefficients must match term by term.
To achieve this, we again employ the matched asymptotic expansion method introduced in Section~\ref{sec_MAE} on both $c_k(p)$ and $d_k(e,y)$.

We first discuss how to expand $c_k(p)$ into a power series in $1/p^{1/3}$. One can perform the same parameter transformation as (\ref{eq_UAE_param_transform}), $(z-\sin{z})=t^3/3$, which rewrites the integral in $c_k(p)$ in the following form,
\begin{align}
    I_p = \int_0^{(3\pi)^{1/3}} \bigg( \sin{(pt^3/3)}A_S(t) + \big( \cos{(pt^3/3)}-1 \big)A_C(t) - B(p,t) \bigg) \dd{t}.
\end{align}
Since the integrand is analytic over the interval $0\le z\le \pi$, the transformed integrand is also analytic on $0\le t\le (3\pi)^{1/3}$. 
However, when considered separately, $A_S(p,t)$, $A_C(p,t)$, and $B(t)$ may be divergent; in other words, their Laurent expansions near $t\to0$ contain leading terms with negative powers. 
One therefore can decompose the integral into five parts,
\begin{align}
    I_p &= \sum_{k=1}^{n_s}\int_0^{(3\pi)^{1/3}} \bigg( \frac{\hat{s}_{-k} \sin{(pt^3/3)}}{t^k} - \sum_{j}\frac{\bar{s}_{k,j}}{t^{\alpha_{k,j}}} \bigg)\dd{t} + \sum_{k=1}^{n_c} \int_0^{(3\pi)^{1/3}} \bigg( \frac{\hat{c}_{-k} \big( \cos{(pt^3/3)} -1 \big) }{t^k} - \sum_{j}\frac{\bar{c}_{k,j}}{t^{\beta_{k,j} }} \bigg)\dd{t} \nonumber\\
    &\qquad + \int_0^{(3\pi)^{1/3}} \sin{(pt^3/3)}A_S^{\mathrm{reg}}(t)\dd{t} + \int_0^{(3\pi)^{1/3}} \big( \cos{(pt^3/3)}-1 \big) A_C^{\mathrm{reg}}(t)\dd{t} - \int_0^{(3\pi)^{1/3}} B^{\mathrm{reg}}(p,t) \dd{t},
\end{align}
where we assume $A_S(t)$ and $A_C(t)$ admits the following Laurent series
\begin{align}
    & A_S(t) = \sum_{k=1}^{n_s}t^{-k}\hat{s}_{-k} + A_S^{\mathrm{reg}}(t), \\
    & A_C(t) = \sum_{k=1}^{n_c}t^{-k}\hat{c}_{-k} + A_C^{\mathrm{reg}}(t), 
\end{align}
The series $\sum_{j}(\bar{s}_{k,j})/t^{\alpha_{k,j}}$ and $\sum_{j}(\bar{c}_{k,j})/t^{\beta_{k,j}}$ represent the singular parts of
$\hat{s}_{-k} \sin{(pt^3/3)}/t^k$ and $\hat{c}_{-k} \big( \cos{(pt^3/3)} -1 \big)/t^k$, respectively. 
In other words, the first two integrals can be interpreted as the regular parts~\cite{Blanchet_2000} of the integrals of $\hat{s}_{-k} \sin{(pt^3/3)}/t^k$ and $\hat{c}_{-k} \big( \cos{(pt^3/3)} -1 \big)/t^k$.
Accordingly, $B^{\mathrm{reg}}(p,t)$ is defined as
\begin{align}
    & B^{\mathrm{reg}}(p,t) := B(p,t) - \sum_{k=1}^{n_s}\sum_{j}\frac{\bar{s}_{k,j}}{t^{\alpha_{k,j}}} - \sum_{k=1}^{n_c}\sum_{j}\frac{\bar{c}_{k,j}}{t^{\beta_{k,j} }}.
\end{align}
After this decomposition, the integrands of all resulting integrals become regular. 

In this way, we can obtain the asymptotic expansions of all coefficients of $\feJ{(p,a,b)}$ in the limit $p\to\infty$. Here we present the results for the case $b=0$, while the coefficients for $b>0$ are shown Appendix~\ref{app_UAERes_Deltae}.
\begin{align}
    & \Delta_e \feJ{(p,1,0)} = 1 + \hat{p}^{1/3}\Delta_e\bigg( -\frac{\Gamma(2/3)}{\pi} + \frac{\Gamma(1/3)}{10\pi}\hat{p}^{-1/3} - \frac{23\Gamma(2/3)}{4200\pi}\hat{p}^{-4/3} + \order{p^{-2}}\bigg) + \nonumber\\
    &\qquad + \hat{p}\Delta_e^3\bigg(-\frac{2}{3^{5/2}} - \frac{3\Gamma(2/3)}{10\pi}\hat{p}^{-2/3} + \frac{51\Gamma(1/3)}{1400\pi}\hat{p}^{-4/3} + \order{p^{-2}} \bigg) + \order{\Delta_e^5}, \\
    & \Delta^{3}_e \feJ{(p,2,0)} = 1 + \hat{p}\Delta_e^3\bigg( -\frac{2}{3^{5/2}} - \frac{3\Gamma(2/3)}{10\pi}\hat{p}^{-2/3} + \frac{51\Gamma(1/3)}{1400\pi}\hat{p}^{-4/3} + \order{p^{-2}} \bigg) + \order{\Delta_e^5},
\end{align}
where $\hat{p}:=\sqrt{3}p/2$. For $\feK{(p,a,0)}$, the logarithmic term $\ln{(1-\cos{z})}$ occurs, one can use
\begin{align}
    \int_0^\infty z^\alpha \ee^{\pm \ii z} \ln{z} \dd{z} = \pdv{\alpha} \int_0^\infty z^\alpha \ee^{\pm \ii z} \dd{z} = \pdv{\alpha} \Gamma(\alpha+1)\ee^{ \pm \ii \pi (\alpha+1) }.
\end{align}
The subsequent steps are the same as those for evaluating the asymptotic expansion of $\feJ{(p,a,b)}$, one obtains
\begin{align}
    & \Delta_e \feK{(p,1,0)} = \ln{2\Delta_e^2} + \hat{p}^{1/3}\Delta_e\bigg( \frac{1}{9\Gamma(1/3)}\big( 2\sqrt{3}(-6+2\ege+\ln{6}) - 4\pi \big) - \frac{\Gamma(1/3)}{180\pi}\big( 9+12\ege+4\sqrt{3}\pi + 6\ln{6} \big)\hat{p}^{-2/3} \nonumber\\
    &\qquad - \frac{9\sqrt{3}}{1400}\hat{p}^{-4/3} + \bigg( \frac{4}{3 \sqrt{3} \Gamma (1/3)}-\frac{\Gamma (1/3)}{15 \pi }\hat{p}^{-2/3} \bigg) \ln{p} + \order{p^{-2}} \bigg) + \frac{1}{2}\Delta_e^2 + \hat{p}\Delta_e^3\bigg( \frac{1}{27\sqrt{3}}\big( -1+4\ege-4\ln{3}+2\ln{2} \big) \nonumber\\
    &\qquad + \frac{1}{3\Gamma(1/3)}\Big( \sqrt{3}\big( -7+4\ege+2\ln{6} \big) - 4\pi \Big)\hat{p}^{-2/3} - \frac{\Gamma(1/3)}{4200\pi}\big( 34\sqrt{3}\pi + 51\ln{6}+159+102\ege \big)\hat{p}^{-4/3} \nonumber\\
    &\qquad + \bigg( \frac{4}{27\sqrt{3}} + \frac{2}{5\sqrt{3}\Gamma(1/3)}\hat{p}^{-2/3} - \frac{17\Gamma(1/3)}{700\pi}\hat{p}^{-4/3} \bigg)\ln{p} + \order{p^{-2}} \bigg) + \frac{1}{4}\Delta_e^4 + \order{\Delta_e^5}, \\
    & \Delta_e^3 \feK{(p,2,0)} = -1 + \ln{2\Delta_e^2} + \frac{1}{2}\Delta^2_e + \hat{p}\Delta_e^3\bigg( \frac{2}{27\sqrt{3}}(-2 + 2\ege - \ln{(9/2)}) + \frac{1}{15\Gamma(1/3)}\Big( \sqrt{3}\big(-5 + 2\ege + \ln{6}\big) -2\pi \Big)\hat{p}^{-2/3} \nonumber\\
    &\qquad - \frac{\Gamma(1/3)}{8400\pi}( 165+204\ege+68\sqrt{3}\pi+102\ln{6} )\hat{p}^{-4/3} +  \bigg(\frac{4}{27 \sqrt{3}}+\frac{2}{5 \sqrt{3} \Gamma (1/3)}\hat{p}^{-2/3}-\frac{17 \Gamma (1/3)}{700 \pi }\hat{p}^{-4/3}\bigg)\ln{p} \nonumber\\
    &\qquad + \order{p^{-2}}\bigg) + \frac{1}{4}\Delta_e^4 + \order{\Delta_e^5},
\end{align}
where $\ege$ is Euler-Mascheroni constant.
We can thus obtain the asymptotic expansions of these integrals in the joint limits $\Delta_e\to0$ and $p\to\infty$. In principle, the UAE method should yield consistent results. However, it is important to note that the UAE expansion is carried out under the condition $p\to\infty$, with $y=p^{2/3}\zeta$ introduced as the uniform asymptotic variable, so that the expressions are organized in terms of a set of basic integrals.
For the regime of interest here, one must further derive the asymptotic expansions of these basic integrals when $y\to0$.

\begin{figure*}[t]
\begin{tabular}{c}
\includegraphics[width=\textwidth]{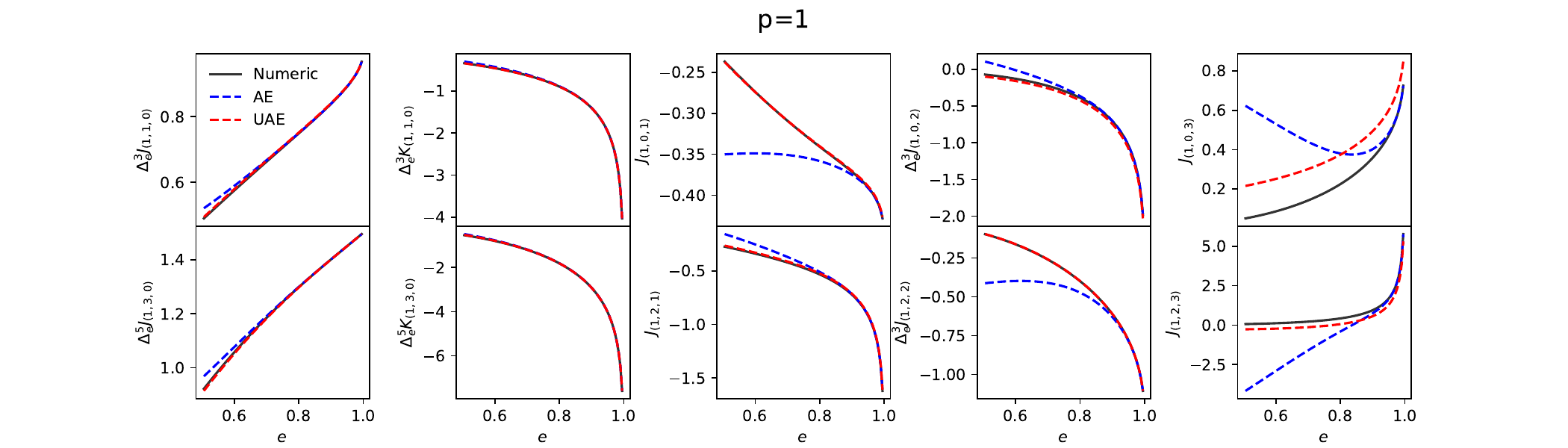} \\
\includegraphics[width=\textwidth]{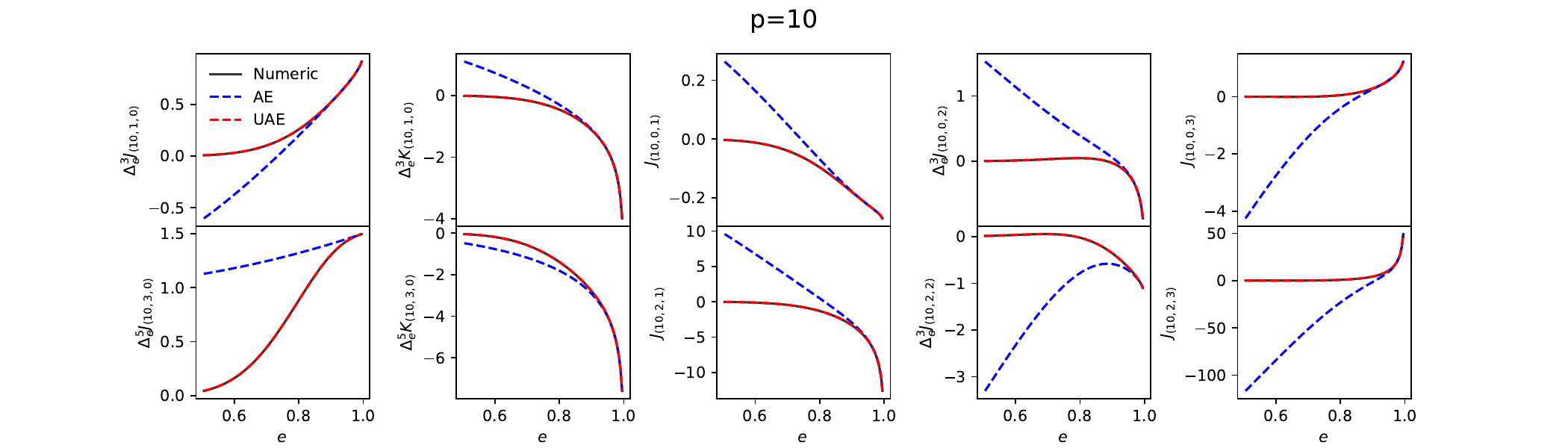}
\end{tabular}
\caption{Comparison of the numerical results (labeled as `Numeric') with the two approximation schemes—the $\Delta_e$ asymptotic expansion (labeled as `AE') and the uniform asymptotic expansion (labeled as `UAE')—for several representative integrals in the range $e>0.5$. The selected examples are $\feJ{(p, a=(1,3), 0 )}$, $\feK{(p, a=(1,3), 0)}$, and $\feJ{(p, a=(0,2), b=(1,2,3))}$. The upper panels correspond to the case $p=1$, while the lower panels show the case $p=10$.}
\label{fig_compare_Jpab_examples}
\end{figure*}

Some of these integrals are directly related to $\Ai$ and $\Gi$, such as $V_0(y)=\Ai(y)$ and $\intW_{0,0}(y) = \Gi(y)$, and can be easily calculated, while others need to be evaluated using the matched asymptotic expansion method introduced in Section.~\ref{sec_MAE}.
We obtain
\begin{align}
    & V_{-1}(y) = \frac{1}{2y^{1/2}} + \frac{\Gamma(-1/3)}{3^{5/6}2\pi} - \frac{1}{9}y - \frac{\Gamma(-5/3)}{3^{1/6}18\pi}y^2 + \frac{1}{9}y^{5/2} + \order{y^3}, \\
    & U_0(y) = -\frac{2\pi+\sqrt{3}(2\ege+\ln{3})}{3^{2/3}\pi}\Gamma(4/3) + y^{1/2} + \frac{(-9+6\ege-2\sqrt{3}\pi+3\ln{3})\Gamma(2/3)}{3^{5/6}6\pi}y + \frac{1}{4}y^2 + \order{y^3}, \\
    & U_{-1}(y) = \frac{\ln{y}+2\ln{2}}{2y^{1/2}} + \frac{140\Gamma(-10/3)(-18+7\ege-2\sqrt{3}\pi+3\ln{3})}{3^{5/6}243\pi} + \frac{-1+4\ege-4\ln{3}}{54}y \nonumber\\
    &\quad - \frac{ 308\Gamma(-14/3) \big(20\pi + \sqrt{3}(3+20\ege + 10\ln{3}) \big) }{3^{2/3}3645\pi} y^2 + \bigg(-\frac{19}{135} + \frac{2}{9}\ln{2} + \frac{1}{9}\ln{y}\bigg)y^{5/2} + \order{y^3}.
\end{align}
Eventually, we obtain the asymptotic expansion of the auxiliary integrals appear in the UAE of $\feJ{(p,a,b)}$,
\begin{align}
& \intV_{0,1}(y) = \frac{3-\ege+\ln{3}-\ln{y^{3/2}}}{3\pi}y^{1/2} + \frac{1}{3^{2/3}18}\bigg( -\frac{9}{\pi}\Gamma(1/3) + \frac{\sqrt{3}}{\Gamma(2/3)} \bigg)y^{3/2} + \bigg( \frac{17}{3^{5/6}60\Gamma(1/3)} - \frac{\Gamma(2/3)}{3^{1/3}4\pi} \bigg)y^{5/2} \nonumber\\
&\qquad + \order{y^{7/2}}, \\
& \intV_{0,2}(y) = \ln{2}\,y^{1/2} - \frac{1}{3^{1/3}\Gamma(1/3)}y - \frac{7}{54}y^4 + \bigg( -\frac{\Gamma(1/3)}{3^{1/6}4\pi} + \frac{127}{3^{2/3}450\Gamma(2/3)} \bigg)y^3 + \order{y^{7/2}}, \\
& \intV_{0,3}(y) = \bigg(\frac{3\pi\ln{2}}{4}-\frac{21\zeta_R(3)}{8\pi}\bigg)y^{1/2} - \frac{1}{3^{1/6}6\Gamma(2/3)}y^{3/2} - \frac{3^{1/6}}{4\Gamma(1/3)}y^{5/2} + \order{y^{7/2}} \\
& \intV_{-1,1}(y) = \frac{\pi}{8y^{1/2}} - \frac{y^{1/2}}{3^{1/3}6\Gamma(2/3)} - \frac{2}{3^{11/6}\Gamma(1/3)}y^{3/2} + \frac{1}{8100\pi}(-713+380\ege+225\pi^2-380\ln{3}+1140\ln{y})y^{5/2} \nonumber\\
&\qquad + \order{y^{7/2}}, \\
& \intV_{-1,2}(y) = \frac{\pi^2}{24y^{1/2}} - \frac{1}{18}y - \frac{1}{3^{5/3}\Gamma(2/3)}y^2 + \frac{1}{1080}(41+10\pi^2-152\ln{2})y^{5/2} + \order{y^{7/2}} ,\\
& \intV_{-1,3}(y) = \frac{\pi^3}{64y^{1/2}} - \frac{1}{3^{5/6}4\Gamma(1/3)}y^{3/2} + \frac{1}{4320\pi}(-2072+320\ege+123\pi^2+15\pi^4-456\pi^2\ln{2}-320\ln{3}+960\ln{y} \nonumber\\
&\qquad +1596\zeta_R(3))y^{5/2} + \order{y^{7/2}}, \\
& \intW_{0,1}(y) = \frac{y^{1/2}}{6} + \frac{1}{3^{2/3}6}\bigg( \frac{3\sqrt{3}\Gamma(1/3)}{\pi} - \frac{1}{\Gamma(2/3)} \bigg)y^3 - \frac{5}{24}y^4 + \bigg( \frac{17}{3^{1/3}60\Gamma(1/3)}-\frac{3^{1/6}\Gamma(2/3)}{4\pi} \bigg)y^{5/2} + \order{y^{7/2}},\\
& \intW_{0,2}(y) = \frac{1}{3^{5/3}\Gamma(1/3)}y + \frac{1}{432\pi}(512-112\ege-45\pi^2+112\ln{3}-336\ln{y})y^2 + \bigg(-\frac{\Gamma(1/3)}{3^{2/3}4\pi} + \frac{127}{3^{1/6}1350\Gamma(2/3)} \bigg)y^3 \nonumber\\
&\qquad +\order{y^{7/2}},\\
& \intW_{0,3}(y) = \frac{1}{3^{2/3}2\Gamma(2/3)}y^{3/2} + \frac{1}{96}(-4-5\pi^2+112\ln{2})y^2 - \frac{3^{2/3}}{4\Gamma(1/3)}y^{5/2} +\order{y^{7/2}},\nonumber\\
& \intW_{-1,0}(y) = \frac{1}{3^{5/6}\Gamma(1/3)} + \frac{-1-2\ege+2\ln{3}-6\ln{y}}{9\pi}y + \bigg( -\frac{\Gamma(1/3)}{3^{2/3}4\pi} + \frac{2}{3^{1/6}15\Gamma(2/3)} \bigg)y^2 \nonumber\\
&\qquad + \bigg( \frac{2}{3^{5/6}7\Gamma(1/3)} - \frac{\Gamma(2/3)}{3^{1/3}12\pi} \bigg)y^3 + \order{y^{7/2}},\\
& \intW_{-1,1}(y) = \frac{1}{3^{2/3}2\Gamma(2/3)}y^{1/2} + \frac{-1+4\ln{2}}{12}y - \frac{2}{3^{4/3}\Gamma(1/3)}y^{3/2} - \frac{19}{810}y^{5/2} + \order{y^{7/2}}, \\
& \intW_{-1,2}(y) = \frac{1}{72\pi}(44-8\ege-3\pi^2+12\pi^2\ln{2}+8\ln{3}-8\ln{y^{3/2}}-24\zeta_R(3))y - \frac{1}{3^{13/6}\Gamma(2/3)}y^2 \nonumber\\
&\qquad - \frac{29}{3^{5/6}180\Gamma(1/3)}y^3 + \order{y^{7/2}},\\
& \intW_{-1,3}(y) = \frac{1}{48}\big(-\pi^2+ 4(6+\pi^2)\ln{2} - 18\zeta_R(3) \big)y - \frac{1}{3^{1/3}4\Gamma(1/3)}y^{3/2} - \frac{1}{27}y^{5/2} + \order{y^{7/2}},
\end{align}
where $\zeta_R$ is the Riemann-zeta function. To distinguish it from the symbol $\zeta$, we add the subscript $R$.
The tilde integrals can be represented by their derivatives through Eq. (\ref{eq_intVDeriv}) and (\ref{eq_intWDeriv}). 

Substituting these expansions into the UAE expressions obtained in Section~\ref{sec_UAE} and then expanding in the limit $\Delta_e\to0$, we find that the resulting coefficients of the $\Delta_e$ expansion agree exactly with those obtained from the direct expansion, up to order $\order{p^{-4/3}}$. 
In this way, we have systematically verified the asymptotic expansions of each integral.

To provide some intuitive insight, we present in FIG.~\ref{fig_compare_Jpab_examples} a comparison for several representative cases, namely $\feJ{(p,a=(1,3), 0)}$, $\feK{(p,a=(1,3), 0)}$, and $\feJ{(p, a=(0,2), b=(1,2,3))}$ with $p=1$ and $p=10$.
We compare the numerical results with those obtained from the two approximation methods. 
It is observed that the UAE method provides consistently high accuracy overall, although the error becomes larger when $p$ is small. 
In contrast, the direct asymptotic expansion agrees well only in the regime $e\to1$. These findings are consistent with expectations.
In practical applications, the UAE contains higher-order information in the $\Delta_e$ expansion, whereas extending the direct expansion method becomes increasingly difficult. 
Therefore, if one aims to obtain higher-order asymptotic information near the $e\to1$ boundary, the UAE approach is more suitable.


\section{Analytic approximation}\label{sec_LEA_gw}

\begin{figure*}[t]
\begin{tabular}{c}
\includegraphics[width=\textwidth]{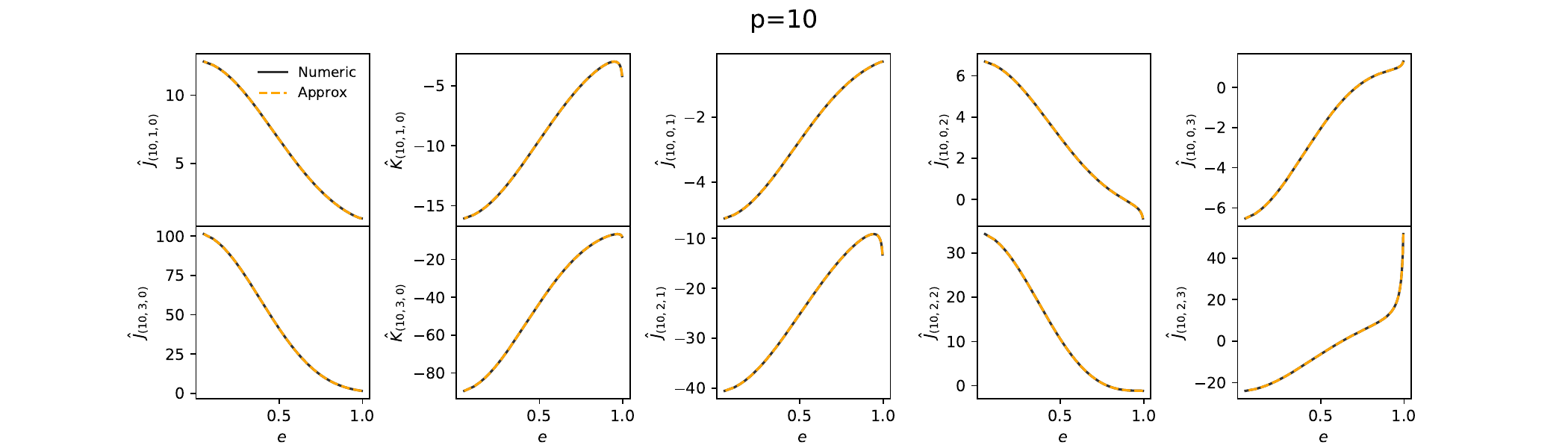} \\
\includegraphics[width=\textwidth]{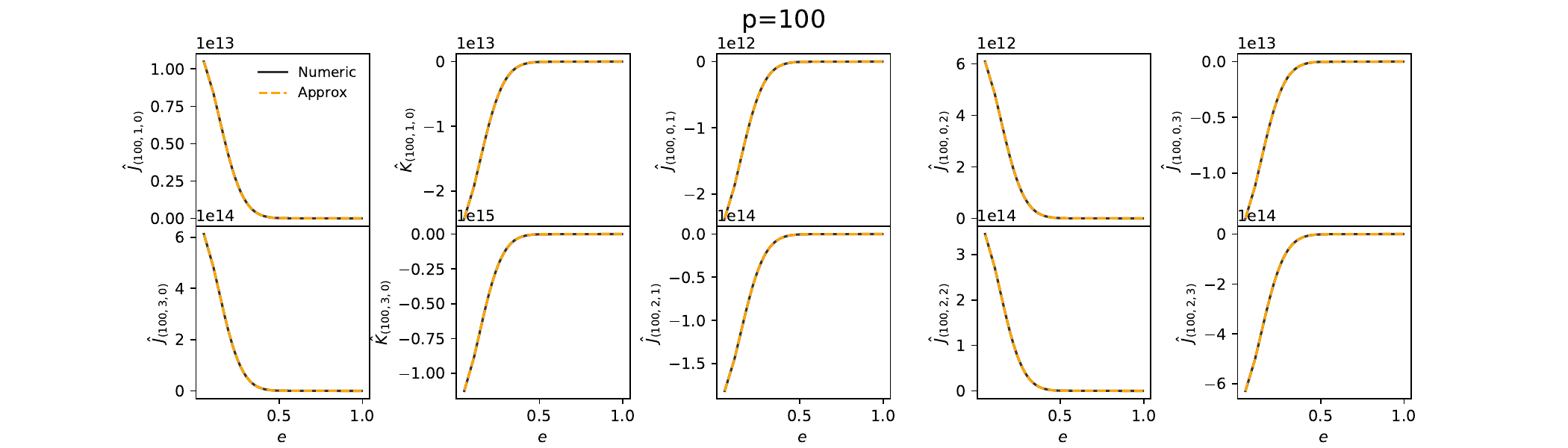}
\end{tabular}
\caption{Comparison of the numerical results (labeled as `Numeric') with the analytic approximated form of the PN-elliptic integrals (\ref{eq_approximated_regI}) (labeled as `Approx'). The selected examples are $\feJ{(p, a=(1,3), 0 )}$, $\feK{(p, a=(1,3), 0)}$, and $\feJ{(p, a=(0,2), b=(1,2,3))}$. The upper panels correspond to the case $p=10$, while the lower panels show the case $p=100$.}
\label{fig_compare_JpabApprox_examples}
\end{figure*}

In the previous section, we derived the asymptotic expansions of these PN-elliptic integrals (\ref{eq_PNEllipticJ}) and (\ref{eq_PNEllipticK}) in the two limits $e\to0$ and $e\to1$. 
Using this asymptotic information, we now attempt to construct an endpoint-constrained and accurate analytic approximation, which can significantly accelerate the computation.

Firstly we factorize the leading asymptotic behavior as
\begin{align}
    & \feJ{(p,a,b)} = \frac{e^p}{\Delta_e^{\mu_{a,b}}}  \hat{J}_{(p,a,b)},\\
    & \feK{(p,a,b)} = \frac{e^p}{\Delta_e^{\mu_{a,b}}}  \hat{K}_{(p,a,b)}, 
\end{align}
where
\begin{align}
    \mu_{a,b} = 
    \begin{cases}
        2a-1 & a>0\text{ and }b=0,2, \\
        2a-4 & a>2\text{ and }b=1,3, \\
        0 & \text{otherwise}
    \end{cases}
\end{align}
The remaining parts, $\hat{J}$ and $\hat{K}$, are then regular over $0\le e\le 1$ after subtracting the potentially occurred logarithmic term.
We adopt the following form as an approximation for $\hat{I}=(\hat{J}, \hat{K})$. 
For simplicity, we will use $\hat{I}$ to denote both integrals of $\hat{J}$ and $\hat{K}$,
\begin{align}
    \hat{I} - \hat{d}^{\ln}_0 \ln\Delta_e \approx 
    \left\{ \begin{aligned}
        & \sum_{n=0}^{N_l} c_ne^{2n} + e^{2N_l+2}\bigg( \sum_{n=0}^{N_r}d_n\Delta_e^n + \sum_{n=1}^{N_r}d_n^{\ln}\ln{\Delta_e}\,\Delta_e^n + \Delta_e^{N_r+1}\delta_{I}\bigg) & p < p_0 \\
        & \bigg[1 + e^{2N_l+2} \bigg( \sum_{n=0}^{N_r}d_n\Delta_e^n + \sum_{n=1}^{N_r}d_n^{\ln}\ln{\Delta_e}\,\Delta_e^n + \Delta_e^{N_r+1}\delta_{I}\bigg) \bigg] \exp\bigg( \sum_{n=0}^{N_l}c_ne^{2n} \bigg) & p \ge p_0.
    \end{aligned}\right. \label{eq_approximated_regI}
\end{align}
Here the coefficients $c_n$ are determined from the expansion of the integral in $e\to0$, while $d_n$ are fixed by the expansion $\Delta_e\to0$. 
The term $\delta_I$ is a polynomial introduced to fit the remaining residual error.
And $\hat{d}^{\ln}_0 \ln\Delta_e$ represents a possible leading-order logarithmic contribution, where the coefficient $\hat{d}^{\ln}$ is determined from the $\Delta_e\to0$ asymptotic expansion. 

\begin{figure}[t]
\begin{tabular}{c}
\includegraphics[width=0.5\textwidth]{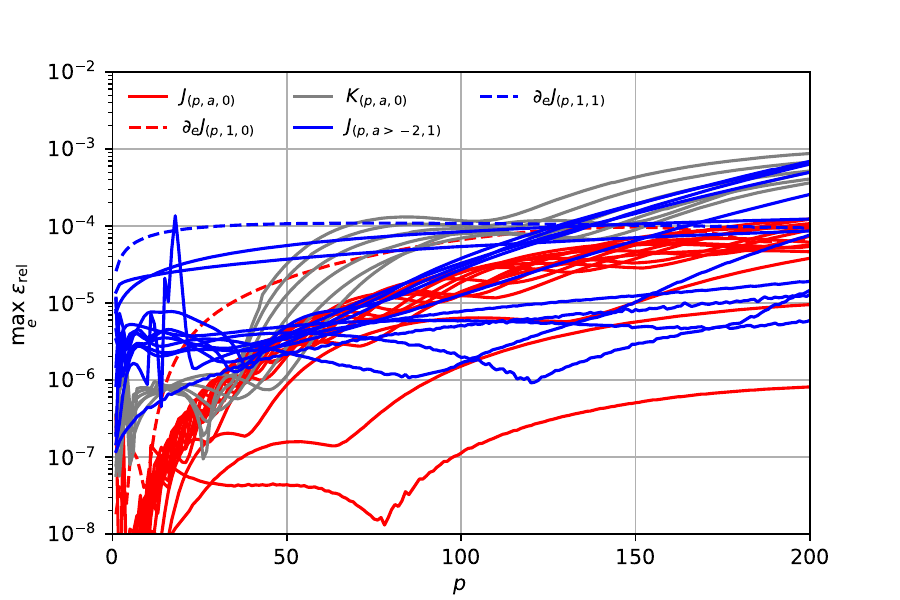}
\end{tabular}
\caption{The maximum relative error among $0<e<1$ between the analytic approximation of the regularized integrals  and the numerical results as a function of $1<p<200$. 
Different types of integrals are distinguished by different colors. Here only the integrals that don't cross zero are shown here.}
\label{fig_JKIApprox_PosOnly_err}
\end{figure}

We find that these regularized integrals $\hat{I}$ exhibit exponential decay in the limit $p\to\infty$. 
The reason for this choice is that, in the limit $p\to\infty$, after factoring out the pre-factor $e^p$, Eqs.~(\ref{eq_feja0n})–(\ref{eq_feka0n}) indicate that the remaining part exhibits a leading-order growth approximately of the form $p^p$. 
Accordingly, we introduce a threshold $p_0$, and for $p\ge p_0$ we construct the approximation based on an exponential ansatz rather than continuing with a power series. 
The parameters $(N_l, N_r, p_0)$, as well as the order of the polynomial $\delta_I$, depend on the specific type of integral. 
In most cases, we adopt $N_l=6$, $N_r=2$, and $p_0=10$ and the order of $\delta_I$ is 6 to 12, depends on the required accuracy.
Because the number of integrals involved is extremely large, and the detailed calculations are too cumbersome to present in full, we just show the main construction strategy here.
And the construction procedure for the derivative term $\partial_e\feJ{(p,1,b)}$ is the same, and we therefore do not repeat it.
The specific format can be found in the open-source code module \texttt{pyPNFourier}.\footnote{https://github.com/Shallyn/pyPNFourier}

The key advantage of this approach is that it preserves the correct asymptotic behavior in both limits $e\to0$ and $e\to1$, without mutual contamination between the two regimes. 
The accuracy can then be systematically improved by increasing the order of the polynomial $\delta_I$. 
After a series of involved calculations, we ultimately construct such endpoint-constrained analytic approximations for all integrals listed in Table~\ref{tab_Integrals} for $p\le 200$.
Since the harmonics waveform scale as $H_p\sim\order{e^p}$, one can make a rough priori estimate of the expected accuracy. For example, $0.9^{200}\sim10^{-10}$, whereas $0.95^{200}\sim10^{-5}$.
However, as $e\to1$, the number of harmonics required increases rapidly. Therefore, whether the range $p\le200$ considered here is sufficient requires careful assessment. In addition, our modeling targets the integrals appearing in $H_{p}$, rather than $H_{p}$ itself. Consequently, combinations such as $\Delta_e^n\feJ{(p,a,b)}$ frequently appear in the calculation of $H_{p}$, implying that the accuracy of $\feJ{(p,a,b)}$ itself is not necessarily identical to the accuracy of $H_{p}$.
Therefore, in practical waveform modeling, a more appropriate strategy would be to apply the two-point constrained approximation directly to $H_{p}$ itself, rather than to the individual integrals. However, since the explicit form of $H_{p}$ depends on the chosen gauge and dynamical variables, in the present work we focus only on approximating the integrals themselves.

Next, we provide a more detailed quantitative analysis from three perspectives: accuracy, computational speed and waveform-level performance. 
This allows us to obtain a more precise estimate of the regime of validity and practical limitations of this approximate model.

\begin{figure*}[t]
\begin{tabular}{cc}
\includegraphics[width=0.5\textwidth]{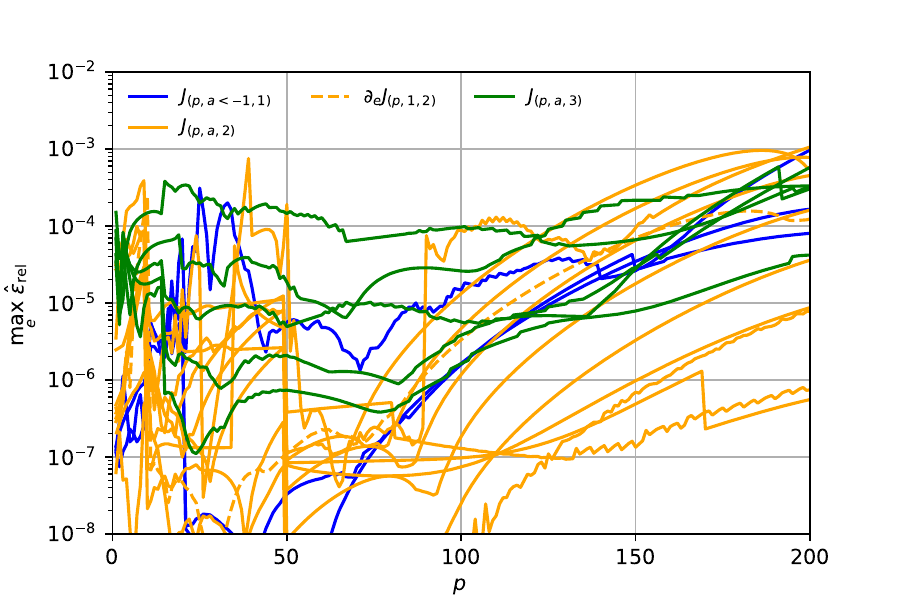}
&\includegraphics[width=0.5\textwidth]{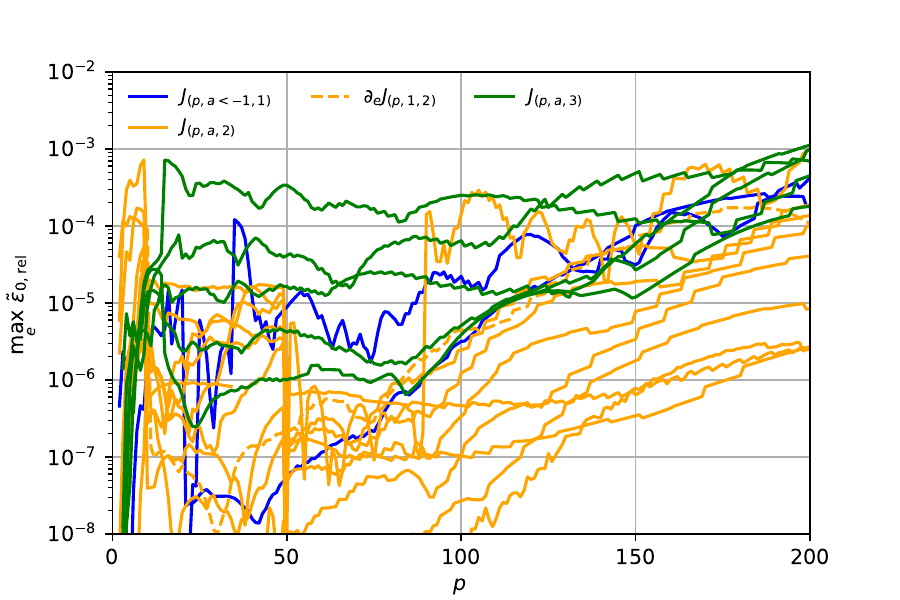}
\end{tabular}
\caption{Left panel: the maximum relative error between the approximate expressions and the numerical results over $0<e<1$, after excluding the neighborhoods of the zero crossings, as a function of $1<p<200$. 
Right panel: the maximum ARMS-amplitude-normalized error near the zero crossings as a function of $1<p<200$.
Different types of integrals are distinguished by different colors.}
\label{fig_JKIApprox_Zero_err}
\end{figure*}

\subsubsection{Accuracy}
In FIG.~\ref{fig_compare_JpabApprox_examples}, we present two illustrative comparisons for $p=10$ and $p=100$, respectively. 
The agreement is visually indistinguishable. 
To quantify the accuracy we evaluate the error between the analytic approximations and the numerical results of $\hat{I}$ for all cases and several worse fitted cases.

It should be noted that, since the regularized integrals $\hat{I}$ can span several orders of magnitude, using only the absolute error is clearly insufficient. 
However, $\hat{I}$ may also cross zero, in which case the relative error can become ill-defined or misleading. 
Therefore, for the purpose of comparing the approximation accuracy, we divide these integrals into two classes. The first class consists of integrals for which $\hat{I}>0$ or $\hat{I}<0$ throughout the entire domain $0<e<1$ and $p\le200$. For this class, we show only the relative error,
$\epsilon_{\mathrm{rel}}:=|\hat{I}_{\mathrm{num}}-\hat{I}_{\mathrm{approx}}|/{|\hat{I}_{\mathrm{num}}|}.$
The second class consists of integrals that cross zero within the parameter domain.

The first class includes the following integrals: all $\feJ{(p,a,0)}$ and $\feK{(p,a,0)}$, as well as $\feJ{(p,a>-2,b>0)}$. 
All other integrals exhibit zero crossings. The maximum relative errors among $0<e<1$ of the approximate expressions for the first class of integrals are shown in FIG.~\ref{fig_JKIApprox_PosOnly_err}. We find that the relative errors of all approximations in this class are controlled below $10^{-3}$. For $p<50$, the accuracy is even higher, with the relative errors generally remaining below $10^{-4}$.

For the remaining integrals, our treatment is as follows. 
Suppose that $\hat{I}(e_0)=0$. We remove a small neighborhood around the zero crossing, $|e-e_0|<\delta$, so that the accuracy in the remaining region can still be quantified using the relative error. 
For the near-zero region, $|e-e_0|<\delta$, we define
$\tilde\epsilon_{0,\mathrm{rel}} =  |\hat{I}_{\mathrm{num}} - \hat{I}_{\mathrm{approx}}| / S_I$, where
\begin{align} 
    &S_I^2:=(2\delta)^{-1}\int_{e_0-\delta}^{e_0+\delta} \hat{I}^2_{\mathrm{num}}(e) \dd{e},
\end{align}
This quantity measures the average root-mean-square (ARMS) amplitude within a small neighborhood of the zero crossing. 
So that $\tilde{\epsilon}_{0,\mathrm{rel}}$ characterizes the error level near the zero crossing relative to the ARMS amplitude in that local region.
We tested three choices, $\delta=(0.005,0.01,0.02)$, and found that the results are not significantly different. Here, we show the case for $\delta=0.01$.
The results are shown in FIG.~(\ref{fig_JKIApprox_Zero_err}). 
We find that the maximum relative errors in the nonzero regions are generally controlled below $10^{-3}$, while the ARMS-amplitude-normalized errors near the zero crossings also remain approximately below $10^{-3}$.

\subsubsection{Computational speed}

\begin{figure*}[t]
\begin{tabular}{c}
\includegraphics[width=1\textwidth]{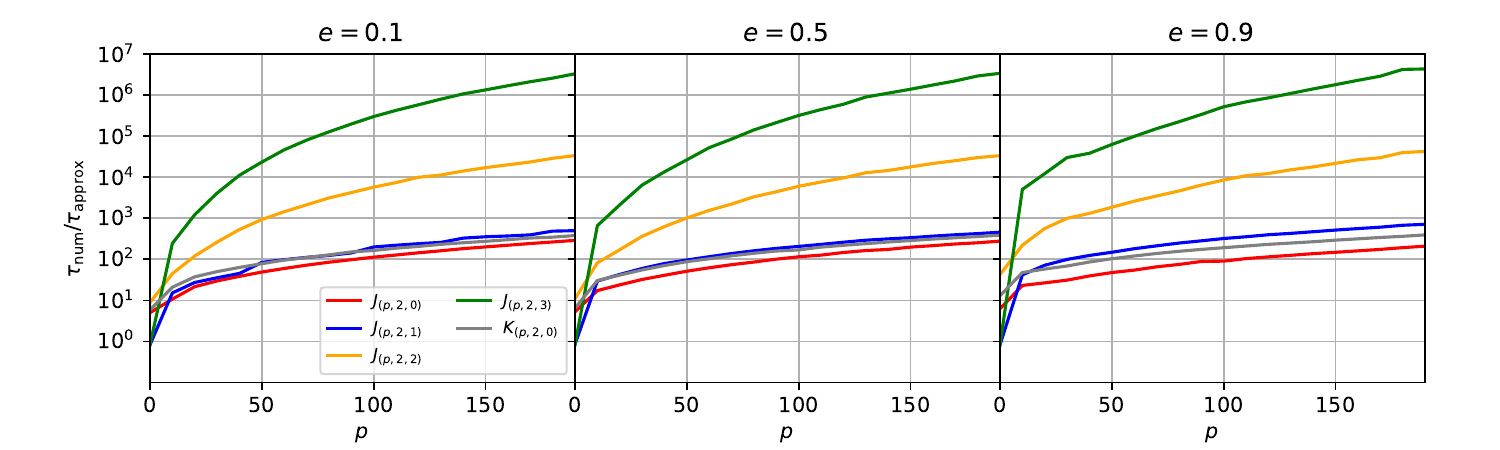}
\end{tabular}
\caption{Ratio between the average numerical evaluation time $\tau_{\mathrm{num}}$ and the average evaluation time of the approximate expressions $\tau_{\mathrm{approx}}$ for several representative integrals and different combinations of $(e,p)$. 
From left to right, the three panels correspond to $e=0.1,0.5,0.9$, respectively. 
The target accuracy of the numerical evaluation is set to $10^{-4}$.
Different integrals are shown using different colors.
}
\label{fig_compspeed}
\end{figure*}

In~\cite{work1}, we computed $\feJ{(p,a,b)}$ and $\feK{(p,a,b)}$ using a Fourier-decomposition method. 
However, as $e\to1$ and $p\to\infty$, the number of Bessel functions required to achieve the target accuracy grows rapidly, leading to a substantial increase in computational cost. 
As can be seen from Eqs.~(37)--(41) of~\cite{work1}, each increase of $b$ by one introduces an additional nested summation. 
Therefore, the slowest case is the family of integrals $\feJ{(p,a,3)}$. 
As a result, without relaxing the accuracy requirement, the cost of computing high-eccentricity waveforms using the previous method becomes prohibitive.

Here, we select $5$ representative integrals, namely $\feJ{(p,a,b)}$ and $\feK{(p,a,0)}$ with $a=2$ and $b=(0,1,2,3)$, as examples to quantitatively compare the computational speed of the exact numerical method in~\cite{work1} with that of the approximate expressions for different combinations of $(p,e)$. 
The comparison is performed on a MacBook Pro laptop equipped with an Apple M3 chip and running macOS Tahoe 26.5.2.

For each integral, we estimate the computational time $\tau$ by taking the median over multiple runs. We then compute the speed-up factor, $\tau_{\mathrm{num}}/\tau_{\mathrm{approx}}$, which quantifies how much more time the numerical method requires relative to the approximate expression for different input-parameter combinations. The target numerical accuracy is set to $10^{-4}$.



Our tests show that, since the approximate expressions are fully analytic, their evaluation time is almost independent of the input parameters. 
The typical computational time is approximately $\tau_{\mathrm{approx}}\lesssim 10^3\mathrm{ns}$. 
The results are shown in FIG.~\ref{fig_compspeed}. 
We find that, for relatively small eccentricities, the overall speedup is not particularly significant. 
However, at larger eccentricities, the speedup becomes more pronounced as $p$ and $e$ increase, reaching as high as $\order{10^6}$ in some cases.
More detailed partial results are shown in Table~\ref{tab_speed}.

\renewcommand{\arraystretch}{1.2}
\setlist[itemize]{nosep,leftmargin=*} 
\begin{longtable}{|c|c|c|c|c||c|c|c|c|c|}
\caption{Comparison of the median computational speeds of the numerical and approximate algorithms for several representative integrals.}\label{tab_speed} \\

\toprule
\hline
Integrals & eccentricity & Fourier index & $\tau_{\mathrm{num}}$ & $\tau_{\mathrm{approx}}$ & Integrals & eccentricity & Fourier index & $\tau_{\mathrm{num}}$ & $\tau_{\mathrm{approx}}$ \\
\hline
\midrule
\endfirsthead



\bottomrule
\hline
\endlastfoot

$\feJ{(p,2,0)}$ & $e=0.1$ & $p=10$ & ${4.75}\mu\mathrm{s}$ & $472\mathrm{ns}$ &
$\feJ{(p,2,1)}$ & $e=0.1$ & $p=10$ & ${6.78}\mu\mathrm{s}$ & $443\mathrm{ns}$\\
\hline
$\feJ{(p,2,0)}$ & $e=0.5$ & $p=10$ & ${7.59}\mu\mathrm{s}$ & $456\mathrm{ns}$ &
$\feJ{(p,2,1)}$ & $e=0.5$ & $p=10$ & ${13.1}\mu\mathrm{s}$ & $442\mathrm{ns}$ \\
\hline
$\feJ{(p,2,0)}$ & $e=0.9$ & $p=10$ & ${10.1}\mu\mathrm{s}$ & $465\mathrm{ns}$ &
$\feJ{(p,2,1)}$ & $e=0.9$ & $p=10$ & ${18.5}\mu\mathrm{s}$ & $520\mathrm{ns}$ \\
\hline
$\feJ{(p,2,0)}$ & $e=0.1$ & $p=100$ & ${49.5}\mu\mathrm{s}$ & $463\mathrm{ns}$ &
$\feJ{(p,2,1)}$ & $e=0.1$ & $p=100$ & ${86.5}\mu\mathrm{s}$ & $449\mathrm{ns}$ \\
\hline
$\feJ{(p,2,0)}$ & $e=0.5$ & $p=100$ & ${50.1}\mu\mathrm{s}$ & $455\mathrm{ns}$ &
$\feJ{(p,2,1)}$ & $e=0.5$ & $p=100$ & ${92.4}\mu\mathrm{s}$ & $477\mathrm{ns}$ \\
\hline
$\feJ{(p,2,0)}$ & $e=0.9$ & $p=100$ & ${41.4}\mu\mathrm{s}$ & $458\mathrm{ns}$ &
$\feJ{(p,2,1)}$ & $e=0.9$ & $p=100$ & ${141}\mu\mathrm{s}$ & $458\mathrm{ns}$ \\
\hline
\hline
$\feJ{(p,2,2)}$ & $e=0.1$ & $p=10$ & ${19.6}\mu\mathrm{s}$ & $456\mathrm{ns}$ &
$\feJ{(p,2,3)}$ & $e=0.1$ & $p=10$ & ${108}\mu\mathrm{s}$ & $449\mathrm{ns}$ \\
\hline
$\feJ{(p,2,2)}$ & $e=0.5$ & $p=10$ & ${36.2}\mu\mathrm{s}$ & $451\mathrm{ns}$ &
$\feJ{(p,2,3)}$ & $e=0.5$ & $p=10$ & ${296}\mu\mathrm{s}$ & $443\mathrm{ns}$ \\
\hline
$\feJ{(p,2,2)}$ & $e=0.9$ & $p=10$ & ${99.1}\mu\mathrm{s}$ & $452\mathrm{ns}$ &
$\feJ{(p,2,3)}$ & $e=0.9$ & $p=10$ & ${2.14}\mathrm{ms}$ & $443\mathrm{ns}$ \\
\hline
$\feJ{(p,2,2)}$ & $e=0.1$ & $p=100$ & ${2.48}\mathrm{ms}$ & $462\mathrm{ns}$ &
$\feJ{(p,2,3)}$ & $e=0.1$ & $p=100$ & ${129}\mathrm{ms}$ & $450\mathrm{ns}$ \\
\hline
$\feJ{(p,2,2)}$ & $e=0.5$ & $p=100$ & ${2.64}\mathrm{ms}$ & $450\mathrm{ns}$ &
$\feJ{(p,2,3)}$ & $e=0.5$ & $p=100$ & ${139}\mathrm{ms}$ & $448\mathrm{ns}$ \\
\hline
$\feJ{(p,2,2)}$ & $e=0.9$ & $p=100$ & ${3.79}\mathrm{ms}$ & $456\mathrm{ns}$ &
$\feJ{(p,2,3)}$ & $e=0.9$ & $p=100$ & ${231}\mathrm{ms}$ & $453\mathrm{ns}$ \\
\hline
\hline
$\feK{(p,2,0)}$ & $e=0.1$ & $p=10$ & ${9.20}\mu\mathrm{s}$ & $495\mathrm{ns}$ &
$\feK{(p,2,0)}$ & $e=0.1$ & $p=100$ & ${73.1}\mu\mathrm{s}$ & $457\mathrm{ns}$ \\
\hline
$\feK{(p,2,0)}$ & $e=0.5$ & $p=10$ & ${13.4}\mu\mathrm{s}$ & $510\mathrm{ns}$ &
$\feK{(p,2,0)}$ & $e=0.5$ & $p=100$ & ${77.6}\mu\mathrm{s}$ & $455\mathrm{ns}$ \\
\hline
$\feK{(p,2,0)}$ & $e=0.9$ & $p=10$ & ${20.7}\mu\mathrm{s}$ & $461\mathrm{ns}$ &
$\feK{(p,2,0)}$ & $e=0.9$ & $p=100$ & ${83.7}\mu\mathrm{s}$ & $458\mathrm{ns}$ \\
\hline
\end{longtable}

We note that, for $e=0.5$, the speedup becomes less significant at large $p$. 
This is because the eccentricity is not very large in this case, so as $p$ increases, these integrals are rapidly suppressed by the factor $(e/2)^p$ and soon satisfy the required accuracy criterion. 
By contrast, as $e\to1$, the suppression associated with the pre-factor $(e/2)^p\to0$ becomes much slower.

\subsubsection{Waveform calculation}

\begin{figure*}[t]
\begin{tabular}{c}
\includegraphics[width=1\textwidth]{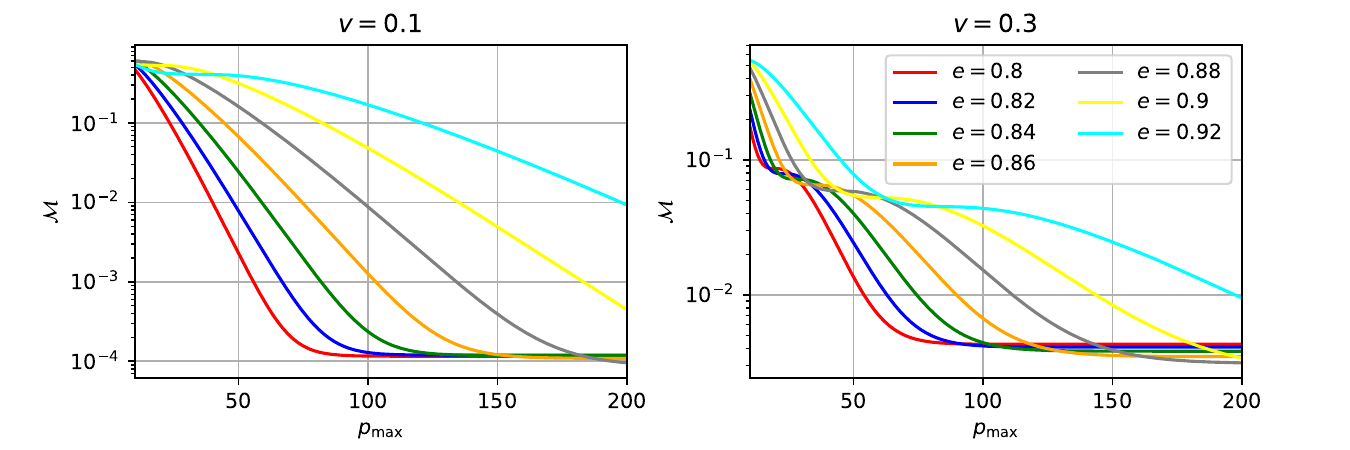}
\end{tabular}
\caption{
Mismatch factors $\mathcal{M}$ between the conservative-orbit waveform $\hat{h}_{22}(e,v,\ell)$ over $l\in(-3\pi,3\pi)$ with $\nu=0.22$ computed using the approximate integrals and the numerical result, for different mode-sum truncation values $p{\max}$. 
The left panel shows the case with $v=0.1$, while the right panel shows the case with $v=0.3$. 
Results for different eccentricities are represented by lines of different colors.
}\label{fig_hhat22_mismatch}
\end{figure*}

Finally, we compare the explicit waveform calculations.
The method proposed in this work only simplifies the evaluation of several difficult integrals. 
These integrals affect only the terms associated with $\tilde{H}_{\ell(-m)p}$ in the formula, and different integrals enter at different PN orders. 
For example, the most difficult integrals to compute, namely the $\feJ{(p,a,3)}$ family, appear only at 3PN order. 
It is therefore conceivable that even if the approximation accuracy of the $\feJ{(p,a,3)}$ integrals is not particularly high, their impact on the final waveform accuracy may still be limited.

\begin{figure*}[t]
\begin{tabular}{c}
\includegraphics[width=1\textwidth]{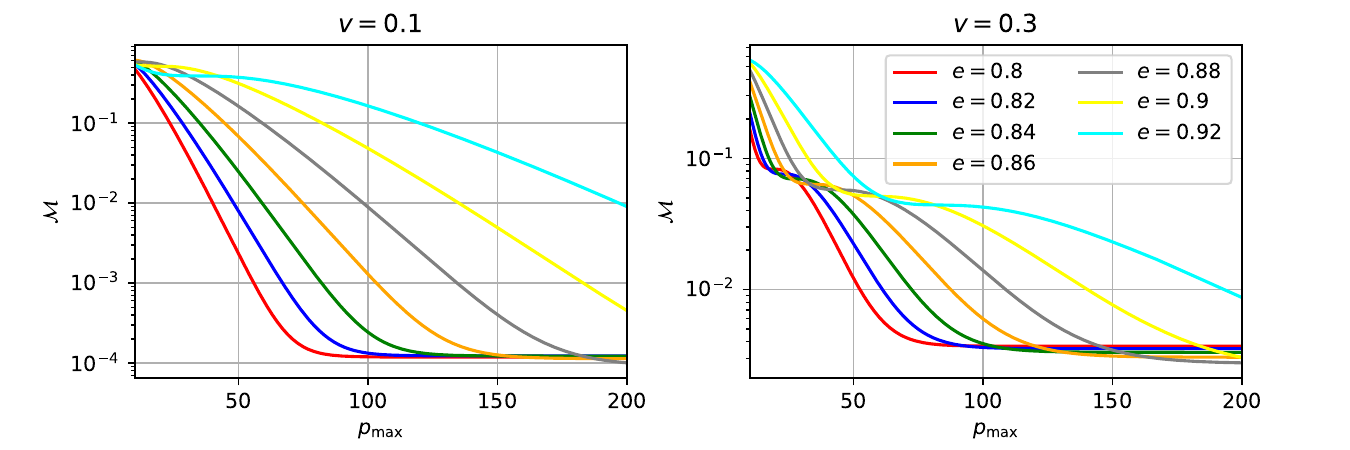}
\end{tabular}
\caption{
Mismatch factors $\mathcal{M}$ between the conservative-orbit waveform $\hat{h}_{22}(e,v,\ell)$ over $l\in(-3\pi,3\pi)$ with $\nu=0.1$ computed using the approximate integrals and the numerical result, for different mode-sum truncation values $p{\max}$. 
The left panel shows the case with $v=0.1$, while the right panel shows the case with $v=0.3$. 
Results for different eccentricities are represented by lines of different colors.
}\label{fig_hhat22_mismatch_q8}
\end{figure*}

For this reason, we do not simply compute the mismatch for a few representative waveform examples. 
Instead, a more comprehensive comparison is required. 
In the following, we use the 3PN-accurate $(\ell,m)=(2,2)$ waveform as a representative case for assessing waveform-level accuracy. 
We temporarily neglect radiation reaction, since the present test does not involve the computation of the dynamical evolution, but focuses instead on the waveform agreement at different eccentricities.

We first sample $\ell\in(-3\pi,3\pi)$ and compute
\begin{align}
    \hat{h}_{22} := \sum_{|p|=0}^{p_{\max}} \big( \tilde{H}_{2(-2)p}^N + v^2\tilde{H}_{2(-2)p}^{\fPN{1}} + v^3\tilde{H}_{2(-2)p}^{\fPN{1.5},\tail} + v^4\tilde{H}_{2(-2)p}^{\fPN{2}} + v^5\tilde{H}_{2(-2)p}^{\fPN{2.5},\tail} + v^6\tilde{H}_{2(-2)p}^{\fPN{3}} + v^6\tilde{H}_{2(-2)p}^{\fPN{3},\tail} \big)\ee^{\ii pl}
\end{align}
with the prescribed numerical accuracy, namely $10^{-10}$.
We tested several cases with $m_1/m_2<20$, but found no significant difference in their accuracy. 
This is because the waveforms used here are only post-Newtonian waveforms and do not involve comparisons with numerical-relativity waveforms. 
In the post-Newtonian waveform, the mass ratio enters polynomially. 
For example, one may write $\hat{H}^{\fPN{1}}=\nu \hat{H}^{\fPN{1},1}+\hat{H}^{\fPN{1},0}$. 
The accuracy of both $\hat{H}^{\fPN{1},1}$ and $\hat{H}^{\fPN{1},0}$ affects the waveform accuracy; however, these terms depend only on the eccentricity, and their structures are also similar. 
Therefore, if the accuracy of the constituent integrals is problematic, it affects $\hat{H}^{\fPN{1},1}$ and $\hat{H}^{\fPN{1},0}$ in a similar manner, leading to only weak differences in waveform accuracy for different mass ratios.
Therefore, we show only the case with $m_1/m_2=2$, corresponding to $\nu=0.22$. 
We then compute $\hat{h}_{22}(p_{\max})$ using the approximate expressions with different summation cutoffs $p_{\max}$, and evaluate its agreement with the numerical result using the standard waveform-comparison quantity, the mismatch factor $\mathcal{M}$ defined via
\begin{align}
    \mathcal{M} := 1 - \frac{\langle h_1|h_2 \rangle}{\sqrt{\langle h_1|h_1\rangle \langle h_2|h_2 \rangle}}.
\end{align}
The definition of the inner product $\langle \cdot | \cdot \rangle$ used here differs from that commonly adopted in data analysis: we use a uniform PSD weight,
\begin{align}
    \langle h_1|h_2 \rangle \propto \Re \int \tilde{h}_1(f)\tilde{h}^*_2(f) \dd{f}.
\end{align}
This is a simplifying choice. 
The main reason is that introducing a PSD weighting would require us to account for the dependence on the frequency, the total mass, and other source parameters, which is beyond the scope of the present work. 
We therefore adopt a uniform weight, which is sufficient for providing a quantitative measure of the waveform accuracy in the present comparison.
Since the comparison is performed only at the waveform level, and the dynamical variables used to compute the waveforms are identical, the inner-product calculation here does not involve any maximization over time or phase shifts.

\begin{figure*}[t]
\begin{tabular}{c}
\includegraphics[width=1\textwidth]{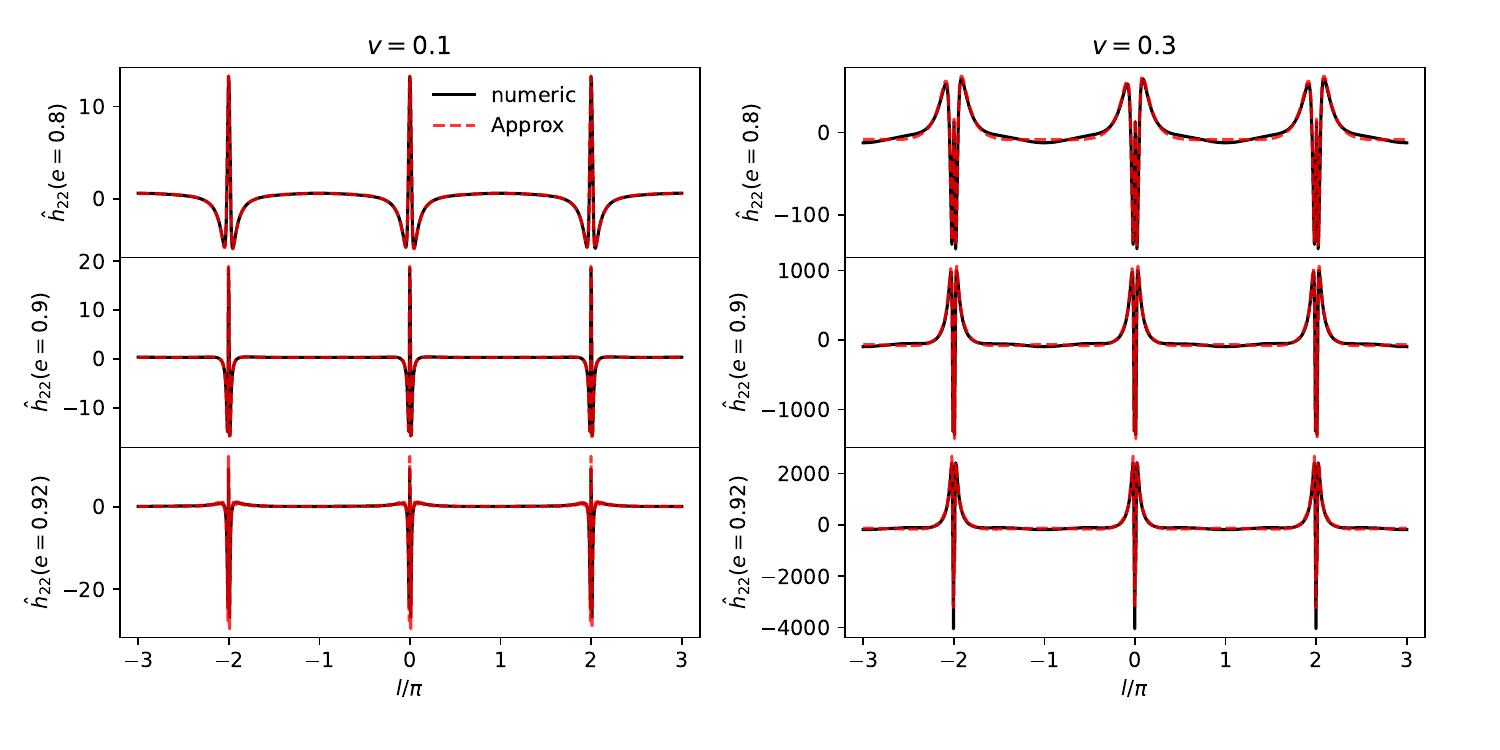}
\end{tabular}
\caption{
Direct comparison of the conservative-orbit waveform $\hat{h}_{22}(e,v,\ell)$ for $e=(0.8,0.9,0.92)$ with $\nu=0.22$. 
The black solid line denotes the numerical result, while the red dashed line denotes the result obtained by computing the mode sum up to $p_{\max}=200$ using the approximate integrals. 
The left panels show the cases with $v=0.1$, and the right panels show the cases with $v=0.3$; the upper panels correspond to $e=0.8$, and the lower panels correspond to $e=0.9$.
}\label{fig_hhat22_compare}
\end{figure*}

We consider seven eccentricities, $e=(0.80,0.82,0.84,0.86,0.88,0.90,0.92)$, and two orbital velocities, $v=(0.1,0.3)$, representing different stages of the orbital evolution. 
The mismatch factor $\mathcal{M}(p_{\max})$ of the approximate waveform $\hat{h}_{22}$ for different cases is shown in FIG.~(\ref{fig_hhat22_mismatch}).
Here we set the precision of the calculated numerical waveform to $10^{-10}$.
We find that, for the early inspiral case with $v=0.1$ and $e\le0.9$, the approximate waveform can keep the mismatch factor below $10^{-3}$ for $p\le200$. 
However, for $v=0.3$, where post-Newtonian effects become more significant, the overall mismatch factor is only at the level of $10^{-2}$. 
The results also indicate that $e=0.9$ is approximately the boundary of the parameter region in which the present level of accuracy can be maintained.
In addition, we observe that, for both $v=0.1$ and $v=0.3$, once $p_{\max}$ becomes sufficiently large, the mismatch no longer decreases. 
This behavior is caused by the intrinsic inaccuracy of the approximate expressions themselves. 
Moreover, the minimum mismatch is higher for $v=0.3$, indicating that the approximate evaluation of the integrals associated with higher PN-order contributions is less accurate.

We also provide a comparison for a different mass ratio in Fig.~\ref{fig_hhat22_mismatch_q8}. This figure is similar to Fig.~\ref{fig_hhat22_mismatch}, except that the mass ratio is set to $\nu=0.1$. 
As can be seen, the results are almost indistinguishable from those for $\nu=0.22$.

In FIG.~(\ref{fig_hhat22_compare}), we provide a more direct comparison between the approximate and numerical results for $\hat{h}_{22}(e,v,l)$. 
For $v=0.1$, the difference is almost indistinguishable, whereas for $v=0.3$, the discrepancy becomes more visible. 
Although the performance is worse for $v=0.3$, which represents a relatively late PN regime, where higher-order PN contributions are substantially enhanced.
In most realistic scenarios, the eccentricity is not expected to remain this large by the time the orbital evolution reaches $v=0.3$.
Based on the above analysis, we conclude that, in practical waveform calculations, the parameter boundary for maintaining an accuracy of $\mathcal{M}\lesssim10^{-3}$ is roughly around $e\sim0.9$. 
More precisely, this level of accuracy is expected for waveforms with initial conditions satisfying $v\lesssim0.1$ and $e\lesssim0.9$.

Finally, we present several explicit calculations of 3PN high-eccentricity gravitational waveform $h_{22}$. 
We consider five families of high-eccentricity waveforms with initial eccentricities $e=(0.7,0.75,0.8,0.85,0.9)$. 
The endpoint of the orbital evolution is set to $v=1/\sqrt{6}$, while the initial conditions are chosen such that the total evolution time is $t=10000M$.

\begin{figure}[t]
\begin{tabular}{c}
\includegraphics[width=0.5\textwidth]{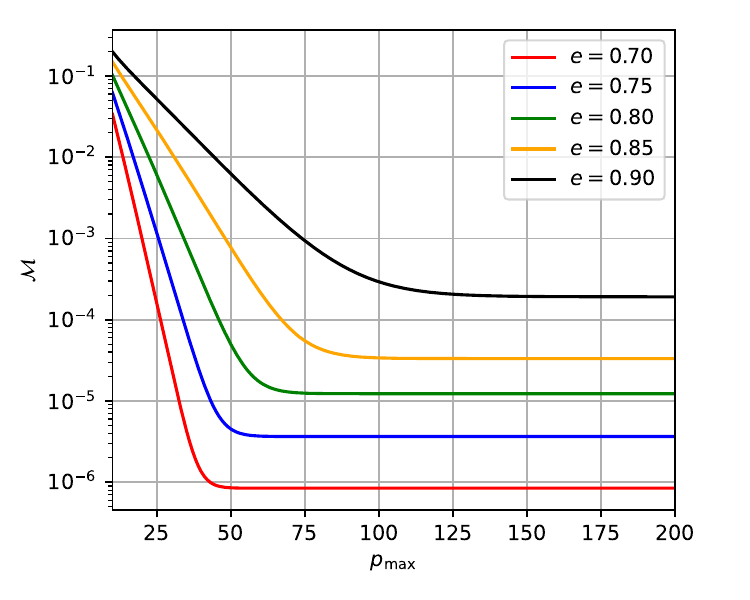}
\end{tabular}
\caption{
Mismatch factor between the 3PN high-eccentricity $h_{22}$ gravitational waveform computed using the approximate integrals and the numerical result, for different mode-sum truncation cutoffs. 
The waveform parameters are $\nu=0.22$, with initial eccentricities $e=(0.7,0.75,0.8,0.85,0.9)$. 
The orbital evolution is terminated at $v=1/\sqrt{6}$, and the initial velocity is chosen such that the total evolution time is $t=10000M$. 
Different eccentricities are represented by different colors.
}
\label{fig_h22_mismatch}
\end{figure}

\begin{figure*}[t]
\begin{tabular}{c}
\includegraphics[width=1\textwidth]{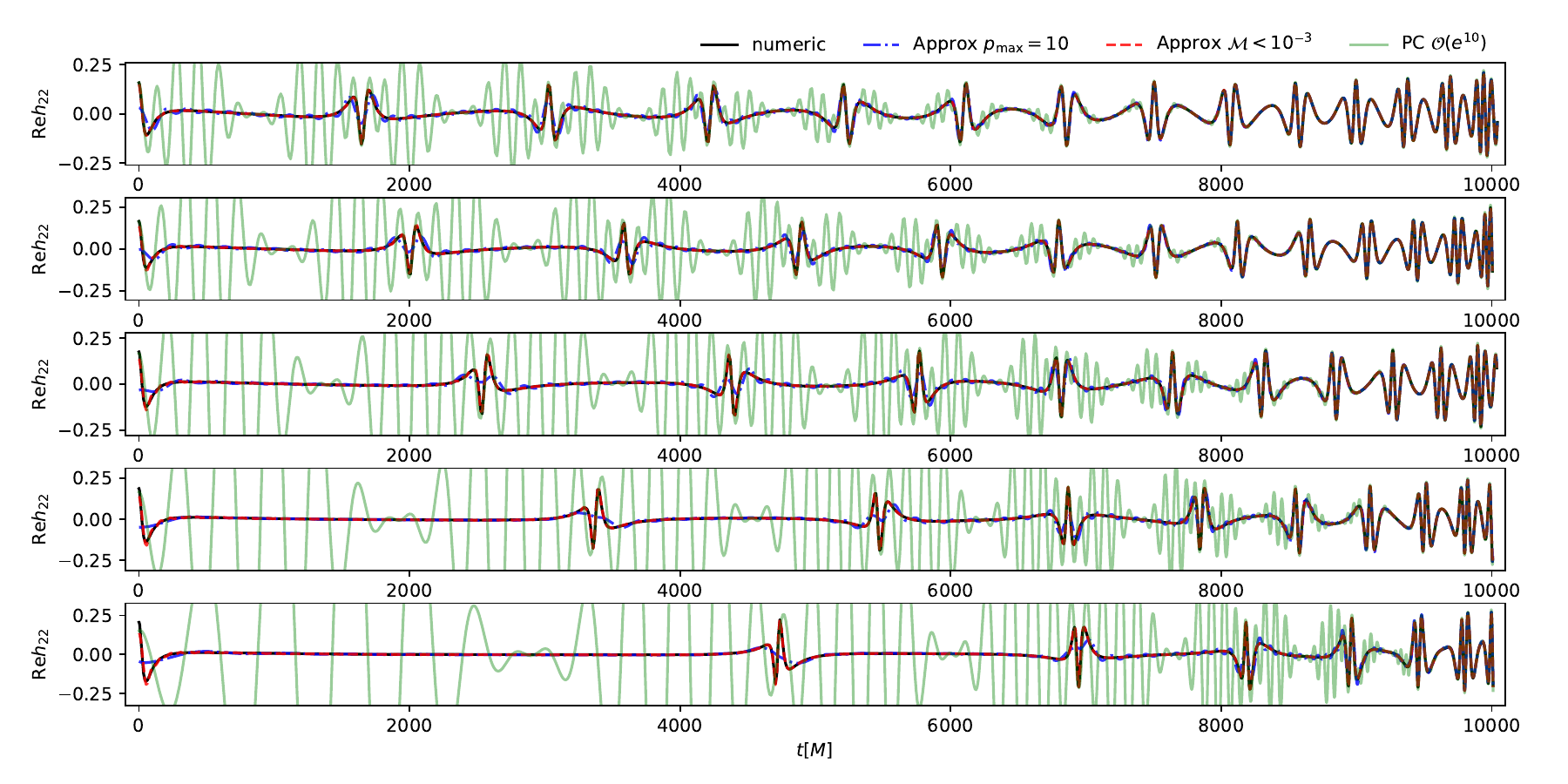}
\end{tabular}
\caption{Comparison of the $h_{22}$ waveforms with different initial eccentricities, $e=(0.7,0.75,0.8,0.85,0.9)$, corresponding to the five panels from top to bottom. The orbital evolution is terminated at $v=1/\sqrt{6}$, the mass ratio is $\nu=0.22$, and the initial velocity is chosen such that the total evolution time is $t=10000M$. The black line denotes the numerical result, the red dashed line denotes the waveform obtained by summing modes using the approximate integrals until the mismatch factor falls below $10^{-3}$, the blue dashed line denotes the waveform obtained by summing modes using the approximate integrals up to $p=10$, and the green line denotes the post-circular expansion through $e^{10}$.
}\label{fig_h22_example}
\end{figure*}

We compare the numerical waveforms with three approximate waveforms: one obtained by summing the modes using the approximate expressions until the accuracy reaches $\mathcal{M}\sim10^{-3}$, one obtained by truncating the mode sum at $p=10$, and one obtained from the post-circular expansion through $e^{10}$. 
The real parts of these waveforms are shown in FIG.~(\ref{fig_h22_example}). 
Similarly, we compute the mismatch factors for waveforms obtained with different numbers of summed modes, and show the results in FIG.~(\ref{fig_h22_mismatch}).

For these cases, the approximate integral expressions perform well in waveform calculations. 
Even for $e_0=0.9$, the mismatch factor can be controlled below $10^{-3}$. 
For smaller eccentricities, the mismatch factor is even lower. 
On the other hand, we find a significant difference between the post-circular waveform and the waveform obtained by summing modes up to $p=10$. 
Since, theoretically, $\hat{H}_{\ell(-m)p}\sim\order{e^p}$, the difference between the two essentially reflects whether higher-order eccentricity contributions to $\hat{H}_{\ell(-m)p}$ are included. 
The comparison shows that, in the high-eccentricity regime, even a mode sum truncated at $p=10$ yields a waveform accuracy much higher than that of a simple post-circular expansion at the same nominal eccentricity order. 
This provides useful guidance for future waveform-modeling work.

\section{Large eccentricity asymptotic expansion of eccentricity enhancement functions}\label{sec_LEA_eef}

In this section, we present another application of the UAE derived in Section~\ref{sec_UAE}, namely the computation of the large-eccentricity asymptotic expansion of the eccentricity enhancement function.

The eccentricity enhancement functions (EEF) would appear during evaluating the orbit-averaged energy and angular momentum fluxes that contributed by tail parts.
Up to 3PN order and in MH coordinate, the tail-contributed, orbit-averaged energy flux $\avg{\Eflux_\tail}$ and angular momentum flux $\avg{\Jflux_\tail}$ in terms of time eccentricity $e\equiv e_t$ are given by~\cite{Arun_2008,Arun_2008_tail,Arun_2009_tail}
\begin{align}
& \avg{\Eflux_{\tail}} = \frac{32\pi\nu^2}{5} v^{13} \Bigg\{ 4\varphi(e) + v^2\Bigg[ 4 \bigg( \varphi_{(0,1)}(e) - \frac{1}{2}\nu\varphi(e) - \frac{21}{1-e^2}\big( \varphi(e) - \varphi_{(1,0)}(e) \big) \bigg) + (1-4\nu)\bigg( \frac{16403}{2016}\beta(e) \nonumber\\
&\quad + \frac{1}{18}\gamma(e)\bigg) \Bigg] + v^3\bigg[ \bigg(-\frac{515063}{11025} + \frac{16\pi^2}{3} - \frac{856}{35}\lgxOxp \bigg)F(e) - \frac{1712}{105}\chi(e) \bigg] + \order{c^{-4}} \Bigg\} ,\\
& \avg{\Jflux_{\tail}} = \frac{32\pi\nu^2}{5} v^{10} \Bigg\{ 4\tilde\varphi(e) + v^2\Bigg[ 4 \bigg( \tilde\varphi_{(0,1)}(e) - \frac{1}{2}\nu\tilde\varphi(e) - \frac{18}{1-e^2}\big( \tilde\varphi(e) - \tilde\varphi_{(1,0)}(e) \big) \bigg) + (1-4\nu)\bigg( \frac{16403}{2016}\tilde\beta(e) \nonumber\\
&\quad + \frac{1}{18}\tilde\gamma(e)\bigg) \Bigg] + v^3\bigg[ \bigg(-\frac{515063}{11025} + \frac{16\pi^2}{3} - \frac{856}{35}\lgxOxp \bigg)\tilde{F}(e) - \frac{1712}{105}\tilde\chi(e) \bigg] + \order{c^{-4}} \Bigg\},
\end{align}
where $x'_0$ is constant~\cite{Blanchet_2008,Blanchet_2000}. The EEF is defined by summing the product of the multipole moments,
\begin{align}
&\varphi(e) = \frac{v^8}{32\nu^2}\sum_{p=1}^\infty p^7 \mathrm{PN}_{0}\Big[\fmodeMp{p}{ij}\fmodeconjMp{p}{ij}\Big],\qquad 
\tilde\varphi(e) = -\ii\frac{v^8}{16\nu^2}\epsilon_{zab}\sum_{p=1}^\infty p^6 \mathrm{PN}_{0}\Big[\fmodeMp{p}{ak}\fmodeconjMp{p}{bk}\Big],\\
&\varphi_{(a,b)}(e) := \frac{v^{8-2b}}{32\nu^2} \sum_{p=1}^\infty p^{7-a}\mathrm{PN}_{b}\Big[\fmodemnMp{p}{ij}{a}\fmodeconjMp{p}{ij}\Big],\qquad \tilde\varphi_{(a,b)}(e) := -\ii\frac{v^{8-2b}}{16\nu^2} \epsilon_{zij}\sum_{p=1}^\infty p^{6-a}\mathrm{PN}_{b}\Big[\fmodemnMp{p}{ik}{a}\fmodeconjMp{p}{jk}\Big], \\
&F(e) = \frac{v^8}{64\nu^2}\sum_{p=1}^\infty p^8 \mathrm{PN}_{0}\Big[\fmodeMp{p}{ij}\fmodeconjMp{p}{ij}\Big],\qquad 
\tilde{F}(e) = -\ii\frac{v^8}{32\nu^2}\epsilon_{zab}\sum_{p=1}^\infty p^7 \mathrm{PN}_{0}\Big[\fmodeMp{p}{ak}\fmodeconjMp{p}{bk}\Big],\\
&\chi(e) = \frac{v^8}{64\nu^2}\sum_{p=1}^\infty p^8 \ln\bigg(\frac{p}{2}\bigg) \mathrm{PN}_{0}\Big[\fmodeMp{p}{ij}\fmodeconjMp{p}{ij}\Big],\qquad
\tilde\chi(e) = -\ii\frac{v^8}{32\nu^2}\epsilon_{zab}\sum_{p=1}^\infty p^7 \ln\bigg(\frac{p}{2}\bigg)\mathrm{PN}_{0}\Big[\fmodeMp{p}{ak}\fmodeconjMp{p}{bk}\Big], \\
&\beta(e) = \frac{20v^{12}}{49209\nu^2(1-4\nu)}\sum_{p=1}^\infty p^9 \mathrm{PN}_{0}\Big[\fmodeMp{p}{ijk}\fmodeconjMp{p}{ijk}\Big],\qquad 
\tilde\beta(e) = -\ii\frac{20v^{12}}{16403\nu^2(1-4\nu)}\epsilon_{zab}\sum_{p=1}^\infty p^8 \mathrm{PN}_{0}\Big[\fmodeMp{p}{ajk}\fmodeconjMp{p}{bjk}\Big], \\
&\gamma(e) = \frac{4v^6}{\nu^2(1-4\nu)} \sum_{p=1}^\infty p^7 \mathrm{PN}_{0}\Big[\fmodeSp{p}{ij}\fmodeconjSp{p}{ij}\Big],\qquad 
\tilde\gamma(e) = -\ii \frac{8v^6}{\nu^2(1-4\nu)} \epsilon_{zab}\sum_{p=1}^\infty p^6 \mathrm{PN}_{0}\Big[\fmodeSp{p}{ak}\fmodeconjSp{p}{bk}\Big],
\end{align}
where
\begin{align}
    \fmodeMp{p}{L} &: = \sum_{m=-\ell}^\ell \fmodeM{p-m}{m}{L}, \\
    \fmodemnMp{p}{L}{s} &: = \sum_{m=-\ell}^\ell m^s\fmodeM{p-m}{m}{L}.
\end{align}
And the symbol $\mathrm{PN}_n$ denotes taking the $n$-th PN term.
The Fourier coefficients of multipole moments $\fmodeM{p}{m}{L},\fmodeS{p}{m}{L}$ can be represented by Bessel functions and PN-elliptic integrals~\cite{Munna_2020}, for example,
\begin{align}
&\fmodeM{p-2}{2}{xx} = \frac{\nu}{v^4e^2p^2}\bigg[ e\sqrt{1-e^2}\Big(1 - \sqrt{1-e^2}p\Big)J'_p(ep) + \frac{1}{2}\Big(2-e^2 -2 (1-e^2)^{3/2}p\Big)J_p(ep) \bigg] + \nonumber\\
&\quad + \frac{\nu}{84v^2e^2p^3(1-e^2)} \Bigg\{ J_p(ep)\bigg[ (1-e^2) \Big(4 (e^2-1)^2 (3 \nu -1) p^3-2 \sqrt{1-e^2} p^2 (11 e^2 (\nu +2)+378 \sqrt{1-e^2}+73 \nu -113) \nonumber\\
&\quad +p (e^2 (17 \nu -57)+2 (630 \sqrt{1-e^2}+67 \nu -195))-756\Big)+1008 \sqrt{1-e^2} p+84 (2 p-9) \bigg] \nonumber\\
&\quad + 2eJ'_p(ep)\Bigg[ \sqrt{1-e^2} \Big(\nu  p (6 (e^2-1)^2 p^2-8 \sqrt{1-e^2} e^2 p-73 \sqrt{1-e^2} p-25 e^2+67) \nonumber\\&\quad +\sqrt{1-e^2} p \big(\sqrt{1-e^2} \big(p (2 (e^2-1) p+23 \sqrt{1-e^2}-378)-111\big)+90 p+819\big)-756\Big)+315 p \Bigg] \Bigg\} + \nonumber\\
&\quad + 42p^3\Bigg[ 3\feJ{(p,-3,1)}(e) + \sqrt{1-e^2}\bigg[] 2p\feJ{(p,-4,1)}(e) - 2\feJ{(p,-3,0)}(e) + \sqrt{1-e^2}\Big( (3-3\sqrt{1-e^2}p)\feJ{(p,-3,1)}(e) \nonumber\\
&\quad + (2+3\sqrt{1-e^2})\feJ{(p,-2,0)}(e) - 12\feJ{(p,-2,1)}(e) + 6(1-e^2) \feJ{(p,-1,1)}(e) \Big) \bigg] \Bigg] + \order{c^{-4}},
\end{align}
Therefore the (Newtonian order) EEF can be represented by the summation of $(J_p(ep))^2$ and its derivatives. 
In practical waveform modeling, only the expansions in the limit $e\to0$ are typically used. 
However, when the orbital eccentricity is large, the errors associated with neglecting the high-eccentricity behavior can accumulate in the phase. 
It is therefore necessary to study the large-eccentricity regime.

In the following subsections we will show how to use our previous expansion to obtain the asymptotic expansion at large eccentricity of these EEF.

\subsection{Asymptotic expansion of Newtonian order EEF}
We define
\begin{align}
X_k(e) := \sum_{p=1}^\infty p^k \big(J_p(ep)\big)^2.
\end{align}
When $k$ is even, $X_k$ admits a closed form, where the closed-form of $F(e)$ and $\tilde{F}(e)$ are well-known
\begin{align}
    &F(e) = \frac{1}{(1-e^2)^{13/2}}\bigg(1+\frac{85 e^2}{6}+\frac{5171 e^4}{192}+\frac{1751 e^6}{192}+\frac{297 e^8}{1024}\bigg), \\
    &\tilde{F}(e) = \frac{1}{(1-e^2)^5}\bigg(1+\frac{229 e^2}{32}+\frac{327 e^4}{64}+\frac{69 e^6}{256}\bigg).
\end{align}
The rest Newtonian order EEF are given by
\begin{align}
&\varphi(e) = \frac{1}{8e^3}\bigg[ e \left(e^2-1\right)^2 X''_3(e)+e \left(1-e^2\right) X''_1(e)+\left(1-e^2\right) X'_1(e)+e \left(2-\frac{4 e^2}{3}\right) X_3(e) \nonumber\\
&\qquad +\left(-2 e^4+5 e^2-3\right) X'_3(e) \bigg], \\
&\tilde\varphi(e) = \frac{\sqrt{1-e^2}}{2e^3}\bigg[ \left(e^2-1\right) e X''_1(e)+\frac{1}{2} e^2 X'_1(e)+\left(e^2-1\right)^2 X'_3(e)+\left(e^2-1\right) e X_3(e) \bigg] = \varphi_{(1,0)}(e), \\
&\tilde\varphi_{(1,0)}(e) = \frac{1}{2e^3}\bigg[ e \left(e^2-1\right)^2 X''_1(e)+e \left(1-e^2\right) X''_{-1}(e)+\left(1-e^2\right) X'_{-1}(e)+e \left(2-\frac{3 e^2}{2}\right) X_1(e) \nonumber\\
&\qquad +\left(-2 e^4+5 e^2-3\right) X'_1(e) \bigg], \\
&\beta(e) = \frac{4}{16403e^5}\bigg[ 30 e \left(e^2-1\right)^4 X''_5(e)-30 \left(5 e^2-11\right) \left(e^2-1\right)^3 X'_5(e)-30 e \left(7 e^2-10\right) \left(e^2-1\right)^2 X_5(e)\nonumber\\
&\qquad -5 e \left(37 e^4-105 e^2+78\right) \left(e^2-1\right) X''_3(e)+12 e \left(6 e^4-15 e^2+10\right) X''_1(e)-5 \left(e^4+39 e^2-66\right) \left(e^2-1\right) X'_3(e) \nonumber\\
&\qquad +12 \left(6 e^4-15 e^2+10\right) X'_1(e)+120 e \left(e^4-3 e^2+2\right) X_3(e) \bigg], \\
&\tilde\beta(e) = \frac{4\sqrt{1-e^2}}{16403e^5}\Big[ 45 e \left(7 e^2-12\right) \left(e^2-1\right)^2 X''_3(e)-120 e \left(3-2 e^2\right)^2 X''_1(e)+180 \left(e^2-1\right)^4 X'_5(e) \nonumber\\
&\qquad +180 e \left(e^2-1\right)^3 X_5(e)-5 \left(89 e^4-369 e^2+360\right) \left(e^2-1\right) X'_3(e)-24 \left(14 e^4-30 e^2+15\right) X'_1(e) \nonumber\\
&\qquad -720 e \left(e^4-3 e^2+2\right) X_3(e) \Big], \\
&\gamma(e) = \frac{1-e^2}{e}\Big( X'_3(e) + eX''_3(e) \Big), \\
&\tilde\gamma(e) = \frac{2(1-e^2)^{3/2}}{e}X'_3(e), \\
&\chi(e) = -(\ln{2})F(e) + \frac{1}{48e^3}\pdv{k}\bigg[ 3 e \left(e^2-1\right)^2 X''_{k-4}(e)-3 e \left(e^2-1\right) X''_{k-6}(e)+\left(3-3 e^2\right) X'_{k-6}(e)\nonumber\\
&\qquad +e \left(6-4 e^2\right) X_{k-4}(e)+\left(-6 e^4+15 e^2-9\right) X'_{k-4}(e) \bigg]\bigg|_{k=8}, \\
&\tilde\chi(e) = -(\ln{2})\tilde{F}(e) + \frac{\sqrt{1-e^2}}{8e^3}\pdv{k}\bigg[ e \left(2 \left(e^2-1\right) X''_{k-5}(e)+2 \left(e^2-1\right) X_{k-3}(e)\right)+e^2 X'_{k-5}(e)\nonumber\\
&\qquad +2 \left(e^2-1\right)^2 X'_{k-3}(e) \bigg]\bigg|_{k=7}.
\end{align}
None of these functions have closed form.
In order to obtain the asymptotic behavior of $X_k(e)$ at $e\to 1$, we can use the UAE of $J_p(ep)$
\begin{align}
p^{1/3}J_p(ep) = a_0(e)\Ai( p^{2/3}\zeta ) + \frac{a_1(e)}{p^{4/3}}\Ai'( p^{2/3}\zeta ) + \order{p^{-2}} ,\label{eq_UAE_of_BesselJ}
\end{align}
where
\begin{align}
    & a_0(e) = h_0, \ a_1(e) = -\frac{h_0}{48\zeta^2}\bigg(5 + \frac{2\zeta^{3/2}(3\Delta_e^2-5)}{\Delta_e^3} \bigg).
\end{align}

The $X_k(e)$ can be rewritten as
\begin{align}
& X_k(e) = \sum_{p=0}^\infty p^{k-2/3} \bigg[ a_0^2\big( \Ai(p^{2/3}\zeta) \big)^2 + 2p^{-4/3}a_0a_1 \Ai(p^{2/3}\zeta) \Ai'(p^{2/3}\zeta) + \order{p^{-2}} \bigg] \nonumber\\
& \qquad := a_0^2 S\big[ \Ai^2 \big](k-2/3;\zeta) + 2a_0a_1S[\Ai\Ai'](k-2;\zeta) + \sum_p\order{p^{k-10/3}},
\end{align}
where, we define
\begin{align}
S[f](\alpha;\zeta) = \sum_{p=1}^\infty p^\alpha f(p^{2/3}\zeta),
\end{align}
where the function $f(x)$ is analytic as $x\to 0$.
If $\alpha>0$, then this summation will actually be dominated by the large $p$, which corresponds to the UAE case we are considering, that is, $p\to\infty,\zeta\to0$.
However, when $\alpha<0$, the contribution of small $p$ cannot be ignored. 
To obtain the correct asymptotic expression, we should employ expansion
\begin{align}
&\sum_{p=0}^\infty p^\alpha f(p^{2/3}\zeta) = \sum_{p=0}^\infty p^{\alpha} f(0) + \zeta\sum_{p=0}^\infty p^{\alpha+2/3} f'(0) + ...  = \nonumber\\
&\quad = \sum_{j=0}^{j_\alpha}  \frac{\zeta^jf^{(j)}(0)}{j!}\zeta_R\bigg(-\alpha-\frac{2}{3}j\bigg) + ...
\end{align}
where $j_\alpha := \lfloor -3(\alpha+1)/2\rfloor$.
The contribution of negative $\alpha$ is given by the first $j_\alpha$ expansion terms.
Thus we define
\begin{align}
\hat{f}_\alpha(x) = f(x) - \sum_{j=0}^{j_\alpha}\frac{x^jf^{(j)}(0)}{j!}.
\end{align}
The summation is rewritten as
\begin{align}
S[f](\alpha;\zeta) = \sum_{j=0}^{j_\alpha}\frac{\zeta^jf^{(j)}(0)}{j!}\zeta_R\left(-\alpha-\frac{2}{3}j\right) + \sum_{p=0}^\infty p^\alpha \hat{f}_\alpha(p^{2/3}\zeta).
\end{align}
Now the summation is dominated by large $p$, which can be converted to integral through Euler–Maclaurin (EM) formula.
\begin{align}
\sum_{k=m}^n f(k) = \int_m^n f(x) \dd{x} + \frac{1}{2}\left( f(n) + f(m) \right) + \sum_{j=1}^{\left[\lambda/2\right]} \frac{B_{2j}}{(2j)!}  \left( f^{(2j-1)}(n) - f^{(2j-1)}(m) \right) + R_\lambda,
\end{align}
where $B_{j}$ is the $j$th Bernoulli number and the remainder term $R_\lambda$ can be estimated as
\begin{align}
|R_\lambda| \le \frac{2\zeta_R(\lambda)}{(2\pi)^\lambda}\int_m^n |f^{(\lambda)}(x)|\dd{x}.
\end{align}
We then substitute the UAE form of $J_p(ep)$ to Euler–Maclaurin formula and carefully choose an integer $\lambda$ that the remainder term $R_\lambda \sim \order{\zeta^{n+1}}$. 
$S[f](\alpha;\zeta)$ in terms of integral reads
\begin{align}
& S[f](\alpha;\zeta) = \sum_{j=0}^{j_\alpha}\frac{\zeta^jf^{(j)}(0)}{j!}\zeta_R\left(-\alpha-\frac{2}{3}j\right) + \frac{3}{2}\zeta^{-3(\alpha+1)/2}I[f]\left(\frac{3\alpha+1}{2};\zeta\right) + \mathcal{B}^\lambda[f](\alpha;\zeta),\label{eq_Sfzetasum}
\end{align}
If $j_\alpha<0$, the first summation term vanishes. 
$\mathcal{B}^\lambda[f](\alpha;\zeta)$ denotes the boundary term and Bernoulli corrections,
\begin{align}
 \mathcal{B}^\lambda[f](\alpha;\zeta) = \frac{1}{2}f(\zeta) - \sum_{j=1}^{\lfloor\lambda/2\rfloor} \frac{B_{2j}}{(2j)!}\dv[2j-1]{p}\eval{\Big[p^\alpha \hat{f}(p^{2/3}\zeta)\Big]}_{p=1} + R_\lambda,
\end{align}
The integral $I[f]$ is given by
\begin{align}
I[f]\left(\frac{3\alpha+1}{2};\zeta\right) = \int_\zeta^\infty x^{(3\alpha+1)/2} \hat{f}_\alpha(x) \dd{x}.
\end{align}
Note that the lowest power of $x^{(3\alpha+1)/2} \hat{f}_\alpha(x) \sim x^{(3\alpha+1)/2 + j_\alpha + 1} + \order{x^{(3\alpha+1)/2+j_\alpha+2}}$. It may be equal to $-1$, in which case the logarithm term would appear. We define
\begin{align}
&I[f]\left(\frac{3\alpha+1}{2};\zeta \right) := I[f]\left(\frac{3\alpha+1}{2} \right) - \sum_{j=0}^{\infty} \frac{ \hat{f}_\alpha^{(j)}(0)}{j!}\left[ \frac{(1-\delta_{(3\alpha+1)/2+j+1})}{(3\alpha+1)/2+j+1}\zeta^{(3\alpha+1)/2+j+1} + \delta_{(3\alpha+1)/2+j+1} \ln\zeta \right],
\end{align}
The integral $I[f]\big((3\alpha+1)/2\big) := \mathop{\FinitePart}\limits_{\zeta\to0} I[f]\big((3\alpha+1)/2;\zeta\big)$ denotes removing the diverge part of the integrand as $\zeta\to0$. 
Inspection of Eq.~(\ref{eq_Sfzetasum}) shows that the integral-related terms in the UAE of the sum scale as $\order{\zeta^{-(3k+1)/2}}\sim\order{\Delta_e^{-(3k+1)}}$. For $k>0$, these terms are therefore strongly divergent, whereas the other two correction terms remain finite and are both of order $\order{\zeta^0}$.
In this work, we retain terms up to $\order{\Delta_e^4}$. For $k\ge3$, the leading divergent contribution is already of order $\order{\Delta_e^{-7}}$, so the non-integral correction terms lie beyond the order retained and can be neglected. Although non-integral corrections may still contribute for $k=1$, the expression for the EEF contains many sums with $k>1$, and these higher-$k$ contributions are the terms that genuinely dominate the UAE. 
For $k>1$, expanding the sum up to $\order{\Delta_e^4}$ yields
\begin{align}
& \frac{\Delta_e^{3k+1}}{2^{k-1}}X_k(e) = \frac{3}{175}\Big[ 350+63k^2\Delta_e^4 - 3k(70+29\Delta_e^2)\Delta_e^2 \Big]I[\Ai^2]\bigg(\frac{3k-1}{2}\bigg) + \frac{3}{35}\Delta_e^4I[\Ai\Ai']\left(\frac{3k-5}{2}\right) + \order{\Delta_e^6} \nonumber\\
& \ = \frac{3^{1-k/2}(k-1)\Gamma\left(\frac{k-1}{2}\right)\Gamma\left(\frac{3(k-1)}{2}\right)}{2800\pi\Gamma(k+1)}\Big[ -20k\Delta_e^4+(k-1)(3k-1)\big(350+3k\Delta_e^2(-70+(-29+21k)\Delta_e^2) \big) \Big] + \order{\Delta_e^6}
\end{align}

\begin{figure*}[t]
\begin{tabular}{c}
\includegraphics[width=1\textwidth]{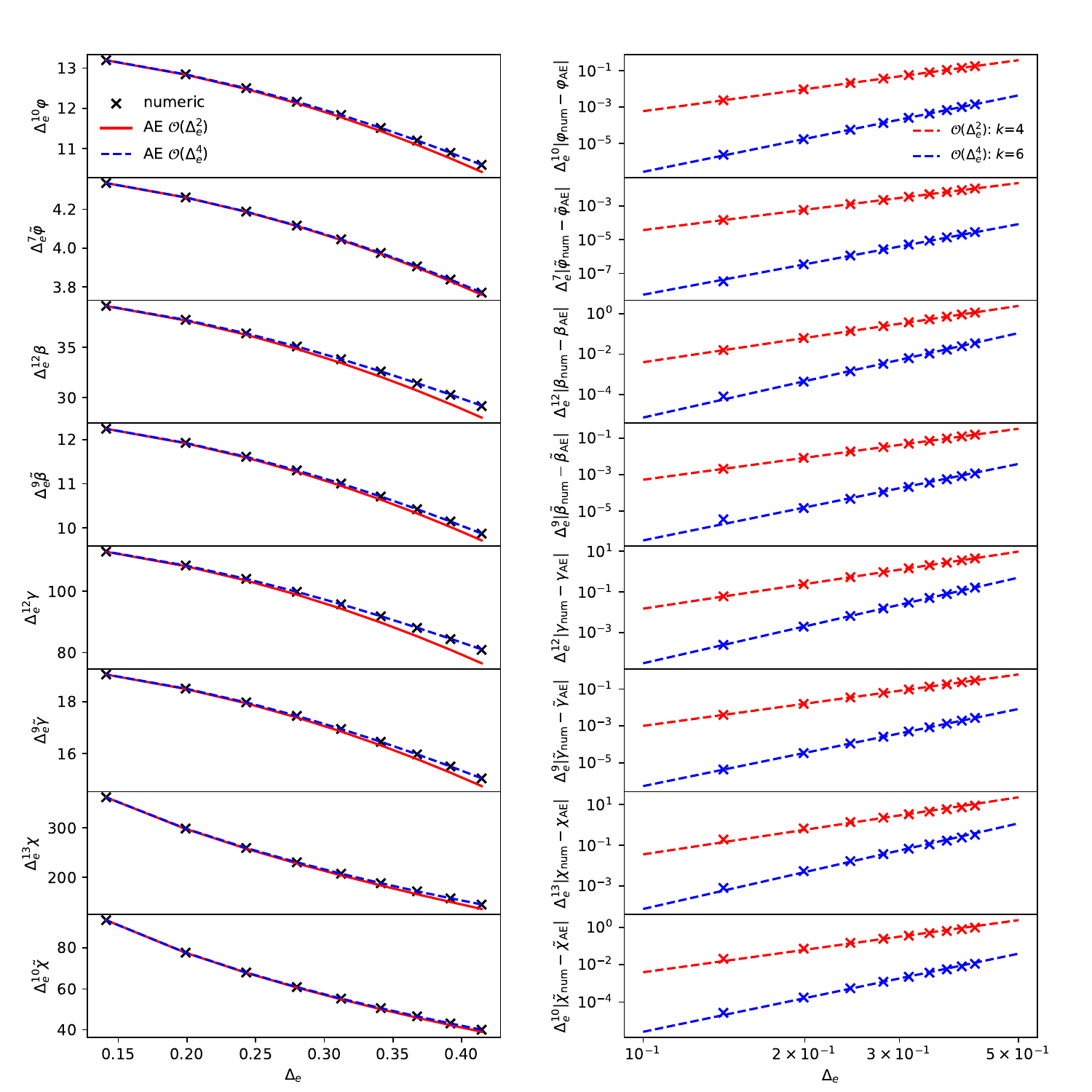}
\end{tabular}
\caption{
Left panel: Comparison between the numerical results for the regularized EEFs (black cross markers) $(\varphi,\tilde\varphi,\beta,\tilde\beta,\gamma,\tilde\gamma,\chi,\tilde\chi)$ and the asymptotic expansions truncated at $\order{\Delta_e^2}$ (red solid line) and $\order{\Delta_e^4}$ (blue dashed line).
Right panel: log-log plot of the absolute error between the numerical result and the asymptotic-expansion result, shown with cross markers. 
The dashed lines denote linear fits with slopes $k=4$ and $k=6$ to the errors between the numerical result and the asymptotic expansions through $\order{\Delta_e^2}$ and $\order{\Delta_e^4}$, respectively.
}\label{fig_eefpn0_example}
\end{figure*}

In particular, we present here the UAE for $k=-1$, although it does not contribute to the EEF expansion considered in this work,
\begin{align}
X_{-1}(e) &= \frac{6^{2/3}2450\Gamma^2(1/3)\zeta_R(5/3)-280\sqrt{3}\pi\zeta_R(3)+6^{1/3}3\Gamma^2(2/3)\zeta(13/3)}{29400\pi^2} + \frac{\Delta_e^2}{2646000\pi^2}\Big[6^{2/3}44100\Gamma(1/3)\zeta_R(5/3) \nonumber\\
&\quad + 35\sqrt{3}\pi\big( -52285+25200\ege-12600\ln{(4/3)} - 368\zeta_R(3) \big) + 6^{4/3}\Gamma^2(2/3)\big(3150\zeta_R(7/3)+37\zeta_R(13/3)\big)\Big] \nonumber\\
&\quad + \frac{\sqrt{3}\Delta_e^2}{\pi}\ln\Delta_e + \order{\Delta_e^4}. 
\end{align}
Substituting these to the EEFs, we find
\begin{align}
& \Delta_e^{10}\varphi(e) = \frac{116200-156240\Delta_e^2+50973\Delta_e^4}{1575\sqrt{3}\pi} + \order{\Delta_e^6}, \\
& \Delta_e^{7}\tilde\varphi(e) = \frac{1680-1442\Delta_e^2+141\Delta_e^4}{70\sqrt{3}\pi} + \order{\Delta_e^6}, \\
& \Delta_e^{12}\beta(e) = \frac{512(148575-267715\Delta_e^2+147051\Delta_e^4)}{344463\sqrt{3}\pi} + \order{\Delta_e^6}, \\
& \Delta_e^{9}\tilde\beta(e) = \frac{128(184100-243180\Delta_e^2+77067\Delta_e^4)}{344463\sqrt{3}\pi} + \order{\Delta_e^6}, \\
& \Delta_e^{12}\gamma(e) = \frac{32(10500-21350\Delta_e^2+13707\Delta_e^4)}{525\sqrt{3}\pi} + \order{\Delta_e^6}, \\
& \Delta_e^{9}\tilde\gamma(e) = \frac{8(7000-10080\Delta_e^2+3627\Delta_e^4)}{525\sqrt{3}\pi} + \order{\Delta_e^4}.
\end{align}
Let $X_k(e)$ take the partial derivative with respect to $k$, one finds
\begin{align}
&\Delta_e^{3k+1}\pdv{k} X_k(e) = -\frac{\Gamma\left(\frac{1+3k}{2}\right)}{3^{k/2}20\sqrt{\pi}\Gamma\left(1+\frac{k}{2}\right)}\Bigg[ \bigg(30-18k\Delta_e^2 + \frac{3k(-107+411k-513k^2+189k^3)\Delta_e^4}{35(k-1)(3k-1)} \bigg) \ln\Delta_e \nonumber\\
&\quad + 5\Bigg( -\frac{2(9k-5)}{(k-1)(3k-1)} +2\psi_\Gamma(k-1) - \psi_\Gamma\bigg(\frac{k-5}{2}\bigg) - 3\psi_\Gamma\bigg(\frac{3k-7}{2}\bigg) + \ln{(3/4)} \Bigg) \nonumber\\
&\qquad + \Delta_e^2\Bigg( \frac{6(1-9k+12k^2)}{(k-1)(3k-1)}  + 3k\bigg( \psi_\Gamma\bigg( \frac{k-1}{2} \bigg) + 3\psi_\Gamma\bigg( \frac{3(k-1)}{2} \bigg) - 2\psi_\Gamma(k+1) + \ln{(4/3)} \bigg) \Bigg) \nonumber\\
&\qquad - \frac{\Delta_e^4}{70(k-1)^2(3k-1)}\Bigg( 1890 k^4-5616 k^3+5544 k^2-2072 k+214 + (k-1) k (189 k^3-513 k^2+411 k-107)\nonumber\\
&\qquad \times \bigg( \psi_\Gamma\bigg(\frac{k-1}{2}\bigg) + 3\psi_\Gamma\bigg(\frac{3(k-1)}{2} - 2\psi_\Gamma(k+1) + \ln{(4/3)} \bigg) \bigg) \Bigg) \Bigg] + \order{\Delta_e^6},
\end{align}
where $\psi_\Gamma(x):=\Gamma'(x)/\Gamma(x)$ is the digamma function.
The EEFs involving logarithm terms are
\begin{align}
& \Delta_e^{13}\big(\chi(e)+(\ln{2})F(e)\big) = \bigg( -\frac{158235}{1024} + \frac{74151}{256}\Delta_e^2 - \frac{86065}{512}\Delta_e^4 \bigg)\ln\Delta_e + \Bigg[ \bigg( -\frac{110845}{3072} + \frac{52745\ege}{1024} + \frac{52745}{2048}\ln{(4/3)} \nonumber\\
&\qquad + \frac{52745}{2048}\psi_\Gamma(3/2) + \frac{158235}{2048}\psi_\Gamma(9/2) \bigg) + \Delta_e^2\bigg( \frac{40565}{1024} - \frac{24717\ege}{256} - \frac{24717}{512}\ln{(4/3)} - \frac{24717}{512}\psi_\Gamma(3/2) \nonumber\\
&\qquad - \frac{74151}{512}\psi_\Gamma(9/2) \bigg) + \Delta_e^4\bigg(  \frac{62211}{10240} + \frac{86065\ege}{1536} + \frac{86065}{3072}\ln{(4/3)} + \frac{315}{512}\psi_\Gamma(1/2) + \frac{89845}{3072}\psi_\Gamma(3/2) \nonumber\\
&\qquad + \frac{84175}{1024}\psi_\Gamma(9/2) \bigg) \Bigg] + \order{\Delta_e^6} ,\\
& \Delta_e^{10}\big(\tilde\chi(e)+(\ln{2})\tilde{F}(e)\big) = \bigg( -\frac{10395}{256} + \frac{13965}{256}\Delta_e^2 - \frac{4545}{256}\Delta_e^4 \bigg)\ln\Delta_e + \Bigg[  -\frac{6195}{512} + \frac{3465\ege}{256} + \frac{3465}{512}\ln{(4/3)} \nonumber\\
&\qquad + \frac{3465}{512}\psi_\Gamma(3/2) + \frac{10395}{512}\psi_\Gamma(9/2) + \Delta_e^2\bigg( -\frac{2441}{512} - \frac{4655\ege}{256} - \frac{4655}{512}\ln{(4/3)} - \frac{595}{256}\psi_\Gamma(1/2) - \frac{7035}{512}\psi_\Gamma(3/2) \nonumber\\
&\qquad - \frac{10395}{512}\psi_\Gamma(9/2) \bigg) + \Delta_e^4\bigg( \frac{5753}{640} + \frac{1515\ege}{256} + \frac{1515}{512}\ln{(4/3)} + \frac{5}{4}\psi_\Gamma(1/2) + \frac{2795}{512}\psi_\Gamma(3/2) + \frac{2625}{512}\psi_\Gamma(9/2) \bigg) \Bigg] \nonumber\\
&\qquad + \order{\Delta_e^6}.
\end{align}
In FIG.~\ref{fig_eefpn0_example}, we present a comparison between the asymptotic expansion in the limit $\Delta_e\to0$ and the corresponding numerical results of $(\varphi,\tilde\varphi,\beta,\tilde\beta,\gamma,\tilde\gamma,\chi,\tilde\chi)$.
The accuracy of the numerical results in this comparison is $10^{-6}$. In addition, we compare the logarithmic slope of the errors between the numerical results and the asymptotic expansions at different truncation orders, and find that the slopes are indeed $4$ and $6$.

\subsection{EEF at 1PN order}

At 1PN order, $\feJ{(p,a,b)}(e)$ would appear. But we can always express the EEF at 1PN using only $J_p(ep),\feJ{{(p,a,b)}}$ and $J'_p(ep)$.
Here we define
\begin{align}
& X_k(a,b,e) = \sum_{p=1}^\infty p^k J_p(ep) \feJ{(p,a,b)}(e), \\
& Y_k(a,b,e) = \sum_{p=1}^\infty p^k \big(J'_p(ep)\big) \feJ{(p,a,b)}(e). \label{eq_generalized_summation}
\end{align}
We express EEF terms $\varphi_{(0,1)}(e)$ and $\tilde\varphi_{(0,1)}(e)$ by $X_k$ and $Y_k$,
\begin{align}
& \varphi_{(0,1)}(e) = \frac{1}{252e^4(-1+e^2)}\Big[e^6 (390-92 \nu )-3 e^2 (1092 \sqrt{1-e^2}+67 \nu -111)-252 \
\sqrt{1-e^2} \nonumber\\
&\quad +e^4 (2646 \sqrt{1-e^2}+293 \nu -891)\Big]X_3(e) + \frac{(1-e^2)^2}{84e^4}\Big[ e^2 (6 \nu -2)-21 (3 \sqrt{1-e^2}+2) \Big]X_5(e) \nonumber\\
&\quad -\frac{1}{168e^3}\bigg[ e^2 \bigg(\frac{819}{\sqrt{1-e^2}}+25 \nu -111\bigg)-\frac{378}{\sqrt{1-e^2}}-67 \nu +111 \bigg] X'_1(e) - \frac{1}{336e^3} \Big[ -6 e^4 (\nu -26) \nonumber\\
&\quad +e^2 (3213 \sqrt{1-e^2}+328 \nu -968)-6 (910 \sqrt{1-e^2}+67 \nu -111) \Big] X'_3(e) + \frac{(e^2-1)^3 (3 \nu -1)}{42 e^3} X'_5(e) \nonumber\\
&\quad + \frac{1}{168} \bigg[\frac{1}{e^2}\bigg(\frac{1890}{\sqrt{1-e^2}}+67 \nu -111\bigg)-\frac{1575}{\sqrt{1-e^2}}-25 \nu +111\bigg]X''_1(e) + \frac{1-e^2}{168e^2}\Big[ e^2 (2 \nu +25) \nonumber\\
&\quad +441 \sqrt{1-e^2}+79 \nu -73 \Big] X''_3(e) + \frac{2-e^2}{4 e^4 \sqrt{1-e^2}}X_6(-4,1,e) + \frac{9 \sqrt{1-e^2} (e^2-2)}{8 e^4}X_6(-3,1,e) \nonumber\\
&\quad + \frac{3 (1-e^2)^{3/2}}{e^4}X_6(-2,1,e) -\frac{3 (1-e^2)^{5/2}}{2 e^4} X_6(-1,1,e) -\frac{3 (e^2-2)}{4 e^3 \sqrt{1-e^2}} Y_5(-3,1,e) -\frac{\sqrt{1-e^2}}{2 e^3}Y_7(-4,1,e) \nonumber\\
&\quad + \frac{3 (1-e^2)^{3/2}}{4 e^3} Y_7(-3,1,e), \\
& \tilde\varphi_{(0,1)}(e) =-\frac{9 (3 e^2-4)}{2 e^2 (e^2-1)} X_1(e) -\frac{\sqrt{1-e^2}}{84 e^4} \Big[e^4 (303-62 \nu )+2 e^2 (441 \sqrt{1-e^2}+73 \nu -71)+42 (\sqrt{1-e^2}-2) \Big]X_3(e) \nonumber\\
&\quad -\frac{9}{e^3}\Big( X'_{-1}(e) + eX''_{-1}(e) \Big) -\frac{1}{168 e^3 (1-e^2)}\Big[(1-e^2)^{5/2} (26 \nu -214)+14 (1-e^2)^{3/2} (2 \nu +33)-42 \sqrt{1-e^2} (\nu +4) \nonumber\\
&\quad -1827 (e^2-1)^2+2436 (e^2-1)-315\Big]X'_1(e) + \frac{(1-e^2)^{3/2}}{84 e^3} \Big[2 e^2 (8 \nu +23)+819 \sqrt{1-e^2}+152 \nu -186\Big]X'_3(e) \nonumber\\
&\quad -\frac{1}{84 e^2}\Big[(1-e^2)^{3/2} (17 \nu -134)+3 \sqrt{1-e^2} (41 \nu -16)-1260 (e^2-1)+315\Big] X''_1(e) + \frac{(1-e^2)^{5/2} (1-3 \nu )}{42 e^2}X''_3(e) \nonumber\\
&\quad -\frac{3 (e^2-2)^2}{4 e^4 (e^2-1)} X_4(-3,1,e) + \frac{3 (e^2-2)}{e^4}X_4(-2,1,e) + \frac{3 (e^4-3 e^2+2)}{2 e^4}X_4(-1,1,e) + \frac{e^2-1}{e^4}X_6(-4,1,e) \nonumber\\
&\quad + \frac{3 (e^2-1)^2}{2 e^4}X_6(-3,1,e) + \frac{1}{e^3}Y_5(-4,1,e)  + \frac{3 (2 e^2-3)}{2 e^3}Y_5(-3,1,e) + \frac{6-6 e^2}{e^3}Y_5(-2,1,e) -\frac{3 (e^2-1)^2}{e^3}Y_5(-1,1,e) 
\end{align}
To study the asymptotic properties of these EEF appear in 1PN or even higher-order post-Newtonian fluxes, one way is to imitate the UAE of the Bessel functions and the elliptic integrals $\feJ{(p,a,1)}(e)$.
To obtain the expansion up to $\order{\Delta_e^4}$, the UAE must be carried out through $\order{p^{-5/3}}$,
\begin{align}
    &p^3h_0^3\feJ{(p,-4,1)} = \calJ_{(-4,0)}(p^{2/3}\zeta) + \frac{\Delta_{\chi,0}(e)}{p^{1/3}}\calJ_{(-4,1)}(p^{2/3}\zeta) + \frac{h_0^3-2}{p^{2/3}\zeta}\calJ_{(-4,2)}(p^{2/3}\zeta) \nonumber\\
    &\qquad + \frac{(h_0^3-2)\Delta_{\chi,0}(e)-4\zeta\Delta_{\chi_1}(e)}{p\zeta}\calJ_{(-4,3)}(p^{2/3}\zeta) + \frac{124-64h_0^3+h_0^6+36h_0^2\zeta}{p^{4/3}\zeta^2}\calJ_{(-4,4)}(p^{2/3}\zeta) \nonumber\\
    &\qquad + \frac{(124-64h_0^3+h_0^6+36h_0^2\zeta)\Delta_{\chi,0}(e)-96\zeta (h_0^3-2)\Delta_{\chi,1}(e) + 384\zeta^2\Delta_{\chi,2}(e)}{p^{5/3}\zeta^2}\calJ_{(-4,5)}(p^{2/3}\zeta) + \order{p^{-2}}, \\
    &p^{7/3}h_0^2\feJ{(p,-3,1)} = \calJ_{(-3,0)}(p^{2/3}\zeta) + \frac{\Delta_{\chi,0}(e)}{p^{1/3}}\calJ_{(-3,1)}(p^{2/3}\zeta) + \frac{h_0^3-2}{p^{2/3}\zeta}\calJ_{(-2,2)}(p^{2/3}\zeta) \nonumber\\
    &\qquad + \frac{(h_0^3-2)\Delta_{\chi,0}(e)-6\zeta\Delta_{\chi_1}(e)}{p\zeta}\calJ_{(-3,3)}(p^{2/3}\zeta) + \frac{36-16h_0^3+h_0^6+12h_0^2\zeta}{p^{4/3}\zeta^2}\calJ_{(-3,4)}(p^{2/3}\zeta) \nonumber\\
    &\qquad + \frac{(36-16h_0^3+h_0^6+12h_0^2\zeta)\Delta_{\chi,0}(e)-32\zeta (h_0^3-2)\Delta_{\chi,1}(e) + 192\zeta^2\Delta_{\chi,2}(e)}{p^{5/3}\zeta^2}\calJ_{(-3,5)}(p^{2/3}\zeta) + \order{p^{-2}}, \\
    &p^{5/3}h_0\feJ{(p,-2,1)} = \calJ_{(-2,0)}(p^{2/3}\zeta) + \frac{\Delta_{\chi,0}(e)}{p^{1/3}}\calJ_{(-2,1)}(p^{2/3}\zeta) + \frac{h_0^3-2}{p^{2/3}\zeta}\calJ_{(-2,2)}(p^{2/3}\zeta) \nonumber\\
    &\qquad + \frac{(h_0^3-2)\Delta_{\chi,0}(e)-12\zeta\Delta_{\chi_1}(e)}{p\zeta}\calJ_{(-2,3)}(p^{2/3}\zeta) + \frac{92-32h_0^3-7h_0^6+36h_0^2\zeta}{p^{4/3}\zeta^2}\calJ_{(-2,4)}(p^{2/3}\zeta) \nonumber\\
    &\qquad + \frac{(92-32h_0^3-7h_0^6+36h_0^2\zeta)\Delta_{\chi,0}(e)-96\zeta (h_0^3-2)\Delta_{\chi,1}(e) + 1152\zeta^2\Delta_{\chi,2}(e)}{p^{5/3}\zeta^2}\calJ_{(-2,5)}(p^{2/3}\zeta) + \order{p^{-2}}, \\
    &p\feJ{(p,-1,1)} = \calJ_{(-1,0)}(p^{2/3}\zeta) + \frac{\Delta_{\chi,0}(e)}{p^{1/3}}\calJ_{(-1,1)}(p^{2/3}\zeta) + \frac{\Delta_{\chi,1}(e)}{p}\calJ_{(-1,3)}(p^{2/3}\zeta) + \frac{\Delta_{\chi,2}(e)}{p^{5/3}}\calJ_{(-1,5)}(p^{2/3}\zeta) + \order{p^{-2}}, \\
    &p^{1/3}h_0^{-1}\feJ{(p,0,1)} = \calJ_{(0,0)}(p^{2/3}\zeta) + \frac{\Delta_{\chi,0}(e)}{p^{1/3}}\calJ_{(0,1)}(p^{2/3}\zeta) + \frac{h_0^3-2}{p^{2/3}\zeta}\calJ_{(0,2)}(p^{2/3}\zeta) \nonumber\\
    &\qquad + \frac{(h_0^3-2)\Delta_{\chi,0}(e)+12\zeta\Delta_{\chi,1}(e)}{p\zeta}\calJ_{(0,3)}(p^{2/3}\zeta) + \frac{20-5h_0^6+12h_0^2\zeta}{p^{4/3}\zeta^2}\calJ_{(0,4)}(p^{2/3}\zeta) \nonumber\\
    &\qquad + \frac{(20-5h_0^6+12h_0^2\zeta)\Delta_{\chi,0}(e)-32\zeta (h_0^3-2)\Delta_{\chi,1}(e) -384\zeta^2\Delta_{\chi,2}(e)}{p^{5/3}\zeta^2}\calJ_{(0,5)}(p^{2/3}\zeta) + \order{p^{-2}},
\end{align}
where
\begin{align}
    & \calJ_{(-4,0)}(y) = 4 y^{1/2} (9 \intV_{-1,0}(y)+4 \tintW_{0,0}(y))-12 y (\pi  \intW_{0,0}(y)-2 \intW_{0,1}(y)), \ 
     \calJ_{(-4,1)}(y) = 4 (7 \intV_{0,0}(y)-3 \tintW_{0,0}(y) y), \nonumber\\
    & \calJ_{(-4,2)}(y) = 16 \intV_{-1,0}(y) y^{3/2}-26 \intV_{0,0}(y) y^{1/2}-20 \pi  \tintV_{0,0}(y)+40 \tintV_{0,1}(y), \ 
    \calJ_{(-4,3)}(y) = 35 \intV_{0,0}(y) y, \nonumber\\
    & \calJ_{(-4,4)}(y) = \frac{1}{48} \left(-23 \intV_{0,0}(y) y^{3/2}-15 \pi  \tintV_{0,0}(y) y+30 \tintV_{0,1}(y) y-35 \pi  \intW_{0,0}(y)+70 \intW_{0,1}(y)-70 \tintW_{-1,0}(y) y^{1/2}\right), \nonumber\\
    & \calJ_{(-4,5)}(y) = \frac{5}{48} \left(3 \intV_{0,0}(y) y^2-22 \tintW_{0,0}(y)\right), \nonumber\\
    & \calJ_{(-3,0)}(y) = 4 \pi  \intW_{0,0}(y)-8 \intW_{0,1}(y)+8 \tintW_{-1,0}(y) y^{1/2},\ 
    \calJ_{(-3,1)}(y) = 10 \tintW_{0,0}(y),\nonumber\\
    & \calJ_{(-3,2)}(y) = 2 y (\pi  \intW_{0,0}(y)-2 \intW_{0,1}(y))-\frac{2}{3} y^{1/2} (9 \intV_{-1,0}(y)+4 \tintW_{0,0}(y)),\
    \calJ_{(-3,3)}(y) = 2 \tintW_{0,0}(y) y-\frac{14 \intV_{0,0}(y)}{3},\nonumber\\
    & \calJ_{(-3,4)}(y) = \frac{1}{24} \left(-8 \intV_{-1,0}(y) y^{3/2}+13 \intV_{0,0}(y) y^{1/2}+10 \pi  \tintV_{0,0}(y)-20 \tintV_{0,1}(y)\right),\ 
    \calJ_{(-3,5)}(y) = -\frac{35 \intV_{0,0}(y) y}{48}, \nonumber\\
    & \calJ_{(-2,0)}(y) = -2 \left(\intV_{0,0}(y) y^{1/2}+\pi  \tintV_{0,0}(y)-2 \tintV_{0,1}(y)\right),\ 
    \calJ_{(-2,1)}(y) = 2 \intV_{0,0}(y) y,\nonumber\\
    & \calJ_{(-2,2)}(y) = \frac{1}{3} \left(-\pi  \intW_{0,0}(y)+2 \intW_{0,1}(y)-2 \tintW_{-1,0}(y) y^{1/2}\right),\ 
    \calJ_{(-2,3)}(y) = -\frac{5 \tintW_{0,0}(y)}{6}, \nonumber\\
    & \calJ_{(-2,4)}(y) = \frac{1}{288} \left(y (6 \intW_{0,1}(y)-3 \pi  \intW_{0,0}(y))+y^{1/2} (9 \intV_{-1,0}(y)+4 \tintW_{0,0}(y))\right), \nonumber\\
    & \calJ_{(-2,5)}(y) = \frac{1}{288} (7 \intV_{0,0}(y)-3 \tintW_{0,0}(y) y), \nonumber\\
    & \calJ_{(-1,0)}(y) = -2 \intV_{-1,0}(y) y^{1/2},\ 
    \calJ_{(-1,1)}(y) = -\intV_{0,0}(y),\ 
    \calJ_{(-1,3)}(y) = 2 \intV_{0,0}(y) y,\ 
    \calJ_{(-1,5)}(y) = 10 \tintW_{0,0}(y), \nonumber\\
    & \calJ_{(0,0)}(y) = 2 \intW_{0,1}(y)-\pi  \intW_{0,0}(y),\ 
    \calJ_{(0,1)}(y) = -\tintW_{0,0}(y),\ 
    \calJ_{(0,2)}(y) = -\frac{\intV_{-1,0}(y) y^{1/2}}{6},\nonumber\\
    & \calJ_{(0,3)}(y) = -\frac{\intV_{0,0}(y)}{12},\ 
    \calJ_{(0,4)}(y) = \frac{1}{192} \left(\intV_{0,0}(y) y^{1/2}+\pi  \tintV_{0,0}(y)-2 \tintV_{0,1}(y)\right),\ \calJ_{(0,5)}(y) = -\frac{\intV_{0,0}(y) y}{192}.
\end{align}
And one can check all these functions are converge as $y\to0$.
To evaluate the UAE of $Y_k(a,b,e)$, we need the UAE of $J'_p(ep)$.
Differentiating (\ref{eq_UAE_of_BesselJ}) with respect to $e$, one finds
\begin{align}
    & J_p'(ep) = \frac{1}{p}\dv{e} J_p(ep) = \frac{1}{p^{2/3}}\bigg( b_0(e)\Ai'(p^{2/3}\zeta) + \frac{b_1(e)}{p^{2/3}}\Ai(p^{2/3}\zeta) + \order{p^{-2}} \bigg),
\end{align}
where
\begin{align}
    & b_0(e) = a_0(e)\zeta'(e),\ b_1(e) = a_0'(e) + a_1(e)\zeta(e)\zeta'(e).
\end{align}
Fortunately, the $k$ index in these 1PN EEF are positive, which would hugely simplify the expressions. 
Thus, both classes of sums, $X_k(a,1,e)$ and $Y_k(a,1,e)$, can be expressed in terms of $S[f](\alpha;\zeta)$,
\begin{align}
    & X_k(-|a|,1,e) \sim S[\Ai \calJ_{(-|a|,0)}]\big(k-2(|a|+1)/3;\zeta\big) \sim \order{\Delta_e^{2|a|-3k-1}}, \\
    & Y_k(-|a|,1,e) \sim S[\Ai' \calJ_{(-|a|,0)}]\big( k-2|a|/3-1 ;\zeta\big) \sim \order{\Delta_e^{2|a|-3k}}.
\end{align}
So the leading divergent orders are $1/\Delta_e^{3k-2|a|-1}$ and $1/\Delta_e^{3k-2|a|}$, respectively. 
Since we again retain the overall expansion only up to $\order{\Delta_e^4}$, it is sufficient to expand only the most divergent term through $\order{\Delta_e^4}$. 
This greatly simplifies the calculation.

\begin{figure*}[t]
\begin{tabular}{c}
\includegraphics[width=1\textwidth]{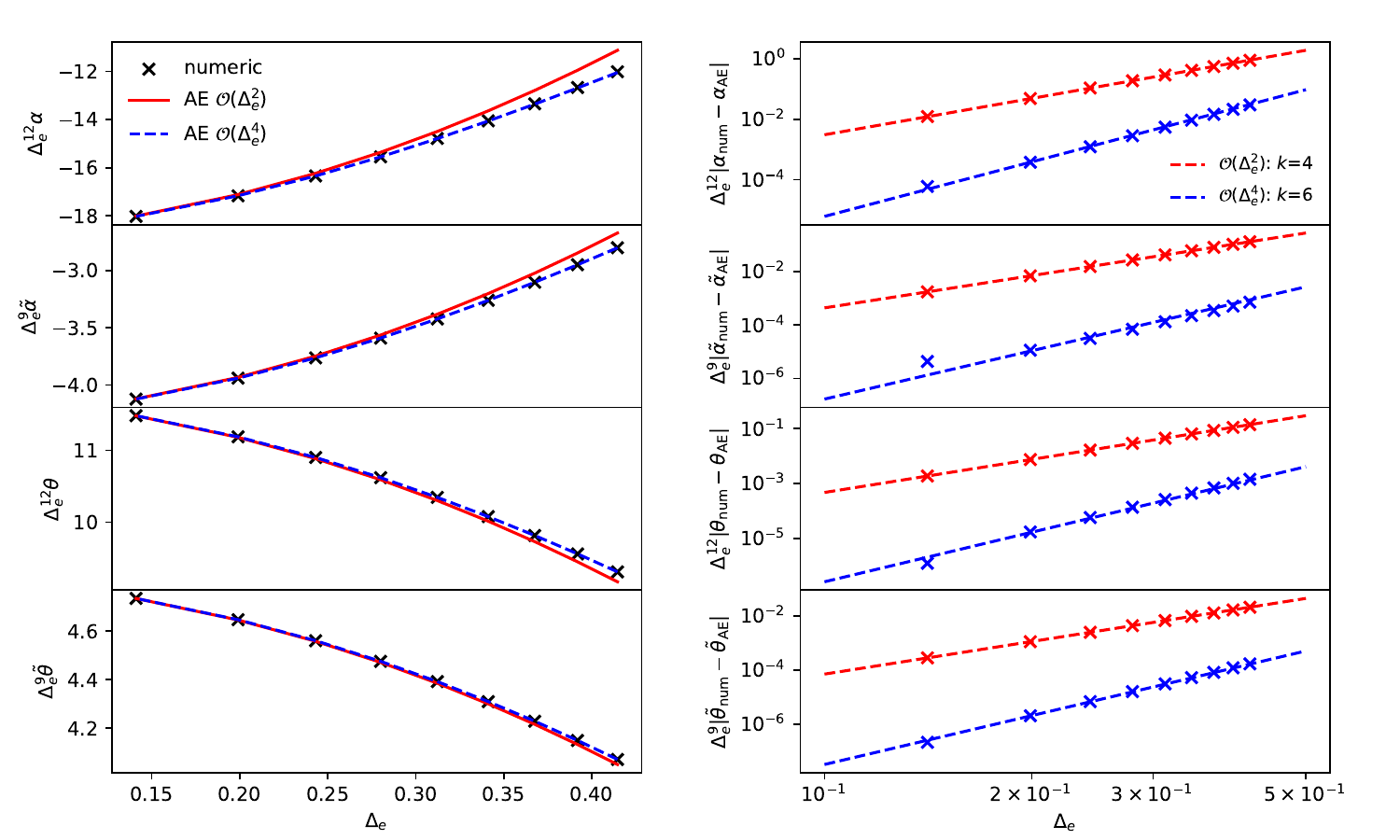}
\end{tabular}
\caption{
Left panel: Comparison between the numerical results for the EEFs (black cross markers) $(\alpha,\tilde\alpha,\theta,\tilde\theta)$ and the asymptotic expansions truncated at $\order{\Delta_e^2}$ (red solid line) and $\order{\Delta_e^4}$ (blue dashed line).
Right panel: log-log plot of the absolute error between the numerical result and the asymptotic-expansion result, shown with cross markers. 
The dashed lines denote linear fits with slopes $k=4$ and $k=6$ to the errors between the numerical result and the asymptotic expansions through $\order{\Delta_e^2}$ and $\order{\Delta_e^4}$, respectively.
}\label{fig_eefpn1_example}
\end{figure*}

One usually rewrite the EEF as
\begin{align}
    & 4 \bigg( \varphi_{(0,1)}(e) - \frac{1}{2}\nu\varphi(e) - \frac{21}{1-e^2}\big( \varphi(e) - \varphi_{(1,0)}(e) \big) \bigg) = -\frac{428}{21}\alpha(e) + \frac{178}{21}\nu \theta(e), \\
    & 4 \bigg( \tilde\varphi_{(0,1)}(e) - \frac{1}{2}\nu\tilde\varphi(e) - \frac{18}{1-e^2}\big( \tilde\varphi(e) - \tilde\varphi_{(1,0)}(e) \big) \bigg) =  -\frac{428}{21}\tilde\alpha(e) + \frac{178}{21} \nu \tilde\theta(e).
\end{align}
The expansion of of $\theta(e), \tilde\theta(e)$ are solely determined by $X_k(e)$. We obtain 
\begin{align}
    & \Delta_e^{12}\theta(e) = \frac{4(2247000 - 2899750\Delta_e^2 + 893637\Delta_e^4)}{140175\sqrt{3}\pi} + \order{\Delta_e^6}, \\
    & \Delta_e^9\tilde\theta(e) = \frac{2(204400-190575\Delta_e^2 + 29944\Delta_e^4)}{15575\sqrt{3}\pi} + \order{\Delta_e^6} .
\end{align}
However, the generalized summations (\ref{eq_generalized_summation}) are involved in $\alpha(e), \tilde\alpha(e)$,

\begin{align}
    &\Delta_e^{12}\alpha(e) = \alpha_0 + \alpha_1\Delta_e^2 + \alpha_2\Delta_e^4 + \order{\Delta_e^6}, \\
    &\Delta_e^9\tilde\alpha(e) = \tilde\alpha_0 + \tilde\alpha_1\Delta_e^2 + \tilde\alpha_2\Delta_e^4 + \order{\Delta_e^6},
\end{align}
where the coefficients $\alpha_0,\alpha_1,\alpha_2,\tilde\alpha_0,\tilde\alpha_1,\tilde\alpha_2$ are constants and can only be expressed in terms of integrals,
\begin{align}
    & \alpha_0 = -\frac{63}{107}I\left[\Ai\,\calJ_{(-4,0)}\right]\left(\frac{9}{2}\right)-\frac{252}{107}I\left[\Ai'\,\calJ_{(-4,0)}\right]\left(\frac{11}{2}\right)+\frac{567}{107}I\left[\Ai\,\calJ_{(-3,0)}\right]\left(\frac{11}{2}\right)+\frac{756}{107}I\left[\Ai'\,\calJ_{(-3,0)}\right]\left(\frac{13}{2}\right) \nonumber\\
    &\qquad -\frac{3024}{107}I\left[\Ai\,\calJ_{(-2,0)}\right]\left(\frac{13}{2}\right)+\frac{3024}{107}I\left[\Ai\,\calJ_{(-1,0)}\right]\left(\frac{15}{2}\right)+\frac{116 \sqrt{3}}{\pi }, \\
    & \alpha_1 = -\frac{63}{107}I\left[\Ai\,\calJ_{(-4,0)}\right]\left(\frac{9}{2}\right)+\frac{252}{107}I\left[\Ai'\,\calJ_{(-4,0)}\right]\left(\frac{11}{2}\right)-\frac{63}{535}I\left[\Ai\,\calJ_{(-4,1)}\right]\left(4\right)-\frac{252}{535}I\left[\Ai'\,\calJ_{(-4,1)}\right]\left(5\right) \nonumber\\
    &\qquad -\frac{189}{535}I\left[\Ai\,\calJ_{(-4,2)}\right]\left(\frac{7}{2}\right)-\frac{756}{535}I\left[\Ai'\,\calJ_{(-4,2)}\right]\left(\frac{9}{2}\right)+\frac{2457}{1070}I\left[\Ai\,\calJ_{(-3,0)}\right]\left(\frac{11}{2}\right)+\frac{189}{107}I\left[\Ai'\,\calJ_{(-3,0)}\right]\left(\frac{7}{2}\right)\nonumber\\
    &\qquad -\frac{4914}{535}I\left[\Ai'\,\calJ_{(-3,0)}\right]\left(\frac{13}{2}\right)+\frac{567}{535}I\left[\Ai\,\calJ_{(-3,1)}\right]\left(5\right)+\frac{756}{535}I\left[\Ai'\,\calJ_{(-3,1)}\right]\left(6\right)+\frac{1701}{535}I\left[\Ai\,\calJ_{(-3,2)}\right]\left(\frac{9}{2}\right)\nonumber\\
    &\qquad  +\frac{2268}{535}I\left[\Ai'\,\calJ_{(-3,2)}\right]\left(\frac{11}{2}\right)+\frac{3024}{107}I\left[\Ai\,\calJ_{(-2,0)}\right]\left(\frac{13}{2}\right)-\frac{1512}{107}I\left[\Ai'\,\calJ_{(-2,0)}\right]\left(\frac{9}{2}\right)-\frac{3024}{535}I\left[\Ai\,\calJ_{(-2,1)}\right]\left(6\right) \nonumber\\
    &\qquad -\frac{9072}{535}I\left[\Ai\,\calJ_{(-2,2)}\right]\left(\frac{11}{2}\right)-\frac{19656}{535}I\left[\Ai\,\calJ_{(-1,0)}\right]\left(\frac{15}{2}\right)+\frac{1512}{107}I\left[\Ai'\,\calJ_{(-1,0)}\right]\left(\frac{11}{2}\right)+\frac{3024}{535}I\left[\Ai\,\calJ_{(-1,1)}\right]\left(7\right) \nonumber\\
    &\qquad -\frac{721544}{1605 \sqrt{3} \pi }, \\
    & \alpha_2 = \frac{162}{535}I\left[\Ai\,\calJ_{(-4,0)}\right]\left(\frac{9}{2}\right)-\frac{9}{2140}I\left[\Ai'\,\calJ_{(-4,0)}\right]\left(\frac{5}{2}\right)-\frac{324}{535}I\left[\Ai'\,\calJ_{(-4,0)}\right]\left(\frac{11}{2}\right)-\frac{513}{2675}I\left[\Ai\,\calJ_{(-4,1)}\right]\left(4\right)\nonumber\\
    &\qquad +\frac{468}{2675}I\left[\Ai'\,\calJ_{(-4,1)}\right]\left(5\right)-\frac{5913}{10700}I\left[\Ai\,\calJ_{(-4,2)}\right]\left(\frac{7}{2}\right)+\frac{1647}{2675}I\left[\Ai'\,\calJ_{(-4,2)}\right]\left(\frac{9}{2}\right)-\frac{297}{5350}I\left[\Ai\,\calJ_{(-4,3)}\right]\left(3\right)\nonumber\\
    &\qquad -\frac{594}{2675}I\left[\Ai'\,\calJ_{(-4,3)}\right]\left(4\right)-\frac{2592}{2675}I\left[\Ai\,\calJ_{(-4,4)}\right]\left(\frac{5}{2}\right)-\frac{10368}{2675}I\left[\Ai'\,\calJ_{(-4,4)}\right]\left(\frac{7}{2}\right)+\frac{189}{1070}I\left[\Ai\,\calJ_{(-3,0)}\right]\left(\frac{5}{2}\right)\nonumber\\
    &\qquad -\frac{77139}{21400}I\left[\Ai\,\calJ_{(-3,0)}\right]\left(\frac{11}{2}\right)+\frac{3483}{2140}I\left[\Ai'\,\calJ_{(-3,0)}\right]\left(\frac{7}{2}\right)+\frac{18063}{5350}I\left[\Ai'\,\calJ_{(-3,0)}\right]\left(\frac{13}{2}\right)+\frac{6021}{5350}I\left[\Ai\,\calJ_{(-3,1)}\right]\left(5\right)\nonumber\\
    &\qquad +\frac{189}{535}I\left[\Ai'\,\calJ_{(-3,1)}\right]\left(3\right)-\frac{2538}{2675}I\left[\Ai'\,\calJ_{(-3,1)}\right]\left(6\right)+\frac{33939}{10700}I\left[\Ai\,\calJ_{(-3,2)}\right]\left(\frac{9}{2}\right)+\frac{567}{535}I\left[\Ai'\,\calJ_{(-3,2)}\right]\left(\frac{5}{2}\right)\nonumber\\
    &\qquad -\frac{8343}{2675}I\left[\Ai'\,\calJ_{(-3,2)}\right]\left(\frac{11}{2}\right)+\frac{4617}{10700}I\left[\Ai\,\calJ_{(-3,3)}\right]\left(4\right)+\frac{1539}{2675}I\left[\Ai'\,\calJ_{(-3,3)}\right]\left(5\right)+\frac{972}{2675}I\left[\Ai\,\calJ_{(-3,4)}\right]\left(\frac{7}{2}\right)\nonumber\\
    &\qquad +\frac{1296}{2675}I\left[\Ai'\,\calJ_{(-3,4)}\right]\left(\frac{9}{2}\right)-\frac{756}{535}I\left[\Ai\,\calJ_{(-2,0)}\right]\left(\frac{7}{2}\right)-\frac{3888}{535}I\left[\Ai\,\calJ_{(-2,0)}\right]\left(\frac{13}{2}\right)+\frac{2916}{535}I\left[\Ai'\,\calJ_{(-2,0)}\right]\left(\frac{9}{2}\right)\nonumber\\
    &\qquad +\frac{5616}{2675}I\left[\Ai\,\calJ_{(-2,1)}\right]\left(6\right)-\frac{1512}{535}I\left[\Ai'\,\calJ_{(-2,1)}\right]\left(4\right)+\frac{19764}{2675}I\left[\Ai\,\calJ_{(-2,2)}\right]\left(\frac{11}{2}\right)-\frac{4536}{535}I\left[\Ai'\,\calJ_{(-2,2)}\right]\left(\frac{7}{2}\right)\nonumber\\
    &\qquad -\frac{648}{535}I\left[\Ai\,\calJ_{(-2,3)}\right]\left(5\right)+\frac{93312}{2675}I\left[\Ai\,\calJ_{(-2,4)}\right]\left(\frac{9}{2}\right)+\frac{756}{535}I\left[\Ai\,\calJ_{(-1,0)}\right]\left(\frac{9}{2}\right)+\frac{36126}{2675}I\left[\Ai\,\calJ_{(-1,0)}\right]\left(\frac{15}{2}\right)\nonumber\\
    &\qquad -\frac{5184}{535}I\left[\Ai'\,\calJ_{(-1,0)}\right]\left(\frac{11}{2}\right)-\frac{10152}{2675}I\left[\Ai\,\calJ_{(-1,1)}\right]\left(7\right)+\frac{1512}{535}I\left[\Ai'\,\calJ_{(-1,1)}\right]\left(5\right)+\frac{486}{2675}I\left[\Ai\,\calJ_{(-1,3)}\right]\left(6\right)\nonumber\\
    &\qquad +\frac{4672579}{37450 \sqrt{3} \pi }, \\
    & \tilde\alpha_0 = \frac{252}{107}I\left[\Ai\,\calJ_{(-4,0)}\right]\left(\frac{9}{2}\right)-\frac{189}{214}I\left[\Ai\,\calJ_{(-3,0)}\right]\left(\frac{5}{2}\right)-\frac{756}{107}I\left[\Ai\,\calJ_{(-3,0)}\right]\left(\frac{11}{2}\right)-\frac{378}{107}I\left[\Ai'\,\calJ_{(-3,0)}\right]\left(\frac{7}{2}\right)\nonumber\\
    &\qquad +\frac{756}{107}I\left[\Ai\,\calJ_{(-2,0)}\right]\left(\frac{7}{2}\right)+\frac{3024}{107}I\left[\Ai'\,\calJ_{(-2,0)}\right]\left(\frac{9}{2}\right)-\frac{756}{107}I\left[\Ai\,\calJ_{(-1,0)}\right]\left(\frac{9}{2}\right)-\frac{3024}{107}I\left[\Ai'\,\calJ_{(-1,0)}\right]\left(\frac{11}{2}\right)\nonumber\\
    &\qquad +\frac{32164}{321 \sqrt{3} \pi }, \\
    & \tilde\alpha_1 = -\frac{504}{535}I\left[\Ai\,\calJ_{(-4,0)}\right]\left(\frac{9}{2}\right)+\frac{126}{107}I\left[\Ai'\,\calJ_{(-4,0)}\right]\left(\frac{5}{2}\right)+\frac{252}{535}I\left[\Ai\,\calJ_{(-4,1)}\right]\left(4\right)+\frac{756}{535}I\left[\Ai\,\calJ_{(-4,2)}\right]\left(\frac{7}{2}\right)\nonumber\\
    &\qquad -\frac{5481}{2140}I\left[\Ai\,\calJ_{(-3,0)}\right]\left(\frac{5}{2}\right)+\frac{2646}{535}I\left[\Ai\,\calJ_{(-3,0)}\right]\left(\frac{11}{2}\right)-\frac{3591}{535}I\left[\Ai'\,\calJ_{(-3,0)}\right]\left(\frac{7}{2}\right)-\frac{189}{1070}I\left[\Ai\,\calJ_{(-3,1)}\right]\left(2\right)\nonumber\\
    &\qquad -\frac{756}{535}I\left[\Ai\,\calJ_{(-3,1)}\right]\left(5\right)-\frac{378}{535}I\left[\Ai'\,\calJ_{(-3,1)}\right]\left(3\right)-\frac{567}{1070}I\left[\Ai\,\calJ_{(-3,2)}\right]\left(\frac{3}{2}\right)-\frac{2268}{535}I\left[\Ai\,\calJ_{(-3,2)}\right]\left(\frac{9}{2}\right)\nonumber\\
    &\qquad -\frac{1134}{535}I\left[\Ai'\,\calJ_{(-3,2)}\right]\left(\frac{5}{2}\right)+\frac{6048}{535}I\left[\Ai\,\calJ_{(-2,0)}\right]\left(\frac{7}{2}\right)-\frac{6048}{535}I\left[\Ai'\,\calJ_{(-2,0)}\right]\left(\frac{9}{2}\right)+\frac{756}{535}I\left[\Ai\,\calJ_{(-2,1)}\right]\left(3\right)\nonumber\\
    &\qquad +\frac{3024}{535}I\left[\Ai'\,\calJ_{(-2,1)}\right]\left(4\right)+\frac{2268}{535}I\left[\Ai\,\calJ_{(-2,2)}\right]\left(\frac{5}{2}\right)+\frac{9072}{535}I\left[\Ai'\,\calJ_{(-2,2)}\right]\left(\frac{7}{2}\right)-\frac{4914}{535}I\left[\Ai\,\calJ_{(-1,0)}\right]\left(\frac{9}{2}\right)\nonumber\\
    &\qquad +\frac{10584}{535}I\left[\Ai'\,\calJ_{(-1,0)}\right]\left(\frac{11}{2}\right)-\frac{756}{535}I\left[\Ai\,\calJ_{(-1,1)}\right]\left(4\right)-\frac{3024}{535}I\left[\Ai'\,\calJ_{(-1,1)}\right]\left(5\right)-\frac{14346 \sqrt{3}}{535 \pi }, \\
    & \tilde\alpha_2 = \frac{63}{535}I\left[\Ai\,\calJ_{(-4,0)}\right]\left(\frac{3}{2}\right)+\frac{108}{535}I\left[\Ai\,\calJ_{(-4,0)}\right]\left(\frac{9}{2}\right)+\frac{27}{107}I\left[\Ai'\,\calJ_{(-4,0)}\right]\left(\frac{5}{2}\right)+\frac{288}{2675}I\left[\Ai\,\calJ_{(-4,1)}\right]\left(4\right)\nonumber\\
    &\qquad +\frac{126}{535}I\left[\Ai'\,\calJ_{(-4,1)}\right]\left(2\right)+\frac{621}{2675}I\left[\Ai\,\calJ_{(-4,2)}\right]\left(\frac{7}{2}\right)+\frac{378}{535}I\left[\Ai'\,\calJ_{(-4,2)}\right]\left(\frac{3}{2}\right)+\frac{594}{2675}I\left[\Ai\,\calJ_{(-4,3)}\right]\left(3\right)\nonumber\\
    &\qquad +\frac{10368}{2675}I\left[\Ai\,\calJ_{(-4,4)}\right]\left(\frac{5}{2}\right)-\frac{130923}{42800}I\left[\Ai\,\calJ_{(-3,0)}\right]\left(\frac{5}{2}\right)-\frac{4779}{5350}I\left[\Ai\,\calJ_{(-3,0)}\right]\left(\frac{11}{2}\right)-\frac{27}{4280}I\left[\Ai'\,\calJ_{(-3,0)}\right]\left(\frac{1}{2}\right)\nonumber\\
    &\qquad +\frac{1917}{10700}I\left[\Ai'\,\calJ_{(-3,0)}\right]\left(\frac{7}{2}\right)-\frac{6669}{10700}I\left[\Ai\,\calJ_{(-3,1)}\right]\left(2\right)+\frac{54}{535}I\left[\Ai\,\calJ_{(-3,1)}\right]\left(5\right)-\frac{4779}{2675}I\left[\Ai'\,\calJ_{(-3,1)}\right]\left(3\right)\nonumber\\
    &\qquad -\frac{7857}{4280}I\left[\Ai\,\calJ_{(-3,2)}\right]\left(\frac{3}{2}\right)+\frac{1539}{2675}I\left[\Ai\,\calJ_{(-3,2)}\right]\left(\frac{9}{2}\right)-\frac{5589}{1070}I\left[\Ai'\,\calJ_{(-3,2)}\right]\left(\frac{5}{2}\right)-\frac{1539}{21400}I\left[\Ai\,\calJ_{(-3,3)}\right]\left(1\right)\nonumber\\
    &\qquad -\frac{1539}{2675}I\left[\Ai\,\calJ_{(-3,3)}\right]\left(4\right)-\frac{1539}{5350}I\left[\Ai'\,\calJ_{(-3,3)}\right]\left(2\right)-\frac{162}{2675}I\left[\Ai\,\calJ_{(-3,4)}\right]\left(\frac{1}{2}\right)-\frac{1296}{2675}I\left[\Ai\,\calJ_{(-3,4)}\right]\left(\frac{7}{2}\right)\nonumber\\
    &\qquad -\frac{648}{2675}I\left[\Ai'\,\calJ_{(-3,4)}\right]\left(\frac{3}{2}\right)+\frac{1944}{535}I\left[\Ai\,\calJ_{(-2,0)}\right]\left(\frac{7}{2}\right)+\frac{27}{535}I\left[\Ai'\,\calJ_{(-2,0)}\right]\left(\frac{3}{2}\right)+\frac{1296}{535}I\left[\Ai'\,\calJ_{(-2,0)}\right]\left(\frac{9}{2}\right)\nonumber\\
    &\qquad +\frac{8424}{2675}I\left[\Ai\,\calJ_{(-2,1)}\right]\left(3\right)+\frac{3456}{2675}I\left[\Ai'\,\calJ_{(-2,1)}\right]\left(4\right)+\frac{24543}{2675}I\left[\Ai\,\calJ_{(-2,2)}\right]\left(\frac{5}{2}\right)+\frac{7452}{2675}I\left[\Ai'\,\calJ_{(-2,2)}\right]\left(\frac{7}{2}\right)\nonumber\\
    &\qquad +\frac{162}{535}I\left[\Ai\,\calJ_{(-2,3)}\right]\left(2\right)+\frac{648}{535}I\left[\Ai'\,\calJ_{(-2,3)}\right]\left(3\right)-\frac{23328}{2675}I\left[\Ai\,\calJ_{(-2,4)}\right]\left(\frac{3}{2}\right)-\frac{93312}{2675}I\left[\Ai'\,\calJ_{(-2,4)}\right]\left(\frac{5}{2}\right)\nonumber\\
    &\qquad +\frac{1701}{5350}I\left[\Ai\,\calJ_{(-1,0)}\right]\left(\frac{9}{2}\right)-\frac{27}{535}I\left[\Ai'\,\calJ_{(-1,0)}\right]\left(\frac{5}{2}\right)-\frac{9558}{2675}I\left[\Ai'\,\calJ_{(-1,0)}\right]\left(\frac{11}{2}\right)-\frac{1458}{535}I\left[\Ai\,\calJ_{(-1,1)}\right]\left(4\right)\nonumber\\
    &\qquad +\frac{216}{535}I\left[\Ai'\,\calJ_{(-1,1)}\right]\left(5\right)-\frac{243}{5350}I\left[\Ai\,\calJ_{(-1,3)}\right]\left(3\right)-\frac{486}{2675}I\left[\Ai'\,\calJ_{(-1,3)}\right]\left(4\right)+\frac{26101 \sqrt{3}}{37450 \pi }.
\end{align}
The numerical results of these coefficients are
\begin{align}
    & \alpha_0 \approx -18.91,\ \alpha_1\approx 45.28,\ \alpha_2\approx 31.16,\ \tilde\alpha_0\approx-4.31,\ \tilde\alpha_1\approx 9.54,\ \tilde\alpha_2\approx - 4.37. \nonumber
\end{align}
Similarly, in FIG.~\ref{fig_eefpn1_example}, we present a comparison between the asymptotic expansion in the limit $\Delta_e\to0$ and the corresponding numerical results of $(\alpha,\tilde\alpha,\theta,\tilde\theta)$.
Since the accuracy of the numerical results is $10^{-6}$, the errors show a slight deviation from the logarithmic slope expected from the corresponding expansion order when they approach the $10^{-6}$ level.
It is expected the summation like $\sum p^k \feJ{(p,a_1,b_1)}\feJ{(p,a_2,b_2)}$ would appear at next PN order, but one can also use this way to evaluate the asymptotic expansion.
Based on the current results, the asymptotic expressions remain reliable at least over the range $\Delta_e\le0.5$.
These results can provide useful reference for future resummation and phenomenological calibration in waveform modeling.

\section{Summary}
With the continuous advancement of GW detection in recent years, data analysis increasingly demands higher accuracy and more refined parameter dependence in waveform templates. 
In this work, we have presented two asymptotic methods for computing the Fourier modes of post-Newtonian binary waveforms in the high-eccentricity regime. 
Based on these developments, we have demonstrated two main applications. 
First, by incorporating new endpoint asymptotic information, we construct an endpoint-constrained analytic approximation that significantly accelerate waveform generation, while also providing practical foundation for future construction of high-eccentricity frequency-domain post-Newtonian templates.
Second, we derive the large-eccentricity asymptotic expansion of the eccentricity enhancement function, which is otherwise difficult to compute, thereby contributing to more accurate template modeling.

Nevertheless, frequency-domain GW modeling still faces several challenges. 
First, the dynamical parameters used in frequency-domain models differ from those in time-domain formulations. 
In particular, frequency-domain approaches require the inclusion of post-adiabatic corrections to properly account for radiation-reaction effects, and the discrepancy between the two approaches becomes more pronounced as the binary approaches merger.
Moreover, the frequency-domain framework relies heavily on analytic Fourier transforms, typically implemented through approximations such as the SPA and shifted uniform asymptotic (SUA) method~\cite{Klein_2013,Klein:2014bua}. 
For eccentric waveforms, this leads to expressions written as sums over Fourier modes $\hat{H}_p$. 
Although the present work provides a substantial acceleration in the computation of these modes, the number of required modes grows rapidly $e\to1$, still causing the computation impractical in this case.

One may envision a smooth transition in the frequency-domain waveform description from bound systems ($e<1$) to scattering systems ($e>1$), but existing methods are clearly insufficient to achieve this. 
These considerations indicate that the construction of frequency-domain GW templates in the high-eccentricity regime, as well as the development of a consistent framework bridging bound and scattering dynamics, will likely require fundamentally new approaches in future work.

\section*{Acknowledgments}
This work was supported in part by the National Key Research and Development Program of China Grant No.
2021YFC2203001 and in part by the NSFC (No.~12475046). XL was supported by the I+D grant PID2023-149018NB-C42 and the Grant IFT Centro de Excelencia Severo Ochoa No CEX2020-001007-S, funded by MCIN/AEI/10.13039/501100011033, the Leonardo Grant for Scientific Research and Cultural Creation 2024 from the BBVA Foundation, and Japan Society for JSPS KAKENHI Grant no. JP23H00110 and JP24K00624.

\begin{appendix}

\section{Asymptotic expansion of PN elliptic integrals}\label{app_UAERes_Deltae}
Here we show the asymptotic expansion results of all elliptic integrals $\feJ{(p,a,b)}$ that $b>0$. 
They are expanded in terms of $\Delta_e$.
Up to $\order{\Delta_e^4}$, the expansion reads
For $\feJ{(p,a,1)}$, we have
\begin{align}
    & \feJ{(p,-2,1)} = -\frac{1}{\pi}\int_0^\pi (\pi-z)(1-\cos{z})^2\sin{p(z-\sin{z})}\dd{z} + \frac{\Delta_e}{\pi}\int_0^\pi (1-\cos{z})\sin{z}\sin{p(z-\sin{z})}\dd{z} \nonumber\\
    &\qquad + \frac{\Delta_e^2}{8\pi} \int_0^\pi (\pi-z)(1-\cos{z})\big[ 4 p \sin {z} (\cos {z}-1) \cos {p (z-\sin {z})}-8 \cos {z} \sin {p (z-\sin {z})} \big]\dd{z}  \nonumber\\
    &\qquad + \frac{\Delta_e^3}{2\pi}\int_0^\pi \bigg( p(1-\cos{z})^2(1+\cos{z})\cos{p(z-\sin{z})} + \frac{1}{3}(1+4\cos{z})\sin{z}\sin{p(z-\sin{z})} \bigg)\dd{z}  \nonumber\\
    &\qquad + \frac{\Delta_e^4}{8\pi}\int_0^\pi (\pi-z)\Big[ \sin {p (z-\sin {z})} \big(p^2 (1-\cos {z}) \cos ^2{z} (\cos {z}+1)+p^2 (1-\cos {z}) (\cos {z}+1)  \nonumber\\
    &\qquad -2 \cos {z} \left(p^2 (1-\cos {z}) (\cos {z}+1)+1\right) \big)+p \sin {z} \left(3 \cos ^2{z}-2 \cos {z}-1\right) \cos {p (z-\sin {z})} \Big]\dd{z} + \order{\Delta_e^5}, \\
    & \feJ{(p,-1,1)} = -\frac{1}{\pi}\int_0^\pi (\pi-z)(1-\cos{z})\sin{p(z-\sin{z})}\dd{z} + \frac{\Delta_e}{\pi}\int_0^\pi \sin {z} \sin {p (z-\sin {z})} \dd{z} \nonumber\\
    &\qquad + \frac{\Delta_e^2}{2\pi}\int_0^\pi (\pi -z) \big(p \sin {z} (\cos {z}-1) \cos {p (z-\sin {z})}-\cos {z} \sin {p (z-\sin {z})}\big)\dd{z} \nonumber\\
    &\qquad + \frac{\Delta_e^3}{\pi}\int_0^\pi \bigg(\frac{1}{2} p\left(1- \cos ^2{z}\right) \cos {p (z-\sin {z})}+\frac{\sin {z} (\cos {z}+1) \sin {p (z-\sin {z})}}{6(1- \cos {z})}\bigg)\dd{z} \nonumber\\
    &\qquad - \frac{\Delta_e^4}{8\pi}\int_0^\pi (\pi-z)\Big[ \sin {p (z-\sin {z})} \left(-\left(p^2 (1-\cos {z}) (\cos {z}+1)\right)+p^2 (1-\cos {z}) \cos {z} (\cos {z}+1)+\cos {z}\right) \nonumber\\
    &\qquad +p \sin {z} (\cos {z}+1) \cos {p (z-\sin {z})} \Big]\dd{z}  + \order{\Delta_e^5},\\
    & \feJ{(p,0,1)} = -\frac{1}{\pi}\int_0^\pi (\pi-z)\sin{p(z-\sin{z})}\dd{z} - \frac{\Delta_e}{\pi}\int_0^\pi \bigg(\frac{\sin {z} \sin {p (z-\sin {z})}}{\cos {z}-1}\bigg)\dd{z} \nonumber\\
    &\qquad - \frac{\Delta_e^2}{2\pi}\int_0^\pi  p (\pi -z) \sin {z} \cos {p (z-\sin {z})}\dd{z} - \frac{\Delta_e^3}{\pi}\int_0^\pi \bigg(-\frac{1}{2} p (\cos {z}+1) \cos {p (z-\sin {z})} \nonumber\\
    &\qquad +\frac{\sin {z} (2 \cos {z}-1) \sin {p (z-\sin {z})}}{6 (\cos {z}-1)^2}\bigg)\dd{z} + \Delta_e^4\Bigg[ -\frac{5p}{24} + \frac{1}{8\pi}\int_0^\pi p (\pi -z) \sin {z} \cos{p (z-\sin {z})} \nonumber\\
    &\qquad -p \sin {z} \sin {p (z-\sin (z))} \Bigg]  + \order{\Delta_e^5}, \\
    & \feJ{(p,1,1)} = -\frac{1}{\pi}\int_0^\pi \frac{(\pi-z)\sin{p(z-\sin{z})}}{1-\cos{z}}\dd{z} + \frac{\Delta_e}{\pi}\int_0^\pi \frac{\sin{z}\sin{p(z-\sin{z})}}{(1-\cos{z})^2}\dd{z} + \Delta_e^2\Bigg[ -\frac{p}{6} + \frac{2p}{3}\ln{\frac{2\Delta_e}{\pi}} \nonumber\\
    &\qquad + \frac{1}{\pi}\int_0^\pi \bigg( -\frac{p (\pi -z) \sin {z} \cos {p (z-\sin {z})}}{2 (1-\cos {z})} + \frac{(\pi -z) \cos {z} \sin {p (z-\sin {z})}}{2 (1-\cos {z})^2} + \frac{2p\pi}{3z} \bigg) \dd{z} \Bigg] + \Delta_e^3\nonumber\\
    &\qquad \times \Bigg[ -\frac{p}{3} - \frac{10p}{9\pi^2}  + \frac{1}{\pi}\int_0^\pi \bigg( \frac{p (\cos {z}+1) \cos {p (z-\sin {z})}}{2 (1-\cos {z})}+\frac{\sin {z} (1-5 \cos {z}) \sin {p (z-\sin {z})}}{6 (1-\cos {z})^3}-\frac{10 p}{9 z^2} \bigg)\dd{z}\Bigg] \nonumber\\
    &\qquad + \Delta_e^4\Bigg[ \frac{53p}{360} + \frac{p}{3\pi^2} + \frac{2p}{5}\ln{\frac{2\Delta_e}{\pi}} + 
    \frac{1}{\pi}\int_0^\pi \bigg(\frac{p (\pi -z) \sin {z} (3 \cos {z}-1) \cos {p (z-\sin {z})}}{8 (\cos {z}-1)^2} \nonumber\\
    &\qquad +\frac{(\pi -z) \sin {p (z-\sin {z})} (p^2 \cos ^4{z}-2 p^2 \cos ^3{z}+(2 p^2-1) \cos {z}-p^2+3 \cos ^2{z})}{8 (\cos {z}-1)^3}+\frac{2 p (\pi  (3 z^2-5)+5 z)}{15 z^3}\bigg)\dd{z}
    \Bigg] \nonumber\\
    &\qquad + \order{\Delta_e^5},  \\
    & \feJ{(p,2,1)}=\frac{2p}{3}\ln{\frac{2\Delta_e}{\pi}} -\frac{p}{3} + \frac{1}{\pi}\int_0^\pi \bigg(\frac{(z-\pi ) \sin {p (z-\sin {z})}}{(\cos {z}-1)^2}+\frac{2 \pi  p}{3 z}\bigg)\dd{z} + \Delta_e\Bigg[ -\frac{4p}{3\pi^2} \nonumber\\
    &\qquad + \frac{1}{\pi}\int_0^\pi \bigg(-\frac{\sin {z} \sin {p (z-\sin {z})}}{(\cos {z}-1)^3}-\frac{4 p}{3 z^2}\bigg)\dd{z} \Bigg] + \Delta_e^2\Bigg[ \frac{17 p}{180}-\frac{p}{3 \pi ^2} + \frac{2p}{5}\ln{\frac{2\Delta_e}{\pi}} \nonumber\\
    &\qquad + \frac{1}{\pi}\int_0^\pi \bigg(-\frac{p (\pi -z) \sin {z} \cos {p (z-\sin {z})}}{2 (\cos {z}-1)^2}+\frac{(z-\pi ) \cos {z} \sin {p (z-\sin {z})}}{(\cos {z}-1)^3}+\frac{2 p \left(\pi  \left(3 z^2+5\right)-5 z\right)}{15 z^3}\bigg)\dd{z} \Bigg] \nonumber\\
    &\qquad + \Delta_e^3\Bigg[ -\frac{146 p}{135 \pi ^2}-\frac{8 p}{27 \pi ^4}-\frac{p}{9} + \frac{1}{\pi}\int_0^\pi \bigg(\frac{p (\cos {z}+1) \cos {p (z-\sin {z})}}{2 (\cos {z}-1)^2}+\frac{\sin {p (z-\sin {z})} (\sin {z}-8 \sin {z} \cos {z})}{6 (\cos {z}-1)^4}\nonumber\\
    &\qquad -\frac{2 p \left(73 z^2+60\right)}{135 z^4}\bigg)\dd{z} \Bigg] + \Delta_e^4\Bigg[-\frac{p}{5 \pi ^2}+\frac{p}{6 \pi ^4}+\frac{1601 p}{10080} + \frac{2p}{7}\ln{\frac{2\Delta_e}{\pi}} + \frac{1}{\pi}\int_0^\pi \bigg(\frac{p (\pi -z) (1-5 \cos {z})\sin {z} }{8 (\cos {z}-1)^3}\nonumber\\
    &\qquad \times \cos {p (z-\sin {z})} -\frac{(\pi -z) \sin {p (z-\sin {z})} \left(p^2 \cos ^4{z}-2 p^2 \cos ^3{z}+2 \left(p^2-1\right) \cos {z}-p^2+8 \cos ^2{z}\right)}{8 (\cos {z}-1)^4} \nonumber\\
    &\qquad +\frac{2 p \left(\pi  \left(5 z^4+7 z^2-35\right)-7 z \left(z^2-5\right)\right)}{35 z^5}\bigg)\dd{z} \Bigg] + \order{\Delta_e^5}, \\
    & \Delta_e^{2}\feJ{(p,3,1)}=-\frac{7p}{12}+\frac{p}{2}\Delta_e + \Delta_e^2\Bigg[ -\frac{2 p}{3 \pi ^2}-\frac{19 p}{90} + \frac{4p}{15}\ln{\frac{2\Delta_e}{\pi}} + \frac{1}{\pi}\int_0^\pi \bigg(\frac{(\pi -z) \sin {p (z-\sin {z})}}{(\cos {z}-1)^3} \nonumber\\
    &\qquad +\frac{4 p \left(\pi  \left(z^2+5\right)-5 z\right)}{15 z^3}\bigg)\dd{z} \Bigg] + \Delta_e^3 \Bigg[ -\frac{14 p}{45 \pi ^2}-\frac{8 p}{9 \pi ^4}+\frac{p}{6} + \frac{1}{\pi}\int_0^\pi \bigg(\frac{\sin {z} \sin {p (z-\sin {z})}}{(\cos {z}-1)^4}-\frac{2 p \left(7 z^2+60\right)}{45 z^4}\bigg)\dd{z} \Bigg] \nonumber\\
    &\qquad + \Delta_e^4\Bigg[ -\frac{3 p}{5 \pi ^2}-\frac{19 p}{420}+\frac{1}{\pi } + \frac{8p}{35}\ln{\frac{2\Delta_e}{\pi}} + \frac{1}{\pi}\int_0^\pi \bigg(\frac{p (\pi -z) \sin {z} \cos {p (z-\sin {z})}}{2 (\cos {z}-1)^3}+\frac{3 (\pi -z) \cos {z} \sin {p (z-\sin {z})}}{2 (\cos {z}-1)^4}\nonumber\\
    &\qquad +\frac{2 p \left(\pi  \left(4 z^2+21\right)-21 z\right)}{35 z^3}\bigg)\dd{z} \Bigg] + \order{\Delta_e^5}, \\
    & \Delta_e^{4}\feJ{(p,4,1)}=-\frac{19p}{36} + \frac{p}{3}\Delta_e - \frac{p}{5}\Delta_e^2 + \frac{5p}{18}\Delta_e^3 + \Delta_e^3\Bigg[ -\frac{17 p}{45 \pi ^2}-\frac{2 p}{9 \pi ^4}-\frac{239 p}{1512} + \frac{4p}{35}\ln{\frac{2\Delta_e}{35}}  \nonumber\\
    &\qquad + \frac{1}{\pi}\int_0^\pi \bigg(\frac{(z-\pi ) \sin {p (z-\sin {z})}}{(\cos {z}-1)^4}+\frac{2 p \left(\pi  \left(18 z^4+119 z^2+420\right)-7 z \left(17 z^2+60\right)\right)}{315 z^5}\bigg)\dd{z} \Bigg] + \order{\Delta_e^5}, \\
    &\Delta_e^{6}\feJ{(p,5,1)}=-\frac{185 p}{288}+\frac{3 p}{8}\Delta_e-\frac{p}{160}\Delta_e^2+\frac{p}{9}\Delta_e^3-\frac{9 p}{70}\Delta_e^4+ \order{\Delta_e^5}, \\
    & \Delta_e^{8}\feJ{(p,6,1)}=-\frac{427 p}{480}+\frac{p}{2}\Delta_e+\frac{1807 p}{7200}\Delta_e^2-\frac{p}{120}\Delta_e^3-\frac{p}{14}\Delta_e^4+ \order{\Delta_e^5}, \\
    & \Delta_e^{10}\feJ{(p,7,1)}=-\frac{637 p}{480}+\frac{35 p}{48}\Delta_e+\frac{147 p}{200}\Delta_e^2-\frac{31 p}{144}\Delta_e^3-\frac{67 p}{672}\Delta_e^4+ \order{\Delta_e^5}, \\
    & \Delta_e^{12}\feJ{(p,8,1)}=-\frac{1397 p}{672}+\frac{9 p}{8}\Delta_e+\frac{4757 p}{2800}\Delta_e^2-\frac{91 p}{144}\Delta_e^3-\frac{12091 p}{39200}\Delta_e^4+ \order{\Delta_e^5}, \\
    & \Delta_e^{14}\feJ{(p,9,1)}=-\frac{24167 p}{7168}+\frac{231 p}{128}\Delta_e+\frac{390181 p}{107520}\Delta_e^2-\frac{95 p}{64}\Delta_e^3-\frac{1232071 p}{1254400}\Delta_e^4+ \order{\Delta_e^5}, \\
    &\Delta_e^{16}\feJ{(p,10,1)}=-\frac{155155 p}{27648}+\frac{143 p}{48}\Delta_e+\frac{2412839 p}{322560}\Delta_e^2-\frac{3707 p}{1152}\Delta_e^3-\frac{6344459 p}{2257920}\Delta_e^4 + \order{\Delta_e^5}.
\end{align}
The asymptotic expansion results of $\feJ{(p,a,2)}$ and $\feJ{(p,a,3)}$ are
\begin{align}
    & \feJ{(p,-1,2)} = -\frac{1}{\pi}\int_0^\pi (\pi-z)^2(1-\cos{z})\cos{p(z-\sin{z})}\dd{z} + \frac{2\Delta_e}{\pi}\int_0^\pi (\pi-z)\sin{z}\cos{p(z-\sin{z})} \dd{z} \nonumber\\
    &\qquad + \frac{\Delta_e^2}{\pi}\int_0^\pi \bigg(\frac{1}{2} \left((\pi -z)^2 (-\cos {z})-2 \cos {z}-2\right) \cos {p (z-\sin {z})}-\frac{1}{2} p (\pi -z)^2 \sin {z} (\cos {z}-1) \sin {p (z-\sin {z})}\bigg)\dd{z} \nonumber\\
    &\qquad + \Delta_e^3\Bigg[ \frac{19}{9} - \frac{4}{3}\ln{\frac{2\Delta_e}{\pi}} + \frac{1}{\pi}\int_0^\pi \bigg(\frac{(\pi -z) \sin {z} (\cos {z}+1) \cos {p (z-\sin {z})}}{3 (1-\cos {z})}+p (\pi -z) \left(\cos ^2{z}-1\right) \sin {p (z-\sin {z})} \nonumber\\
    &\qquad -\frac{4 \pi }{3 z}\bigg)\dd{z} \Bigg] + \Delta_e^4\Bigg[ \frac{1}{4} + \frac{2}{3\pi^2} + \frac{1}{\pi}\int_0^\pi \bigg(\frac{\cos {p (z-\sin {z})}}{24 (\cos {z}-1)} \Big[\left(\left(6 p^2+3\right) z^2-6 \pi  \left(2 p^2 z+z\right)+\pi ^2 \left(6 p^2+3\right)+4\right) \cos {z} \nonumber\\
    &\qquad -3 p^2 (\pi -z)^2+3 p^2 (\pi -z)^2 \cos ^4{z}-6 p^2 (\pi -z)^2 \cos ^3{z}-\left(3 z^2-6 \pi  z+3 \pi ^2+4\right) \cos ^2{z}+8\Big] \nonumber\\
    &\qquad + \frac{1}{8} p \sin {p (z-\sin {z})} \left(\sin {z} \left(z^2+\left(z^2+\pi ^2+4\right) \cos {z}-2 \pi  z+\pi ^2+4\right)-\pi  z \sin (2 z)\right)+\frac{2}{3 z^2}\bigg)\dd{z} \Bigg] + \order{\Delta_e^5}, \\
    & \feJ{(p,0,2)} =  -\frac{1}{\pi}\int_0^\pi (\pi-z)^2\cos{p(z-\sin{z})} + \Delta_e\Bigg[4 - 4\ln{\frac{2\Delta_e}{\pi}} + \frac{1}{\pi}\int_0^\pi \bigg( \frac{2 (\pi -z) \sin {z} \cos {p (z-\sin {z})}}{1-\cos {z}}-\frac{4 \pi }{z}\bigg)\dd{z} \Bigg] \nonumber\\
    &\qquad \Delta_e^2\Bigg[ 1 + \frac{4}{\pi^2} + \frac{1}{\pi}\int_0^\pi \bigg(\frac{(\cos {z}+1) \cos {p (z-\sin {z})}}{\cos {z}-1}+\frac{1}{2} p (\pi -z)^2 \sin {z} \sin {p (z-\sin {z})}+\frac{4}{z^2}\bigg)\dd{z} \Bigg] \nonumber\\
    &\qquad + \Delta_e^3\Bigg[ -\frac{1}{6}-\frac{2}{3\pi^2}-\frac{4}{3}\ln{\frac{2\Delta_e}{\pi}}  + \frac{1}{\pi}\int_0^\pi \bigg(-\frac{(\pi -z) \sin {z} (2 \cos {z}-1) \cos {p (z-\sin {z})}}{3 (\cos {z}-1)^2} \nonumber\\
    &\qquad -p (\pi -z) (\cos {z}+1) \sin {p (z-\sin {z})}-\frac{4 \left(\pi  \left(z^2-1\right)+z\right)}{3 z^3}\bigg)\dd{z} \Bigg] + \Delta_e^4\Bigg[ \frac{2}{3}-\frac{8}{9\pi^4} + \frac{26}{9\pi^2} \nonumber\\
    &\qquad + \frac{1}{\pi}\int_0^\pi \bigg(\frac{\cos {p (z-\sin {z})}}{24 (\cos {z}-1)^2} \Big[3 p^2 (\pi -z)^2-3 p^2 (\pi -z)^2 \cos ^4{z}+6 p^2 (\pi -z)^2 \cos ^3{z}+\left(8-6 p^2 (\pi -z)^2\right) \cos {z} \nonumber\\
    &\qquad +16 \cos ^2{z}-8\Big]  + \frac{p \sin {p (z-\sin {z})} }{8 (\cos {z}-1)}\Big[\sin {z} \left(-z^2+\left(z^2-4\right) \cos {z}-4\right)+8 z \left(\cos ^2{z}-1\right)+\pi ^2 \sin {z} (\cos {z}-1) \nonumber\\
    &\qquad -2 \pi  (\cos {z}-1) (z \sin {z}+4 \cos {z}+4)\Big] +\frac{2 \left(13 z^2-12\right)}{9 z^4}\bigg)\dd{z} \Bigg] + \order{\Delta_e^5},\\
    & \Delta_e\feJ{(p,1,2)}=-\frac{\pi ^2}{3}+\Delta_e\Bigg[ 2-4\ln{\frac{2\Delta_e}{\pi}} - \frac{1}{\pi}\int_0^\pi \bigg(-\frac{(\pi -z)^2 \cos {p (z-\sin {z})}}{\cos {z}-1}-\frac{2 \pi  (\pi -2 z)}{z^2}\bigg)\dd{z} \Bigg] +\Delta_e^2\Bigg[\frac{5}{3}+\frac{4}{\pi ^2} \nonumber\\
    &\qquad - \frac{1}{\pi}\int_0^\pi \bigg(-\frac{2 (\pi -z) \sin {z} \cos {p (z-\sin {z})}}{(\cos {z}-1)^2}+\frac{8 (\pi -z)}{z^3}\bigg)\dd{z} \Bigg]+\Delta_e^3\Bigg[ -\frac{5}{6}+\frac{8}{3 \pi ^4}-\frac{4}{3 \pi ^2} - \frac{4}{3}\ln{\frac{2\Delta_e}{\pi}} \nonumber\\
    &\qquad - \frac{1}{\pi}\int_0^\pi \bigg(\frac{\left((\pi -z)^2 (-\cos {z})+2 \cos {z}+2\right) \cos {p (z-\sin {z})}}{2 (\cos {z}-1)^2}+\frac{p (\pi -z)^2 \sin {z} \sin {p (z-\sin {z})}}{2 (\cos {z}-1)} \nonumber\\
    &\qquad +\frac{2 \left(2 \pi  z-\pi ^2+4\right) \left(z^2-3\right)}{3 z^4}\bigg)\dd{z} \Bigg]  +\Delta_e^4\Bigg[ \frac{148}{135}-\frac{8}{9 \pi ^4}+\frac{26}{9 \pi ^2} - \frac{1}{\pi}\int_0^\pi \bigg(-\frac{(\pi -z) \sin {z} (5 \cos {z}-1) \cos {p (z-\sin {z})}}{3 (\cos {z}-1)^3} \nonumber\\
    &\qquad -\frac{p (\pi -z) (\cos {z}+1) \sin {p (z-\sin {z})}}{\cos {z}-1}+\frac{4 (\pi -z) \left(13 z^2-24\right)}{9 z^5}\bigg)\dd{z} \Bigg] + \order{\Delta_e^5}, \\
    & \Delta_e^{3}\feJ{(p,2,2)} = 2-\frac{\pi ^2}{3}+2\Delta_e-2\Delta_e^2+\Delta_e^3\Bigg[ 1+\frac{4}{3 \pi ^2} - \frac{1}{\pi}\int_0^\pi \bigg(\frac{(\pi -z)^2 \cos {p (z-\sin {z})}}{(\cos {z}-1)^2} \nonumber\\
    &\qquad -\frac{2 \left(6 z^2-2 \pi  \left(z^2+6\right) z+\pi ^2 \left(z^2+6\right)\right)}{3 z^4}\bigg)\dd{z} \Bigg]  +\Delta_e^4\Bigg[ -\frac{11}{180}+\frac{4}{3 \pi ^4}+\frac{2}{3 \pi ^2} - \frac{1}{\pi}\int_0^\pi \bigg(\frac{2 (\pi -z) \sin {z} \cos {p (z-\sin {z})}}{(\cos {z}-1)^3} \nonumber\\
    &\qquad +\frac{4 (\pi -z) \left(z^2+12\right)}{3 z^5}\bigg)\dd{z} \Bigg], \\
    & \Delta_e^{5}\feJ{(p,3,2)}=\frac{15}{4}-\frac{\pi ^2}{2}+\frac{3}{2}\Delta_e+\Delta_e^2\Bigg(\frac{\pi ^2}{6}-\frac{9}{4}\Bigg)+\frac{2}{3}\Delta_e^3-\Delta_e^4 + \order{\Delta_e^5}, \\
    &\Delta_e^{7}\feJ{(p,4,2)}=\frac{245}{36}-\frac{5 \pi ^2}{6}+\frac{5}{3}\Delta_e+\Delta_e^2\Bigg(\frac{\pi ^2}{2}-\frac{61}{12}\Bigg)+\frac{1}{18}\Delta_e^3-\frac{1}{3}\Delta_e^4 + \order{\Delta_e^5}, \\
    & \Delta_e^{9}\feJ{(p,5,2)}=\frac{7175}{576}-\frac{35 \pi ^2}{24}+\frac{35}{16}\Delta_e+\Delta_e^2\Bigg(\frac{5 \pi ^2}{4}-\frac{1145}{96}\Bigg)-\frac{85}{144}\Delta_e^3+\Delta_e^4\Bigg(\frac{205}{192}-\frac{\pi ^2}{8}\Bigg)+ \order{\Delta_e^5}, \\
    & \Delta_e^{11}\feJ{(p,6,2)}=\frac{36883}{1600}-\frac{21 \pi ^2}{8}+\frac{63}{20}\Delta_e+\Delta_e^2\Bigg(\frac{35 \pi ^2}{12}-\frac{39403}{1440}\Bigg)-\frac{413}{240}\Delta_e^3+\Delta_e^4\Bigg(\frac{5749}{960}-\frac{5 \pi ^2}{8}\Bigg)+ \order{\Delta_e^5}, \\
    & \feJ{(p,-1,3)} = \frac{1}{\pi}\int_0^\pi (\pi-z)^3(1-\cos{z})\sin{p(z-\sin{z})}\dd{z} - \frac{3\Delta_e}{\pi}\int_0^\pi (\pi-z)^2\sin{z}\sin{p(z-\sin{z})}\dd{z} \nonumber\\
    &\qquad - \frac{\Delta_e^2}{\pi}\int_0^\pi \bigg(\frac{1}{2} p (\pi -z)^3 \sin {z} (\cos {z}-1) \cos {p (z-\sin {z})}+\frac{1}{2} (\pi -z) \left(\left(z^2-2 \pi  z+\pi ^2+6\right) \cos {z}+6\right) \nonumber\\
    &\qquad \times \sin {p (z-\sin {z})}\bigg)\dd{z} + \frac{\Delta_e^3}{\pi}\int_0^\pi \bigg(\frac{3}{2} p (\pi -z)^2 \left(\cos ^2{z}-1\right) \cos {p (z-\sin {z})}+\frac{ \sin {z} (\cos {z}+1) \sin {p (z-\sin {z})}}{2 (\cos {z}-1)}\nonumber\\
    &\qquad \times (z^2-2 \pi  z+\pi ^2+2)\bigg)\dd{z} + \frac{\Delta_e^4}{\pi}\int_0^\pi \bigg(\frac{1}{8} p \left(-\left(\left(z^2+12\right) z\right)+3 \pi  \left(z^2+4\right)-3 \pi ^2 z+\pi ^3\right) \nonumber\\
    &\qquad \times \sin {z} (\cos {z}+1) \cos {p (z-\sin {z})}+\frac{(\pi -z) \sin {p (z-\sin {z})} }{8 (\cos {z}-1)} \Big[p^2 (\pi -z)^2-p^2 (\pi -z)^2 \cos ^4{z} \nonumber\\
    &\qquad +2 p^2 (\pi -z)^2 \cos ^3{z}-2 p^2 (\pi -z)^2 \cos {z}+(\pi -z)^2 \cos ^2{z}+4 \cos ^2{z}-(\pi -z)^2 \cos {z}-4 \cos {z}-8\Big] \bigg)\dd{z} + \order{\Delta_e^5}, \\
    & \feJ{(p,0,3)} = \frac{1}{\pi}\int_0^\pi (\pi-z)^3\sin{p(z-\sin{z})}\dd{z} - \frac{3\Delta_e}{\pi}\int_0^\pi \frac{(\pi-z)^2\sin{z}\sin{p(z-\sin{z})}}{1-\cos{z}} \dd{z} \nonumber\\
    &\qquad + \frac{\Delta_e^2}{\pi}\int_0^\pi \bigg(\frac{1}{2} p (\pi -z)^3 \sin {z} \cos {p (z-\sin {z})}-\frac{3 (\pi -z) (\cos {z}+1) \sin {p (z-\sin {z})}}{\cos {z}-1}\bigg)\dd{z} \nonumber\\
    &\qquad + \frac{\Delta_e^3}{\pi}\int_0^\pi \bigg(-\frac{3}{2} p (\pi -z)^2 (\cos {z}+1) \cos {p (z-\sin {z})}+\frac{\sin {z}  \sin {p (z-\sin {z})}}{2 (\cos {z}-1)^2} \big[-(\pi -z)^2+2 (\pi -z)^2 \cos {z} \nonumber\\
    &\qquad -2 \cos {z}-2\big] \bigg)\dd{z} + \Delta_e^4\Bigg[ -\frac{14}{3}\ln{\frac{2\Delta_e}{\pi}} + \frac{139p}{18} + \frac{5p\pi^2}{24} + \frac{1}{\pi}\int_0^\pi \bigg(\frac{p (\pi -z) \sin {z} \cos {p (z-\sin {z})}}{8 (\cos {z}-1)}\Big[-z^2 \nonumber\\
    &\qquad +\left(z^2-2 \pi  z+\pi ^2-12\right) \cos {z}+2 \pi  z-\pi ^2-12\Big]+\frac{(\pi -z) \sin {p (z-\sin {z})}}{8 (\cos {z}-1)^3}\Big[\left(\cos ^2{z}-1\right) \nonumber\\
    &\qquad \times \left(p^2 (\pi -z)^2 (\cos {z}-1)^3+8\right)+8 \sin {z} \sin (2 z)\Big]-\frac{14 \pi  p}{3 z}\bigg)\dd{z} \Bigg] + \order{\Delta_e^5}, \\
    & \feJ{(p,1,3)}=\frac{1}{\pi}\int_0^\pi \bigg(-\frac{(\pi -z)^3 \sin {p (z-\sin {z})}}{\cos {z}-1}\bigg)\dd{z}+\frac{\Delta_e}{\pi}\int_0^\pi \bigg(-\frac{3 (\pi -z)^2 \sin {z} \sin {p (z-\sin {z})}}{(\cos {z}-1)^2}\bigg)\dd{z} \nonumber\\
    &\qquad +\Delta_e^2\Bigg[ \frac{1}{6} p \left(-4 \left(6+\pi ^2\right) \ln {\frac{2 \Delta_e }{\pi }}-24 \zeta_R
   (3)+\pi ^2+36\right) + \frac{1}{\pi}\int_0^\pi \bigg(-\frac{p (\pi -z)^3 \sin {z} \cos {p (z-\sin {z})}}{2 (\cos {z}-1)} \nonumber\\
   &\qquad -\frac{(\pi -z) \left(\left(z^2-2 \pi  z+\pi ^2-6\right) \cos {z}-6\right) \sin {p (z-\sin {z})}}{2 (\cos {z}-1)^2}-\frac{2 \pi  \left(6+\pi ^2\right) p}{3 z}\bigg)\dd{z} \Bigg]+\Delta_e^3\Bigg[ \frac{\pi ^2 p}{3}+\frac{8 p}{3 \pi ^2}+\frac{104 p}{9} \nonumber\\
   &\qquad -\frac{20}{3} p \ln {\frac{2 \Delta_e }{\pi }} + \frac{1}{\pi}\int_0^\pi \bigg(\frac{3 p (\pi -z)^2 (\cos {z}+1) \cos {p (z-\sin {z})}}{2 (\cos {z}-1)}+\frac{\sin {z} \left((\pi -z)^2-5 (\pi -z)^2 \cos {z}+2 \cos {z}+2\right) }{2 (\cos {z}-1)^3}\nonumber\\
   &\qquad \times \sin {p (z-\sin {z})}+\frac{2 p \left(-10 \pi  z+5 \pi ^2+4\right)}{3 z^2}\bigg)\dd{z} \Bigg]+\Delta_e^4\Bigg[ -\frac{12 p \zeta_R (3)}{5}-\frac{53 \pi ^2 p}{360}+\frac{8 p}{3 \pi ^2}+\frac{197 p}{180} -\frac{2p\pi^2}{5}\ln{\frac{2\Delta_e}{\pi}} \nonumber\\
   &\qquad + \frac{1}{\pi}\int_0^\pi \bigg(-\frac{p (\pi -z) \left(\sin {z} \left(z^2+3 \left(z^2-2 \pi  z+\pi ^2-4\right) \cos ^2{z}-2 \pi  z+\pi ^2+12\right)-2 (\pi -z)^2 \sin (2 z)\right) \cos {p (z-\sin {z})}}{8 (\cos {z}-1)^3} \nonumber\\
   &\qquad -\frac{(\pi -z) \sin {p (z-\sin {z})}}{8 (\cos {z}-1)^3}\Big[\left(\left(2 p^2-1\right) z^2+\pi  \left(2-4 p^2\right) z+\pi ^2 \left(2 p^2-1\right)-20\right) \cos {z}-p^2 (\pi -z)^2+p^2 (\pi -z)^2 \cos ^4{z}\nonumber\\
   &\qquad -2 p^2 (\pi -z)^2 \cos ^3{z}+\left(3 z^2-6 \pi  z+3 \pi ^2-28\right) \cos ^2{z}+8\Big]-\frac{2 p \left(\pi ^3 \left(3 z^2-5\right)+15 \pi ^2 z-40 z+40 \pi \right)}{15 z^3}\bigg)\dd{z} \Bigg] \nonumber\\
   &\qquad + \order{\Delta_e^5}, \\
    & \feJ{(p,2,3)}=\frac{\pi ^2 p}{3}-2 p-4p\zeta_R(3)-\frac{2p\pi^2}{3}\ln{\frac{2\Delta_e}{\pi}} + \frac{1}{\pi}\int_0^\pi \bigg(\frac{(\pi -z)^3 \sin {p (z-\sin {z})}}{(\cos {z}-1)^2}-\frac{2 \pi ^3 p}{3 z}\bigg)\dd{z} +\Delta_e\Bigg[ 10p-8p\ln{\frac{2\Delta_e}{\pi}}  \nonumber\\
    &\qquad + \frac{1}{\pi}\int_0^\pi \bigg(\frac{3 (\pi -z)^2 \sin {z} \sin {p (z-\sin {z})}}{(\cos {z}-1)^3} +\frac{4 \pi  p (\pi -2 z)}{z^2}\bigg)\dd{z} \Bigg]+\Delta_e^2\Bigg[ -\frac{12 p \zeta (3)}{5}-\frac{17 \pi ^2 p}{180}+\frac{4 p}{\pi ^2}+\frac{23 p}{15} \nonumber\\
    &\qquad - \frac{2p}{5}(\pi^2+4)\ln{\frac{2\Delta_e}{\pi}} + \frac{1}{\pi}\int_0^\pi \bigg(-\frac{3 p (\pi -z)^2 (\cos {z}+1) \cos {p (z-\sin {z})}}{2 (\cos {z}-1)^2}+\frac{\sin {z} \sin {p (z-\sin {z})}}{2 (\cos {z}-1)^4}\Big[ -(\pi -z)^2 \nonumber\\
    &\qquad +8 (\pi -z)^2 \cos {z}-2 \cos {z}-2\Big] +\frac{2 p \left(44 z^2-2 \pi  \left(73 z^2+60\right) z+\pi ^2 \left(73 z^2+60\right)+120\right)}{45 z^4}\bigg)\dd{z} \Bigg]
    +\Delta_e^3\Bigg[ \frac{\pi ^2 p}{9}+\frac{8 p}{45 \pi ^2} \nonumber\\
    &\qquad +\frac{16 p}{9 \pi ^4}+\frac{1021 p}{135} -\frac{292p}{45}\ln{\frac{2\Delta_e}{\pi}} + \frac{1}{\pi}\int_0^\pi \bigg(-\frac{3 p (\pi -z)^2 (\cos {z}+1) \cos {p (z-\sin {z})}}{2 (\cos {z}-1)^2}+\frac{\sin {z} \sin {p (z-\sin {z})}}{2 (\cos {z}-1)^4} \nonumber\\
    &\qquad \times \Big[-(\pi -z)^2+8 (\pi -z)^2 \cos {z}-2 \cos {z}-2\Big] + \frac{2 p \left(44 z^2-2 \pi  \left(73 z^2+60\right) z+\pi ^2 \left(73 z^2+60\right)+120\right)}{45 z^4}\bigg)\dd{z} \Bigg] \nonumber\\
    &\qquad +\Delta_e^4\Bigg[ -\frac{12 p \zeta_R (3)}{7}-\frac{1601 \pi ^2 p}{10080}+\frac{343 p}{90 \pi ^2}+\frac{2 p}{9 \pi
   ^4}+\frac{649 p}{1890}-\frac{2p}{35}(5\pi^2+12)\ln{\frac{2\Delta_e}{\pi}} + \frac{1}{\pi}\int_0^\pi \bigg(\frac{p (\pi -z) \sin {z}  }{8 (\cos {z}-1)^3}\nonumber\\
   &\qquad \times \cos {p (z-\sin {z})}\Big[-z^2+\left(5 z^2-10 \pi  z+5 \pi ^2-12\right) \cos {z}+2 \pi  z-\pi ^2-12\Big] + \frac{(\pi -z) \sin {p (z-\sin {z})} }{8 (\cos {z}-1)^5} \nonumber\\
   &\qquad\times \Big[(\cos {z}-1) \big(2 \left(\left(p^2-1\right) z^2-2 \pi  \left(p^2-1\right) z+\pi ^2 \left(p^2-1\right)-8\right) \cos {z}-p^2 (\pi -z)^2+p^2 (\pi -z)^2 \cos ^4{z} \nonumber\\
   &\qquad -2 p^2 (\pi -z)^2 \cos ^3{z}+8 \left(z^2-2 \pi  z+\pi ^2-3\right) \cos ^2{z}+8\big)+8 \sin {z} \sin (2 z)\Big]-\frac{2 p }{315 z^5} \Big[-189 \pi ^2 z \left(z^2-5\right) \nonumber\\
   &\qquad -7 z \left(149 z^2+60\right)+9 \pi ^3 \left(5 z^4+7 z^2-35\right)+\pi  \left(108 z^4+413 z^2+420\right)\Big] \bigg)\dd{z} \Bigg].
\end{align}

Some of the expansion coefficients can only be represented in terms of integrals. Here we provide the asymptotic expansions of these coefficients in the limit $p\to\infty$,


\section{The uniform asymptotic expansion of PN elliptic integrals} \label{app_UAE_Integrals}
\subsection{$\feJ{(p,a,0)}$}\label{app_UAE_Integrals_Jpa0}
\begin{align}
    & \feJ{(p,1,0)} = p^{1/3}h_0^2\Bigg[ V_{-1}(y) + \frac{(h_0^3-2)}{6p^{2/3}\zeta}V_0(y) + \order{p^{-4/3}} \Bigg], \\
    & \feJ{(p,2,0)} = ph_0^3\Bigg[ \frac{1}{2y}\big( V_{-1}(y) - V_0'(y) \big) + \frac{h_0^3-2}{4p^{2/3}\zeta}V_{-1}(y) + \order{p^{-4/3}} \Bigg], \\
    & \feJ{(p,3,0)} = p^{5/3}h_0^4\Bigg[ -\frac{1}{4y^2}\big( V'_0(y) - V_{-1}(y) + yV'_{-1}(y) \big) - \frac{h_0^3-2}{6p^{2/3}y\zeta}\big( V'_0(y) - V_{-1}(y) \big)  + \order{p^{-4/3}} \Bigg], \\
    & \feJ{(p,4,0)} = p^{7/3}h_0^5\Bigg[ \frac{1}{24y^3}\big( -5V'_0(y) + 5V_{-1}(y) + y(-5V'_{-1}(y) + 2yV_0(y)) \big) \nonumber\\
    &\qquad - \frac{5(h_0^3-2)}{48p^{2/3}y^2\zeta}\big( V'_0(y) - V_{-1}(y) + yV'_{-1}(y) \big)  + \order{p^{-4/3}} \Bigg], \\
    & \feJ{(p,5,0)} = p^3h_0^6\Bigg[ \frac{1}{192y^4}\Big( -35V'_0(y) + 7y\big( -5V'_{-1}(y) + 2yV_0(y) \big) + (6y^3+35)V_{-1}(y) \Big) \nonumber\\
    &\qquad + \frac{h_0^3-2}{48p^{2/3}\zeta}\Big( -5V'_0(y) + 5V_{-1}(y) + y\big( -5V'_{-1}(y) + 2yV_0(y) \big) \Big) + \order{p^{-4/3}} \Bigg], \\
    & \feJ{(p,6,0)} = p^{11/3}h_0^7\Bigg[ \frac{h_0^7}{1920y^5}\Big[ 63y\big( -5V'_{-1}(y) + 2yV_0(y) \big) + 35(2y^3+9)V_{-1}(y) - (16y^3+315)V'_0(y) \Big] \nonumber\\
    &\qquad + \frac{7(h_0^3-2)}{2304p^{2/3}y^4\zeta}\Big[ -35V'_0(y) + 7y\big( -5V'_{-1}(y) + 2yV_0(y) \big) + (6y^3+35)V_{-1}(y) \Big] + \order{p^{-4/3}} \Bigg], \\
    & \feJ{(p,7,0)} = p^{13/3}h_0^8\Bigg[ -\frac{1}{23040y^6}\Big[ -5(166y^3+693)V_{-1}(y) + (236y^3+3465)V'_0(y) + 3y\big( -462yV_0(y) \nonumber\\
    &\quad + 5(4y^3+231)V'_{-1}(y) \big) \Big] - \frac{h_0^3-2}{2880p^{2/3}y^5\zeta}\Big[ 63y\big( 5V'_{-1}(y) - 2yV_0(y) \big) - 35(2y^3+9)V_{-1}(y) + (16y^3+315)V'_0(y) \Big] \nonumber\\
    &\qquad + \order{p^{-4/3}} \Bigg], \\
    & \feJ{(p,8,0)} = p^5h_0^9\Bigg[ \frac{1}{322560y^7}\Big[ -315y(4y^3+143)V'_{-1}(y) + 6y^2(32y^3+3003)V_0(y) + 35(322y^3+1287)V_{-1}(y) \nonumber\\
    &\qquad - (3548y^3+45045)V'_0(y) \Big] - \frac{h_0^3-2}{30720p^{2/3}y^6\zeta}\Big[ -5(166y^3+693)V_{-1}(y) + (236y^3+3465)V'_0(y) \nonumber\\
    &\qquad + 3y\big( -462yV_0(y) + 5(4y^3+231)V'_{-1}(y) \big) \Big]  + \order{p^{-4/3}} \Bigg], \\
    & \feJ{(p,9,0)} = p^{17/3}h_0^{10}\Bigg[ \frac{1}{1032192y^8}\Big[ -(11624y^3+135135)V'_0(y) + 7(24y^6+4970y^3+19305)V_{-1}(y) \nonumber\\
    &\qquad + y\big( 22y(44y^3+2457)V_0(y) - 35(136y^3 + 3861)V'_{-1}(y) \big) \Big] - \frac{h_0^3-2}{387072p^{2/3}y^7\zeta}\Big[ 315y(4y^3+143)V'_{-1}(y) \nonumber\\
    &\qquad - 6y^2(32y^3+3003)V_0(y) - 35(322y^3+1287)V_{-1}(y) + (3548y^3+45045)V'_0(y) \Big]  + \order{p^{-4/3}} \Bigg], \\
    &\feJ{(p,10,0)} = p^{19/3}h_0^{11}\Bigg[ \frac{1}{92897280y^9}\Big[ 385(72y^6+7838y^3+29835)V_{-1}(y) - (3072y^6 + 1048520y^3 + 11486475)V'_0(y) \nonumber\\
    &\qquad + y\big( -385(1208y^3+29835)V'_{-1}(y) + 2y(53236y^3 + 2297295)V_0(y) \big) \Big]  - \frac{11(h_0^3-2)}{12386304p^{2/3}y^8\zeta} \Big[ (11624y^3+ 135135) \nonumber\\
    &\qquad \times V'_0(y) - 7(24y^6 + 4970y^3 + 19305)V_{-1}(y) + y\big( -22y(44y^3+2457)V_0(y) + 35(136y^3+3861)V'_{-1}(y) \big) \Big] - \nonumber\\
    &\qquad  + \order{p^{-4/3}}\Bigg], \\
    & \feJ{(p,11,0)} = p^7h_0^{12}\Bigg[ \frac{1}{371589120y^{10}}\Big[ 7(21024y^6+1663090y^3+6235515)V_{-1}(y) - (23568y^6+4159012y^3 \nonumber\\
    &\qquad +43648605)V'_0(y) + y\big( 26y(18248y^3+671517)V_0(y) - 7(432y^6+277420y^3+6235515)V'_{-1}(y) \big) \Big] \nonumber\\
    &\qquad - \frac{h_0^3-2}{92897280p^{2/3}y^9\zeta}\Big[ -385(72y^6 + 7838y^3 + 29835)V_{-1}(y) + (3072y^6 + 1048520y^3 + 11486475)V'_0(y) \nonumber\\
    &\qquad + y\big( 385(1208y^3 + 29835)V'_{-1}(y) - 2y(53236y^3 + 2297295)V_0(y) \big) \Big]  + \order{p^{-4/3}} \Bigg], \\
    & \feJ{(p,12,0)} = p^{23/3}h_0^{13}\Bigg[ \frac{1}{8174960640y^{11}}\Big[ -3003y(48y^6+14540y^3+305235)V'_{-1}(y) + 6y^2(2048y^6 + 1852760y^3 \nonumber\\
    &\qquad + 61108047)V_0(y) + 1001(3808y^6 + 247110y^3 + 915705)V_{-1}(y) - (722000y^6+90222132y^3+916620705)V'_0(y)\Big] \nonumber\\
    &\qquad - \frac{13(h_0^3-2)}{4459069440p^{2/3}y^{10}\zeta}\Big[ -7(21024y^6 + 1663090y^3 + 6235515)V_{-1}(y) + (23568y^6 + 4159012y^3 + 43648605)V'_0(y) \nonumber\\
    &\qquad + y\big( -26y(18248y^3+671517)V_0(y) + 7(432y^6 + 277420y^3 + 6235515)V'_{-1}(y) \big) \Big]  + \order{p^{-4/3}} \Bigg], \\
    & \feJ{(p,13,0)} = p^{25/3}h_0^{14}\Bigg[ \frac{1}{28028436480y^{12}}\Big[ -(3029827y^6 + 304088928y^3 + 3011753745)V'_0(y) + 11(846y^9 + 1317512y^6 \nonumber\\
    &\qquad + 74580870y^3 + 273795795)V_{-1}(y) + 3y\big( -77V'_{-1}(y)(3216y^6 + 654160y^3 + 13037895)V'_{-1}(y) + 2y(15856y^6 \nonumber\\
    &\qquad + 6597292y^3 + 200783583)V_0(y) \big) \Big] - \frac{h_0^3-2}{7007109120p^{2/3}y^{11}\zeta}\Big[ 3003y(48y^6+14540y^3+30235)V'_{-1}(y) \nonumber\\
    &\qquad - 6y^2(2048y^6+1852760y^3+61108047)V_0(y) - 1001(3808y^6+247110y^3+915705)V_{-1}(y) + (722000y^6 \nonumber\\
    &\qquad + 90222132y^3 + 916620705)V'_0(y) \Big]  + \order{p^{-4/3}} \Bigg], \\
    &\feJ{(p,14,0)} = p^9h_0^{15}\Bigg[ \frac{1}{5101175439360y^{13}}\Big[ 5005(864 y^9 + 564616 y^6 + 28905270 y^3 + 105306075)V_{-1}(y) - (294912 y^9 \nonumber\\
    &\qquad + 630885520 y^6 + 54318264000 y^3 + 527056905375)V'_0(y) + 3y\big( -5005(35102025 + 1834640 y^3 + 11632 y^6)V'_{-1}(y) \nonumber\\
    &\qquad + 2y(4478352 y^6 + 1228039540 y^3 + 35137127025)V_0(y) \big) \Big] - \frac{h_0^3-2}{22422749184p^{2/3}y^{12}\zeta}\Big[ (3029872y^6 + 304088928y^3 \nonumber\\
    &\qquad + 3011753745)V'_0(y) - 11(864y^9 + 1317512y^6 + 74580870y^3 + 273795795)V_{-1}(y) + 3y\big( 77(3216y^6 + 654160y^3 \nonumber\\
    &\qquad + 13037895)V'_{-1}(y) - 2y(15856y^6 + 6597292y^3 + 200783583)V_0(y) \big) \Big] + \order{p^{-4/3}} \Bigg].
\end{align}

\subsection{$\feK{(p,a,0)}$}\label{app_UAE_Integrals_Kpa0}
\begin{align}
    & \feK{(p,1,0)} = p^{1/3}h_0^2\Bigg[ U_{-1}(y) - \ln{h_0}V_{-1}(y) + \frac{h_0^3-2}{12p^{2/3}\zeta}\big( U_0(y) - V_0(y) - 2\ln{h_0}V_0(y) \big) \nonumber\\
    &\qquad + \ln{p}\bigg[ -\frac{2}{3}V_{-1}(y) - \frac{h_0^3-2}{9p^{2/3}\zeta}V_0(y) \bigg] + \order{p^{-4/3}} \Bigg], \\
    & \feK{(p,2,0)} = ph_0^3\Bigg[ -\frac{1}{2y}\Big[ U'_0(y) + V'_0(y) - U_{-1}(y) + \ln{h_0}\big( -V'_0(y) + V_{-1}(y) \big) \Big] + \frac{h_0^3-2}{12p^{2/3}\zeta}\big( 3U_{-1}(y) - V_{-1}(y) \nonumber\\
    &\qquad - 3\ln{h_0}V_{-1}(y) \big) + \ln{p}\bigg[ \frac{1}{3y}\big( V'_0(y) - V_{-1}(y) \big) - \frac{h_0^3-2}{6p^{2/3}\zeta}V_{-1}(y) \bigg] + \order{p^{-4/3}} \Bigg], \\
    & \feK{(p,3,0)} = p^{5/3}h_0^4\Bigg[ -\frac{1}{8y^2}\Big[ 2U'_0(y) + 2V'_0(y) - 2U_{-1}(y) - 2yU'_{-1}(y) + yV'_{-1}(y) - 2\ln{h_0}\big( V'_0(y) - V_{-1}(y) + yV'_{-1}(y) \big) \Big] \nonumber\\
    &\qquad - \frac{h_0^3-2}{24p^{2/3}y\zeta}\Big[ 4U'_0(y) + 3V'_0(y) - 4U_{-1}(y) + V_{-1}(y) - 4\ln{h_0}\big( V'_0(y) - V_{-1}(y) \big) \Big] \nonumber\\
    &\qquad + \ln{p}\bigg[ \frac{1}{6h^2}\big( V'_0(y) - V_{-1}(y) + yV'_{-1}(y) \big) + \frac{h_0^3-2}{9p^{2/3}y\zeta}\big( V'_0(y) - V_{-1}(y) \big) \bigg] + \order{p^{-4/3}} \Bigg], \\
    & \feK{(p,4,0)} = p^{7/3}h_0^5\Bigg[ \frac{1}{144y^3}\Big[ 30 U'_0(y) + 28 V'_0(y) - 30 U_{-1}(y) + 2 V_{-1}(y) + 30 y U'_{-1}(y) + 13 y V'_{-1}(y) - 12 y^2 U_0(y) \nonumber\\
    &\qquad - 16 y^2 V_0(y) - 6 \ln{h_0} \big(5 V'_0(y) - 5 V_{-1}(y) + 5 V'_{-1}(y) y - 2 y^2V_0(y) \big) \Big] + \frac{h_0^3-2}{96p^{2/3}y^2\zeta}\Big[ 10 U'_0(y) + 8 V'_0(y) - 10 U_{-1}(y) \nonumber\\
    &\quad + 2 V_{-1}(y) + 10  yU'_{-1}(y) + 3 yV'_{-1}(y)  - 10 \ln{h_0} (V'_0(y) - V_{-1}(y) + yV'_{-1}(y) ) \Big] \nonumber\\
    &\qquad + \ln{p}\bigg[ \frac{1}{36y^3}\Big[5 V'_0(y) - 5 V_{-1}(y) + y \big(5 V'_{-1}(y) - 2 yV_0(y) \big)\Big] + \frac{5(h_0^3-2)}{72p^{2/3}y^2\zeta}\big( V'_0(y) - V_{-1}(y) + yV'_{-1}(y) \big) \bigg] + \order{p^{-4/3}} \Bigg], \\
    & \feK{(p,5,0)} = p^3h_0^6\Bigg[ \frac{1}{2304y^4}\Big[ -420 U'_0(y) - 377 V'_0(y) + 420 U_{-1}(y) - 43 V_{-1}(y) - 420 y U'_{-1}(y)  - 167  y V'_{-1}(y) \nonumber\\
    &\qquad + 168y^2 U_0(y)  + 218 y^2V_0(y)  + 72y^3 U_{-1}(y)  + 54y^3 V_{-1}(y)  + 12 \ln{h_0} \big(35 V'_0(y) + 7 y \big(5 V'_{-1}(y) - 2 yV_0(y) \big) \nonumber\\
    &\qquad - V_{-1}(y) (35 + 6 y^3) \big) \Big] - \frac{h_0^3-2}{288p^{2/3}y^3\zeta}\Big[ 30 U'_0(y) + 23 V'_0(y) - 30 U_{-1}(y) + 7 V_{-1}(y) + 30 y U'_{-1}(y)  + 8y V'_{-1}(y)  \nonumber\\
    &\qquad - 12 y^2 U_0(y)  - 14  y^2V_0(y)  - 6 \ln{h_0} \big(5 V'_0(y) - 5 V_{-1}(y) + 5  yV'_{-1}(y) - 2 y^2 V_0(y) \big) \Big] + \ln{p}\bigg[ \frac{1}{288y^4}\Big( 35 V'_0(y) \nonumber\\
    &\qquad + 7 y \big(5 V'_{-1}(y) - 2y V_0(y) \big) - (35 + 6 y^3) V_{-1}(y) \Big) + \frac{h_0^3 - 2}{72p^{2/3}y^3\zeta}\Big( 5 V'_0(y) - 5 V_{-1}(y) + y \big(5 V'_{-1}(y) - 2 y V_0(y) \big) \Big) \bigg] \nonumber\\
    &\qquad + \order{p^{-4/3}} \Bigg], \\
    & \feK{(p,6,0)} = p^{11/3}h_0^7\Bigg[ \frac{1}{115200y^5}\Big[ 16545 V'_0(y) - 18900 U_{-1}(y) + 2355 V_{-1}(y) + 18900  yU'_{-1}(y) + 7095 yV'_{-1}(y)  - 7560y^2 U_0(y) \nonumber\\
    &\qquad - 9642 y^2 V_0(y)  + 1472 y^3 V'_0(y)  - 4200 y^3 U_{-1}(y)  - 2870 y^3 V_{-1}(y)  + 60 U'_0(y) (315 + 16 y^3) - 60 \ln{h_0} \Big(63 y \big(5 V'_{-1}(y) \nonumber\\
    &\qquad - 2 y V_0(y) \big) - 35 V_{-1}(y) (9 + 2 y^3) + V'_0(y) (315 + 16 y^3) \Big) \Big] + \frac{h_0^3-2}{27648p^{2/3}y^4\zeta}\Big[ -2940 U'_0(y) - 2219 V'_0(y) \nonumber\\
    &\qquad + 2940 U_{-1}(y) - 721 V_{-1}(y) - 2940 y U'_{-1}(y)  - 749 y V'_{-1}(y)  + 1176 y^2 U_0(y)  + 1358 y^2 V_0(y)  + 504 y^3 U_{-1}(y) \nonumber\\
    &\qquad + 306 y^3 V_{-1}(y)  + 84 \ln{h_0} \Big(35 V'_0(y) + 7 y \big(5 V'_{-1}(y) - 2y V_0(y)  \big) - V_{-1}(y) (35 + 6 y^3) \Big) \Big] + \nonumber\\
    &\qquad + \ln{p}\bigg[ \frac{1}{2880y^5}\Big( 63 y \big(5 V'_{-1}(y) - 2  y V_0(y) \big) - 35 (9 + 2 y^3) V_{-1}(y)  + (315 + 16 y^3) V'_0(y)  \Big) + \frac{7(h_0^3-2)}{3456p^{2/3}y^4\zeta}\Big( 35 V'_0(y) \nonumber\\
    &\qquad + 7 y \big(5 V'_{-1}(y) - 2y V_0(y) \big) - (35 + 6 y^3) V_{-1}(y)  \Big) \bigg] + \order{p^{-4/3}} \Bigg].
\end{align}

\subsection{$\feJ{(p,a,b)}$}\label{app_UAE_Integrals_Jpab}


\end{appendix}

\bibliography{refs}

\end{document}